\documentclass[aps,prb,reprint]{revtex4-1}
\usepackage{graphicx}
\usepackage{amssymb}
\usepackage{amsmath}
\usepackage{slashed}
\usepackage{subcaption}

\usepackage{lipsum}
\usepackage[dvipsnames]{xcolor}
\usepackage{xspace} 
\usepackage{perpage} 
\usepackage{bm}
\usepackage{lineno}
\usepackage[normalem]{ulem}
\usepackage[toc,page]{appendix}
\MakePerPage{footnote} 
\usepackage{leftidx} 

\usepackage[figuresleft]{rotating}

\usepackage[T1]{fontenc}      
\usepackage{ae} 

\usepackage{braket}
\usepackage{verbatim}
\usepackage{grffile}
\usepackage{dsfont}

\setcitestyle{super}

\newcommand*{\citen}{}
\DeclareRobustCommand*{\citen}[1]{%
  \begingroup
    \romannumeral-`\x 
    \setcitestyle{numbers}%
    \cite{#1}%
  \endgroup
}

\newcommand{\im}{\ensuremath{\mathrm{i}}}

\newcommand{\Nt}{\ensuremath{N_\mathrm{leg}^\mathrm{t}}\xspace}
\newcommand{\NT}{\ensuremath{N_\mathrm{leg}^\mathrm{T}}\xspace}

\newcommand{\Lt}{\ensuremath{L_\mathrm{leg}^\mathrm{t}}\xspace}
\newcommand{\LT}{\ensuremath{L_\mathrm{leg}^\mathrm{T}}\xspace}

\newcommand{\muB}{\ensuremath{\mu_\mathrm{B}}}
\newcommand{\tstep}{\ensuremath{t_\mathrm{step}}}

\newcommand{\Ngate}{\ensuremath{N_\mathrm{gate}}}

\newcommand{\Vgate}{\ensuremath{\mu_\mathrm{gate}}}

\newcommand{\ta}{\ensuremath{\mathrm{t_\mathrm{hop}}/a^2}}

\newcommand{\Dmu}{\ensuremath{\Delta \mu \ }}

\begin{document}

\title{Topological Quantum Computing Using Nanowire Devices}
\author{C.~Tutschku$^{1}$}
\author{R.~W.~Reinthaler$^1$}
\author{C.~Lei$^{2,3}$}
\author{A.~H.~MacDonald$^3$}
\author{E.~M.~Hankiewicz$^1$}

\affiliation{\vspace{.2cm}
$^1$Faculty of Physics and Astrophysics and W\"urzburg-Dresden Cluster of Excellence ct.qmat, University of W\"urzburg, Germany }
\affiliation{$^2$Department of Modern Physics, University of Science and Technology of China, Hefei, China }
\affiliation{$^3$Department of Physics, University of Texas at Austin, USA \\}

\begin{abstract}
\vspace{-0.6cm}
The boundary of topological superconductors might lead to the appearance of Majorana edge modes, whose non-trivial exchange statistics can be used for topological quantum computing. 
In  branched nanowire networks one can exchange Majorana states by time-dependently tuning topologically non-trivial parameter regions. 
In this work, we simulate the exchange of four Majorana modes in T-shaped junctions made out of  p-wave superconducting Rashba wires. 
We derive concrete experimental predictions for (quasi-)adiabatic braiding times and
determine geometric conditions for successful Majorana exchange processes. 
Contrary to the widespread opinion, we show for the first time that in the adiabatic limit the gating time needs to be smaller than the inverse of the squared superconducting order parameter and scales linearly with the gating potential.
Further, we show how to circumvent the formation of additional Majorana modes in branched nanowire systems, arising at wire intersection points of narrow junctions. Finally, we propose a multi qubit setup, which allows for universal and in particular topologically protected quantum computing.
\end{abstract}

\maketitle

\section{Introduction} 

\vspace{-0.25cm}

\textbf{T}opological \textbf{q}uantum \textbf{c}omputing (\textbf{TQC}) represents a fault tolerant quantum computation scheme in which unitary quantum gates are realized by the braiding of so-called anyons, satisfying non-abelian exchange statistics \cite{Kitaev20032}.
These operations are only dependent on the topological class of the braiding group element, leading to TQC algorithms, protected by the energy gap of the system. Such quantum gates remain robust to disturbance and local noise if braiding procedures are (quasi-)adiabatic and if external perturbations do not close this gap  \cite{sankar}.

A special representative of non-abelian anyons are Majorana fermions (\textbf{MF}s), showing up as quasi-particle excitations in condensed matter physics \cite{Kitaev20032}.
Exemplary, such states are  predicted in $\nu\!=\!5/2$ fractional quantum Hall states or in topologically non-trivial systems proximity-coupled to bulk superconductors, e.g. at the edge of 2D topological insulators (\textbf{TI}s), in nanowires made from 3D TIs, in magnetized helical spin chains, or  in semiconducting Rashba nanowires \cite{Fu08, Fu09, Cook11, Choy11, Kitaev06, Greiter09, Moore91, readgreen, kitaevoriginalmajoranapaper, ivanov, sau10, aliceaprojection, Lutchyn10, Oreg10, majoranavorkommen, milestones,He17}.
Since 2012, Majorana modes in Rashba wires are of particular interest, caused by their experimental evidence in InSb nanowires \cite{mourik}. Up to this date, several other  experiments showed an indication of Majorana fermions in solid state systems 
\cite{Rokhinson12,Deng12, Heiblum, Churchill13, 
Nadj-Perge14,cmarcus,Pawlak16,Deng16,Zhang16,Feldman17,Suominen17,Hell17,Zhang17,Nichele17,Zhang18,Grivnin18,Marcus18}.
In contrast to recent approaches of (teleportation or measurement based) Majorana braiding procedures on parallel or hexagonal nanowire structures \cite{onedimensionalexchange, Vijay16, Plugge16,Landau16,Molenkamp16, Plugge17,Karzig17, Litinski17, Litinski172,Hankiewicz17,Molenkamp17}, we  simulate in this work spatial Majorana braiding procedures in 
 $p$-wave superconducting nanowires, arranged as triple T-junctions \cite{beenaker,aliceareview,flensberg,tudor}. 
While we consider isolated geometries at zero temperature, the analysis of coupling those systems to parity-conserving thermal baths can be found in Refs.~\citen{DiVincenzo15,DiVincenzo15long}. 
In particular, we derive geometric and adiabatic requirements for successfully exchanging multiple Majorana modes in branched nanowire devices. 
We analytically and numerically evaluate the functional dependence of the adiabatic exchange time on the  superconducting order parameter as well as on the gating potential. 
Moreover, we propose a full nanowire-based setup which is suitable for universal quantum computing.
Further, we analytically and numerically solve the problem of 
 additional MF modes, forming at wire intersection points in branched nanowire based devices \cite{Stanescu18,Sarma18}. 
In general, the formation of such modes prevents any successful topological qubit operation. We solve this serious problem by locally gating the wire intersection points.

This work is structured as follows: In Sec.\,II, we introduce the Bogoliubov-de Gennes (\textbf{BdG}) Hamiltonian, characterizing our exchange geometry. Moreover, we present various, different Majorana exchange protocols, defining TQC algorithms. In Sec.\,III, we discuss the exchange statistics of four MF modes, specifying a topological single qubit.
In Sec.\,IV, we show that for narrow nanowire junctions, additional MF modes form at wire intersection points, and rigorously prove how to solve this effect.
In Sec.\,V, we determine adiabatic time scales and geometric conditions for successful MF exchange protocols, enabling different single qubit operations. Moreover, we compare these results to existent adiabatic limits on MF exchange processes \cite{Scheurer13, Wu14, Cheng14, Karzig15, otherbraidingex, Knapp16, Hell162, Rahmani17,Schmitteckert17, Bauer2018}.
In Sec.\,VI, we discuss the concept of universal TQC, introduce projective measurements and propose a nanowire setup, defining a Majorana based multi qubit system. Finally, we summarize our paper in Sec.\,VII.

\begin{figure}[t]
\centering
\includegraphics[width=.48\columnwidth]{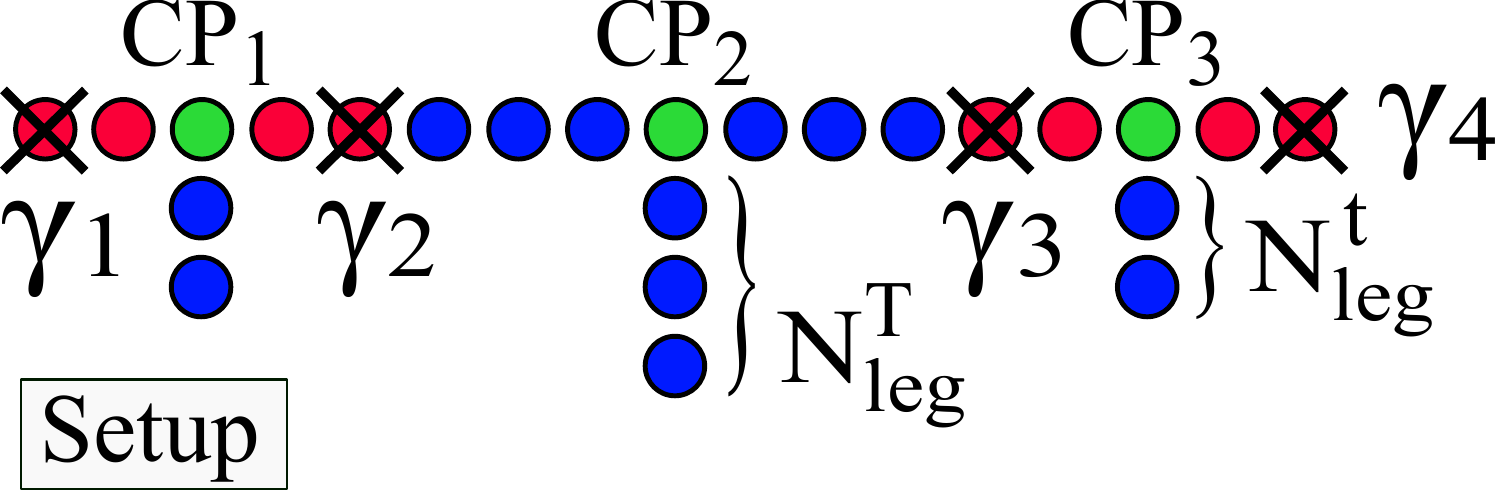}
\hfill
\includegraphics[width=.48\columnwidth]{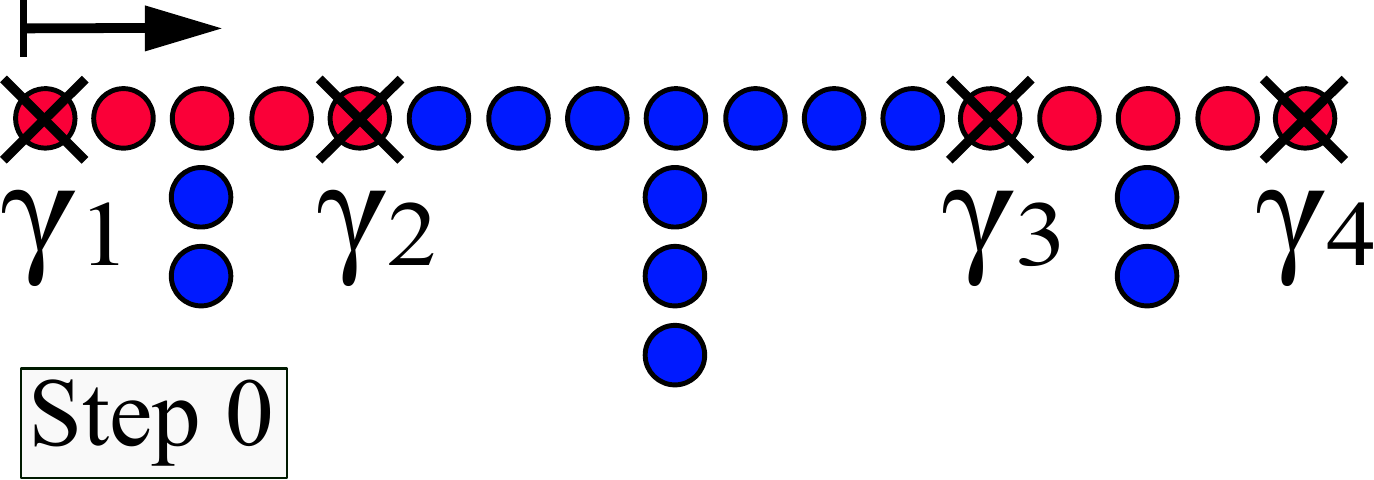}

\vspace{.2cm}
\includegraphics[width=.48\columnwidth]{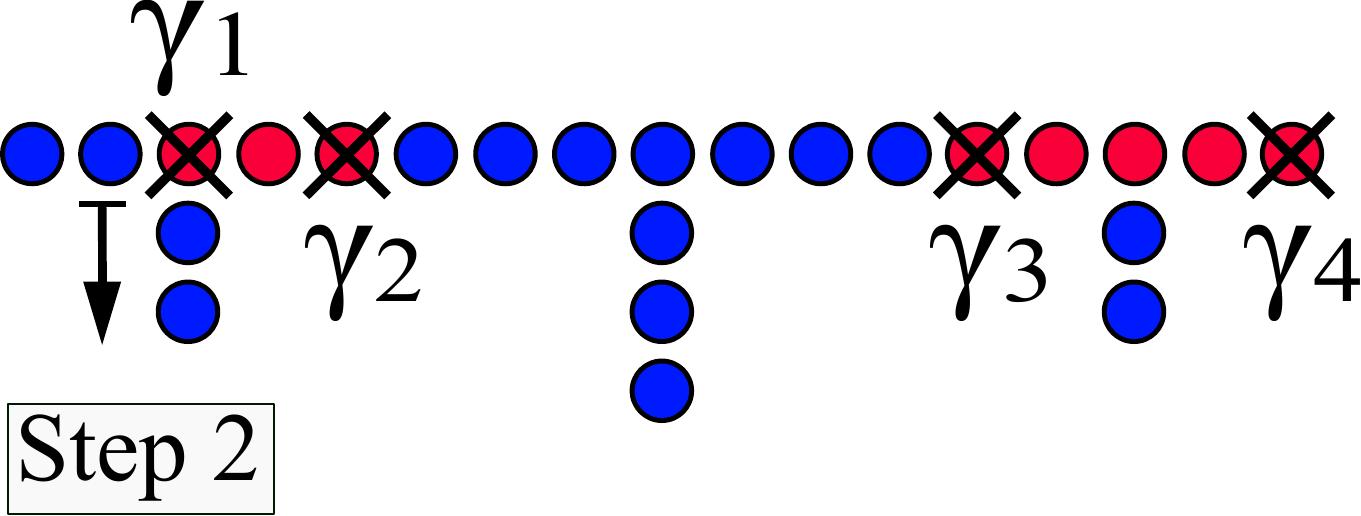}
\hfill
\includegraphics[width=.48\columnwidth]{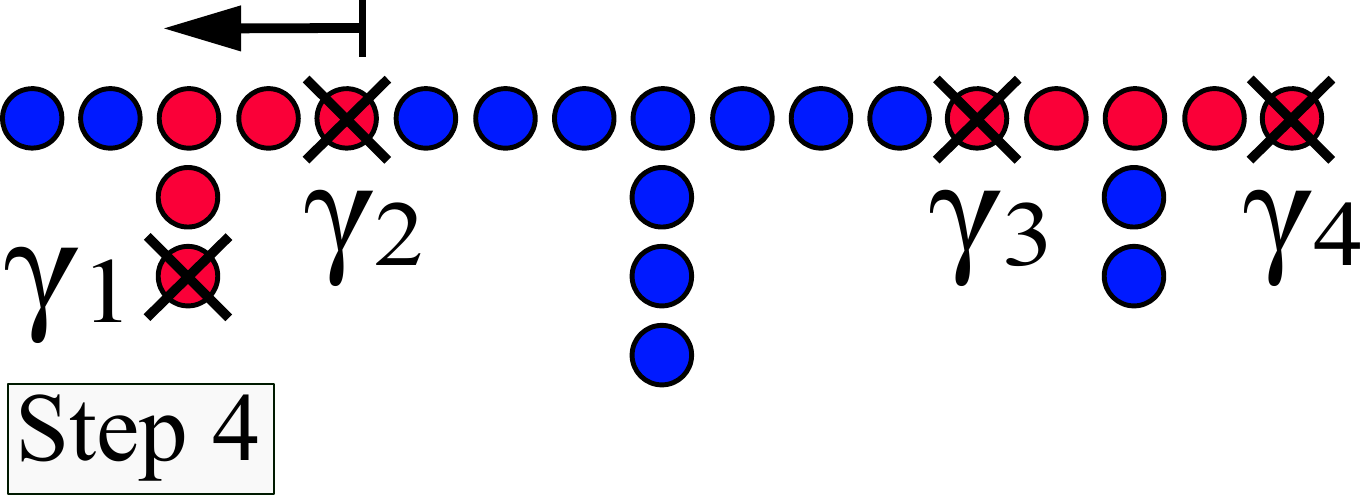}

\vspace{.2cm}
\includegraphics[width=.48\columnwidth]{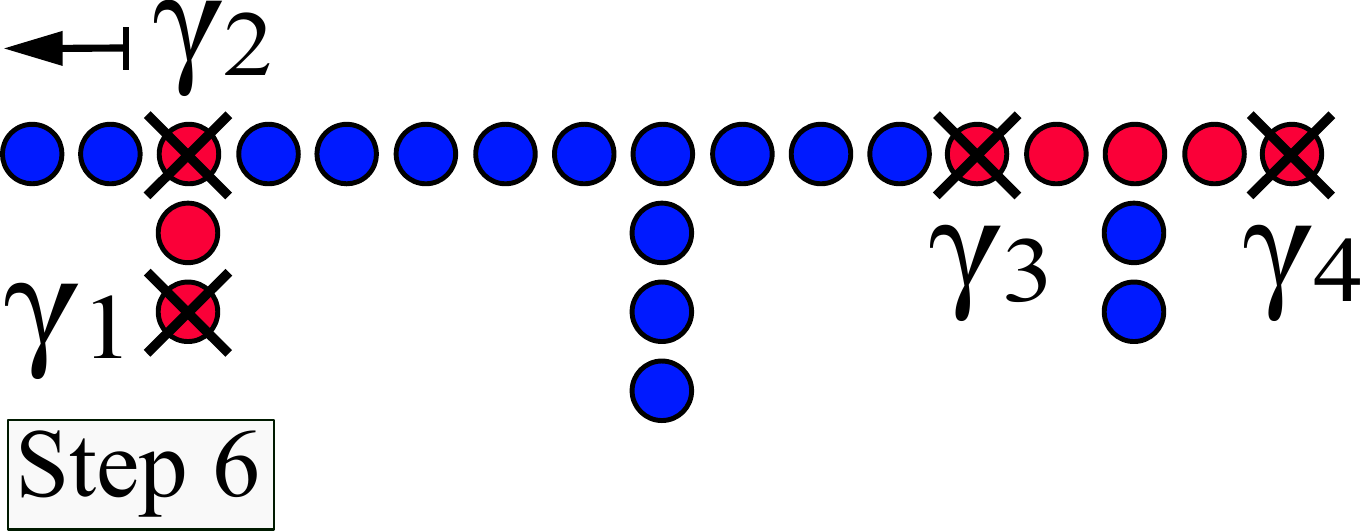}
\hfill
\includegraphics[width=.48\columnwidth]{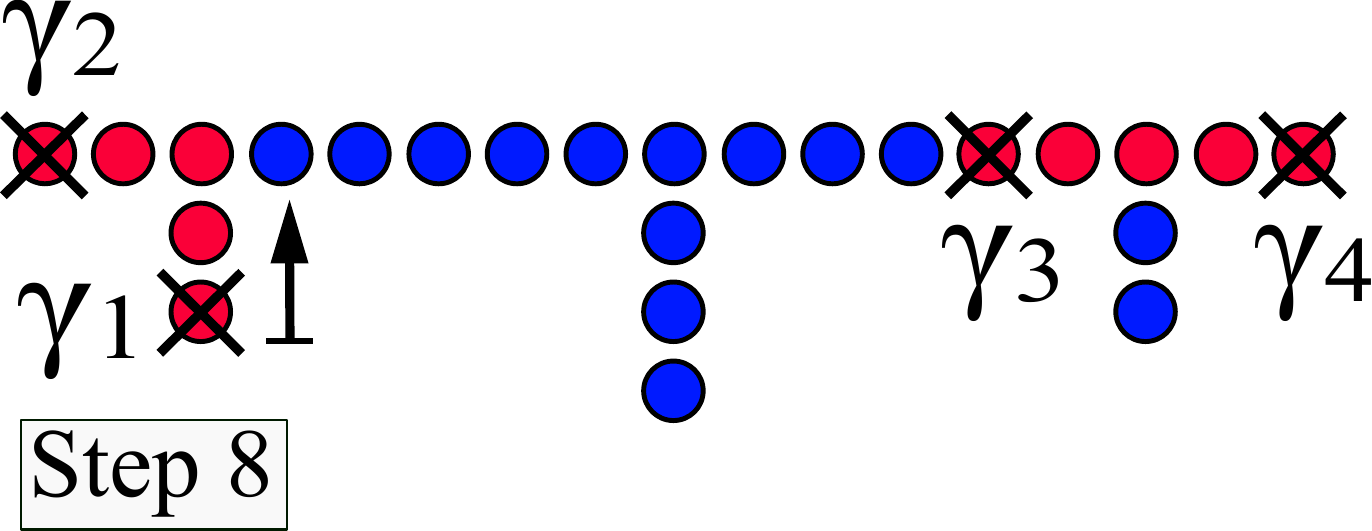}

\vspace{.2cm}
\includegraphics[width=.48\columnwidth]{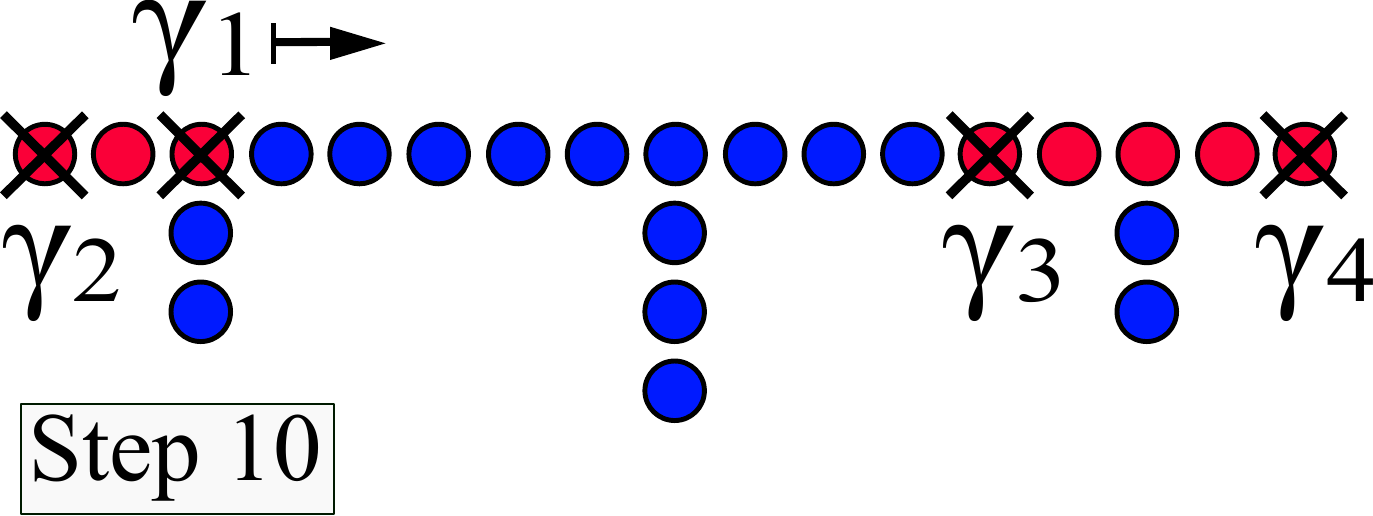}
\hfill
\includegraphics[width=.48\columnwidth]{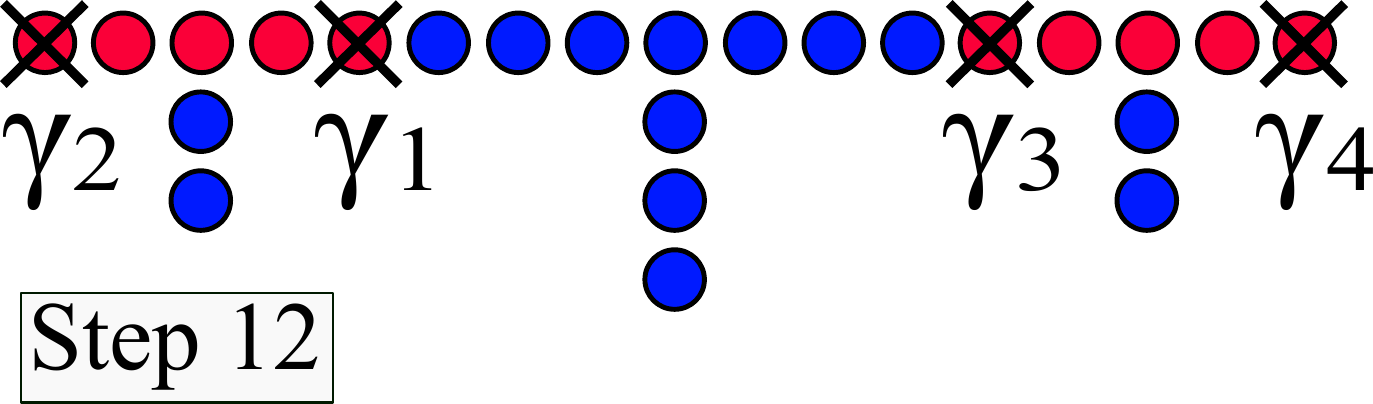}
\caption{\textbf{Setup:} Structure of a tTt device, consisting of 4 nanowires. The chemical potential on each site can be controlled by a "keyboard" gate. Inner and outer T/t structures are characterized by the amount of lattice sites $\NT$ and $\Nt$. Moreover, $\mathrm{CP}_{1,2,3}$ denote nanowire intersection points and red (blue) regions encode topologically non-trivial (trivial) junction parts with associated MF modes $\gamma_\mathrm{1,2,3,4}$. \mbox{\textbf{Step 0-12}} characterizes the exchange protocol $\gamma_1\!\leftrightarrow\!\gamma_2$.
With $\Nt\!=\!2$, $\NT\!=\!3$ and $\Ngate\!=\!1$ the exchange is realized in 12 steps, corresponding to the different keyboard gate configurations.}
\label{outerexchange}
\vspace{-0.2cm}
\end{figure}

\vspace{-0.4cm}

\section{Model $\&$ Setup} \label{ModelandSetup}

\vspace{-0.25cm}

\noindent
Majorana edge modes are predicted at the boundaries of topological superconductors and are associated to local densities which exponentially decay into the bulk of the system \cite{aliceareview,aliceanature}. 
In general, these modes hybridize and form highly nonlocal Dirac fermionic states (\textbf{HMF}s). If $\gamma_1$ and $\gamma_2$ represent second quantized MF operators, the corresponding HMF modes are given by $\psi^{(\dagger)}=(\gamma_1 \pm \im \gamma_2)/2$. Their energy $\Delta_\mathrm{hyb}$  decreases exponentially with the spatial separation of $\gamma_{1,2}$.
Experimentally, it is possible to engineer an effective $p \,$-wave superconductor by using a quasi one-dimenional nanowire
with strong spin-orbit (\textbf{SO}) interaction. The additional ingredients for such a construction are a magnetic field, aligned along the axis of the nanowire and perpendicular to the SO field, and a proximitized $s \, $-wave superconductor \cite{mourik,aliceanature}. 
In the basis $[ c_\uparrow^\dagger (k), c_\downarrow^\dagger (k), c_\downarrow (-k), -c_\uparrow (-k) ]$,
T-junctions based on nanowires are described by the BdG Hamiltonian:
%
\begin{align} \label{rashbananowire}
H_\mathrm{BdG} & =  \left[ \left(\frac{\hbar^2}{2 m_\mathrm{eff}} (k_x^2+k_y^2 ) - \mu \right) \tau_3 +  \Delta_\mathrm{sc}  \tau_1 \right] \otimes \sigma_0 \\  &  \quad + h_\mathrm{Z} \, \tau_0 \otimes \sigma_3  + \alpha \, \tau_3 \otimes
\left[  k_x \sigma_2 + k_y  \sigma_1  \right] \nonumber \, .
\end{align}
Here, $\tau_{0,1,2,3}$ are Pauli matrices acting on the electron-hole subspace, while $\sigma_{0,1,2,3}$ are Pauli matrices, acting on spin degrees of freedom. $ \Delta_\mathrm{sc}  = \vert \Delta_\mathrm{sc} \vert \mathrm{e}^{\im \phi} $ is the $s $-wave order parameter and we choose $\phi \! =\! 0$~($\phi \!=\! \pi/2$) for the superconducting phase in horizontal $\vec{e}_\mathrm{x}$ (vertical $\vec{e}_\mathrm{y}$) direction. Furthermore, $\mu$ is the chemical potential, $h_\mathrm{Z} = g \muB B/2$ the Zeeman energy, $g$ the Land\`{e} $g$-factor, $\muB$ the Bohr magneton, $\alpha$ the Rashba parameter and  $m_\mathrm{eff}$ the effective electron mass in the nanowire. If one guarantees $\vert \mu \vert < \mu_\mathrm{c} \equiv \sqrt{h_\mathrm{Z}^ 2- \vert \Delta_\mathrm{sc} \vert^2}$, the system is in a topologically non-trivial phase, hosting localized MF edge modes, whereas for $\vert \mu \vert > \mu_\mathrm{c}$ it is topologically trivial \cite{aliceanature}. For our simulations, we use input parameters, related to experiments in  InSb nanowires of length $L \!=\!1.6 \mu\mathrm{m}$\cite{mourik}. In particular, we choose $g= 50$, $\vert \Delta_\mathrm{sc} \vert= 250 \mu \mathrm{eV}$, $\alpha= 0.02 \mathrm{eVnm}$, $m_\mathrm{eff}= 0.015 \mathrm{m}_\mathrm{e}$ and $ B = 0.25 \mathrm{T}$. Also, we checked that all our results hold for different parameter variations, justifying the robustness of our calculations.

\begin{figure}[t]
\centering
\includegraphics[width=.48\columnwidth]{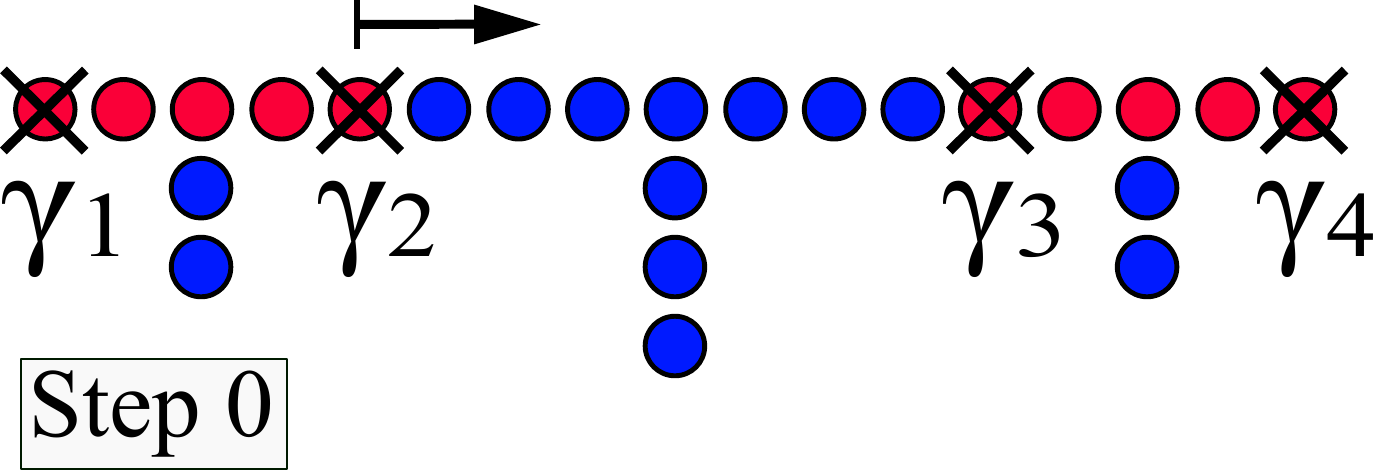}
\hfill

\vspace{.2cm}
\includegraphics[width=.48\columnwidth]{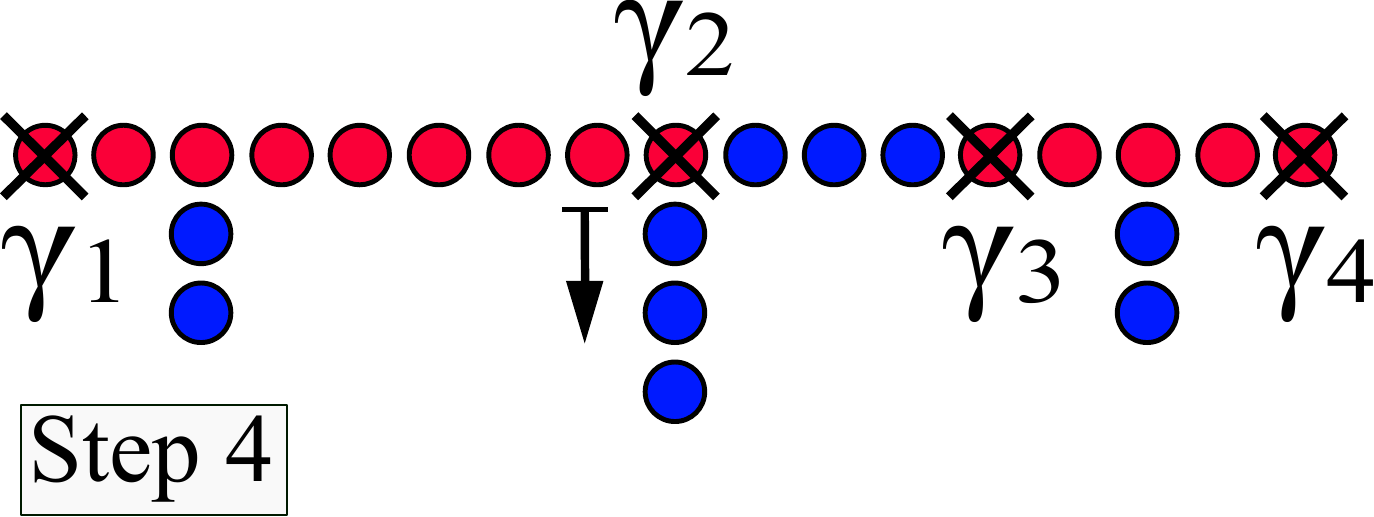}
\hfill
\includegraphics[width=.48\columnwidth]{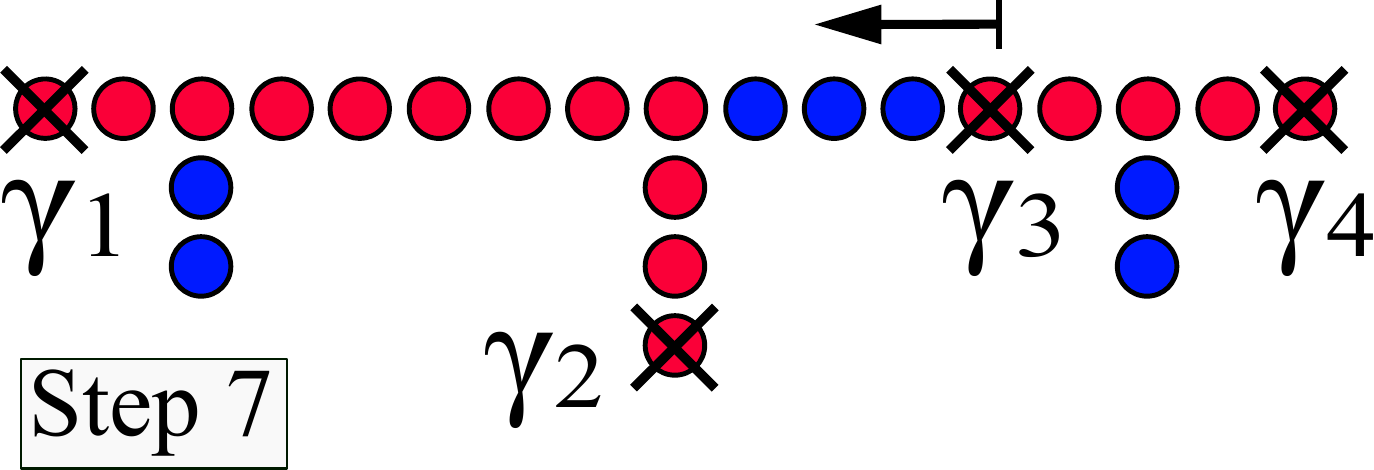}

\vspace{.2cm}
\includegraphics[width=.48\columnwidth]{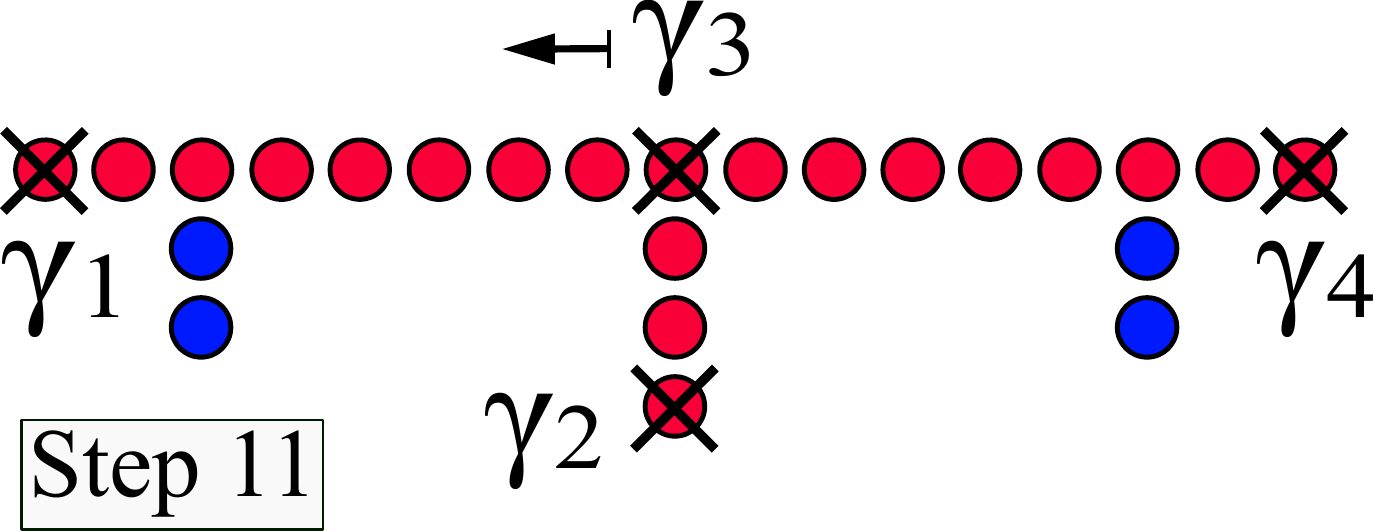}
\hfill
\includegraphics[width=.48\columnwidth]{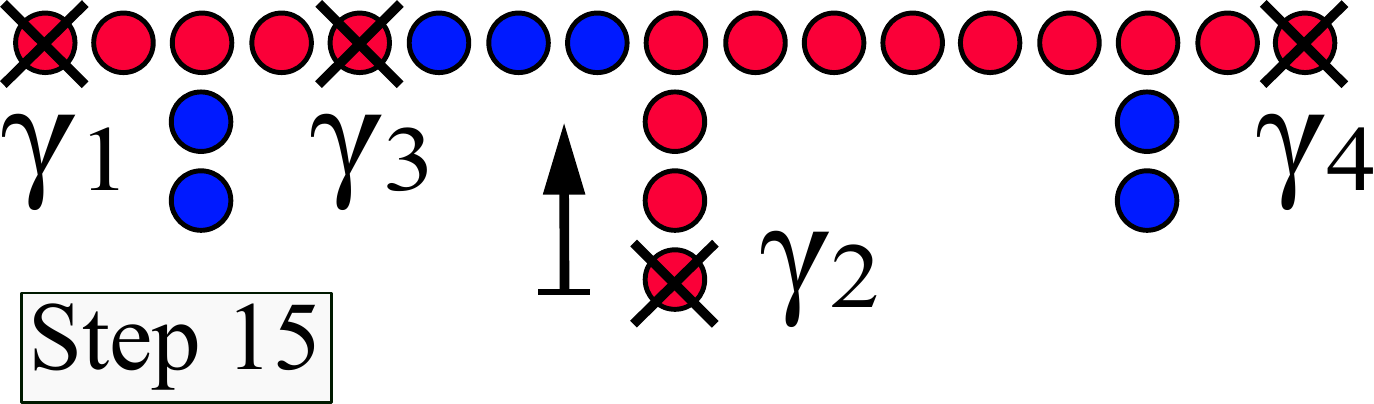}

\vspace{.4cm}
\includegraphics[width=.48\columnwidth]{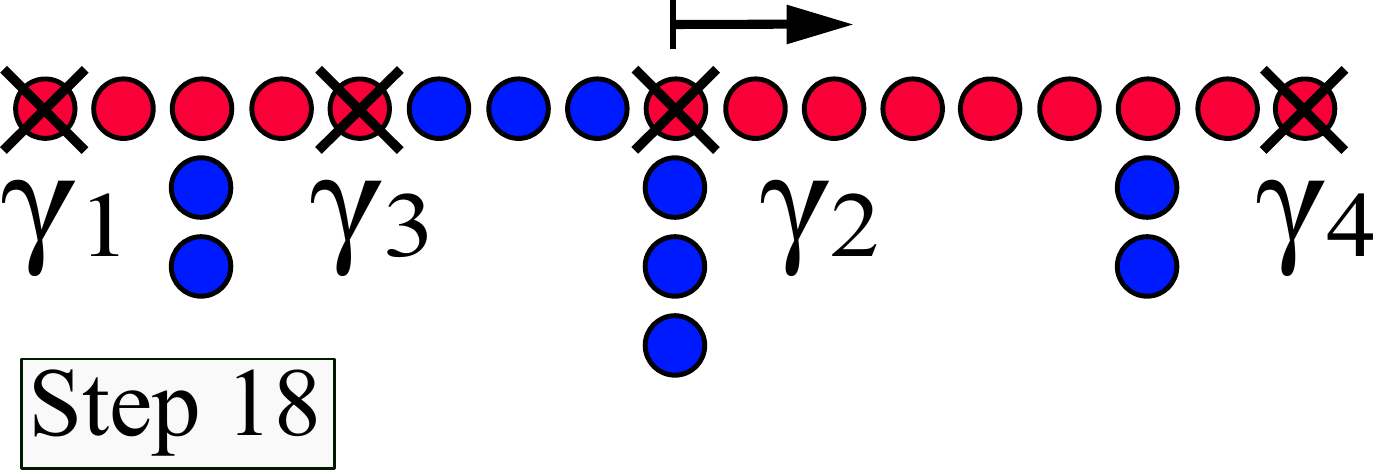}
\hfill
\includegraphics[width=.48\columnwidth]{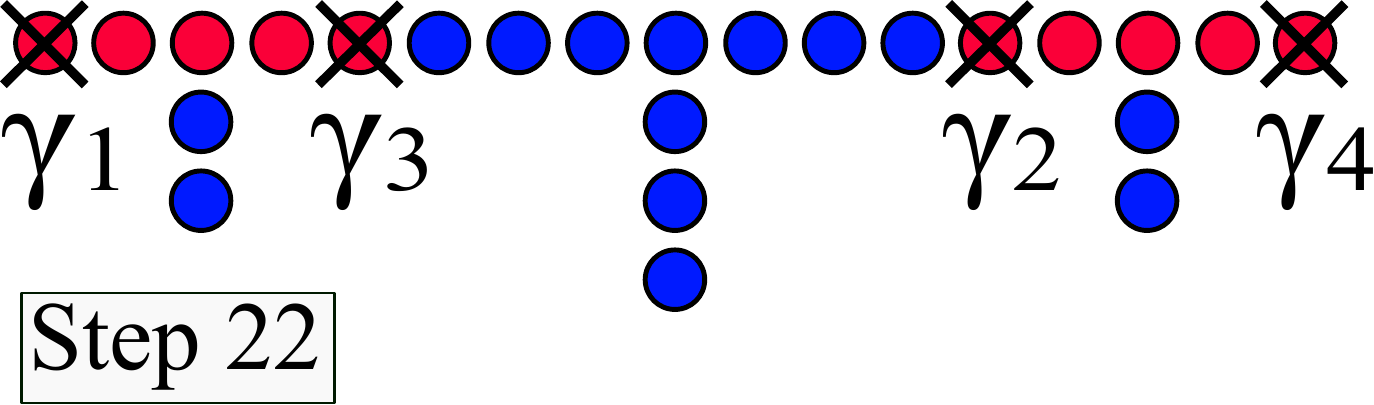}
\caption{Exchange protocol  $\gamma_2\!\leftrightarrow\!\gamma_3$. With $\Nt\!=\!2$, $\NT\!=\!3$ and $\Ngate\!=\!1$ the braiding is realized in 22 exchange steps, corresponding to different keyboard gate configurations. Red (blue) regions encode topologically non-trivial (trivial) junction parts with associated Majorana modes $\gamma_\mathrm{1,2,3,4}$.}
\label{innerexchange}
\end{figure}

\noindent
Majorana single qubit states are generated by at least four MF modes. Shifting these modes spatially, enables the construction of topological gates, using their non-abelian braiding statistics. 

\renewcommand{\arraystretch}{1.2}
\begin{table*}[t!]
\begin{center}
\begin{tabular}{|c||c|c||c|c|}
\hline
Braiding & 
\begin{tabular}{l}
$\psi_1 \equiv \frac{1}{2} \left(  \gamma_1 + \im \gamma_2 \right)$ \\
 $\psi_2 \equiv \psi_1^\dagger = \frac{1}{2} \left(  \gamma_1 - \im \gamma_2 \right)$ \\
\end{tabular}
& 
\begin{tabular}{l}
$ \psi_3 \equiv \frac{1}{2} \left(  \gamma_3 + \im \gamma_4 \right)$ \\
$ \psi_4 \equiv \psi_3^\dagger = \frac{1}{2} \left(  \gamma_3 - \im \gamma_4 \right)$
\end{tabular}
& $\left\vert \left(\kappa_{ij} (T) \right)_{\alpha,\beta} \right\vert $ & $\mathrm{Arg} \left[ \left(\kappa_{ij} (T) \right)_{\alpha,\beta} \right]$ $\left[ \frac{2}{\pi} \right]$\\
\hline
\hline
$\gamma_1 \leftrightarrow \gamma_2$  &  \begin{tabular}{l}
$\psi_1 \overset{B_{12}}{\longrightarrow} \im \psi_1 $\\
$\psi_2 \overset{B_{12}}{\longrightarrow} - \im \psi_2  $\\
\end{tabular}
&  \begin{tabular}{l} $  \psi_3 \overset{B_{12}}{\longrightarrow} \psi_3 $ \\ $
\psi_4 \overset{B_{12}}{\longrightarrow}  \psi_4$ \\
\end{tabular}  & $\delta_{\alpha, \beta}$ & $\begin{pmatrix}
1 & 0 & 0 & 0 \\
0 & -1 & 0 & 0 \\
0 & 0 & 0 & 0 \\
0 & 0 & 0 & 0 \\
\end{pmatrix}$ \\
\hline
$\gamma_3 \leftrightarrow \gamma_4$  &
\begin{tabular}{l} $  
\psi_1 \overset{B_{34}}{\longrightarrow} \psi_1$  \\
$\psi_2 \overset{B_{34}}{\longrightarrow}  \psi_2 $ \\
\end{tabular}
&
\begin{tabular}{l} $
 \psi_3 \overset{B_{34}}{\longrightarrow} \im \psi_3$  \\
 $\psi_4 \overset{B_{34}}{\longrightarrow}  - \im \psi_4$ \\
\end{tabular}
& $\delta_{\alpha, \beta}$ &  $\begin{pmatrix}
0 & 0 & 0 & 0 \\
0 & 0 & 0 & 0 \\
0 & 0 & 1 & 0 \\
0 & 0 & 0 & -1 \\
\end{pmatrix}$ \\ 
\hline
$\gamma_2 \leftrightarrow \gamma_3$  &
\begin{tabular}{l} 
$\psi_1 \overset{B_{23}}{\longrightarrow} \frac{1}{2} \left(\gamma_1 - \im \gamma_3 \right)$  \\
$\psi_2 \overset{B_{23}}{\longrightarrow}  \frac{1}{2} \left(\gamma_1 + \im \gamma_3 \right)$
\end{tabular} 
&
\begin{tabular}{l} 
$ \psi_3 \overset{B_{23}}{\longrightarrow} \frac{1}{2} \left(\gamma_2 + \im \gamma_4 \right) $ \\
$\psi_4 \overset{B_{23}}{\longrightarrow} \frac{1}{2} \left(\gamma_2 - \im \gamma_4 \right) $
\end{tabular}
& $\dfrac{1}{2}$
& $ \begin{pmatrix}
0 & 0 & -1 & -1 \\
0 & 0 & 1 & 1 \\
-1 & 1 & 0 & 2 \\
-1 & 1 & 2 & 0 \\
\end{pmatrix}$ \\
\hline
\end{tabular}
\caption{Exchange statistics of the hybridized MF states $\psi_{1,2,3,4}$, generated by the four MF modes $\gamma_\zeta \ (\zeta \in \lbrace 1,2,3,4 \rbrace)$. The MF braiding operations are mediated by $B_{l,l+1} \equiv \left( 1+ \gamma_l \gamma_{l+1}  \right) / \sqrt{2}$ ($l \in \lbrace 1,2,3 \rbrace$) and the overlap of $\psi_\beta(t=0)$ and $\psi_\alpha(T)$
with $ \lbrace \alpha,\beta \rbrace \in \lbrace 1,2,3,4 \rbrace$, is characterized by $(\kappa_{i,j})_{\alpha,\beta}(T)$ (cf.~Eq.~\eqref{kappaTTT}). Here,  $\lbrace i,j \rbrace \in \lbrace 1,2,3,4 \rbrace$ specify the MF braiding procedure \cite{ivanov}.}
\label{exchangestatistics}
\end{center}
\end{table*}

\noindent
In this work, we will analyze the exchange of MF modes in so-called \textbf{tTt} junctions, by using a common tight-binding (\textbf{TB}) approach with effective lattice constant $a$. Fig.\;\ref{outerexchange}, shows the structure of such a device, consisting of four connected nanowires, where
\textbf{CP1}, \textbf{CP2} and \textbf{CP3} denote the  three nanowire intersection points. The chemical potential on each site can be controlled independently via a "keyboard" gate. Topologically non-trivial junction parts, characterized by $\mu_\mathrm{nontrivial}$, can be shifted in the topologically trivial region, by applying a certain chemical potential $\mu_\mathrm{gate}$. The device is characterized by two different stub lengths. $\Nt$ counts the lattice points of the outer \textbf{t} structures, which are initially in the the topologically non-trivial region, illustrated in red. In comparison,  $\NT$ characterizes the geometry of the central \textbf{T} structure, which is initially in the topologically trivial phase, denoted in blue. Due to the four topological boundaries in the system, one generates in total four MF modes, encoded by $\gamma_{1,2,3,4}$. Majorana exchange processes are achieved by applying a time and site dependent potential $\mu_{\mathrm{gate},i}(t)$. While the amount of neighbored lattice points which are shifted simultaneously, is defined by $\Ngate$, $\tstep \equiv t_j-t_{j-1}$ characterizes  the time scale on which $\mu_{\mathrm{gate},i}$ is ramped up/down in an exchange step $j$. The latter parameter, $\tstep$, needs to be adjusted to ensure (quasi-)adiabatic braiding protocols. In particular, the chemical potential on site $i$ in exchange step $j$ is superimposed by the smooth gating potential $(t \in [t_{j-1};t_{j}] )$
\begin{align} 
\mu_{\mathrm{gate},i} \left( t  \right) = \Vgate \times \begin{cases}  \sin^2\left( \dfrac{t-t_{j-1}}{\tstep} \dfrac{\pi}{2} \right) \quad  \quad \quad \text{up}\\[1.2em] 1- \sin^2\left( \dfrac{t-t_{j-1}}{\tstep} \dfrac{\pi}{2} \right) \quad \  \text{down}
\end{cases} \nonumber
\end{align}

\noindent
The gating potential is kept constant ($0$ or $\Vgate$) in all other exchange steps. 
We distinguish two different types of braiding. The protocol for an exchange $\gamma_1 \leftrightarrow \gamma_2$ is visualized in Fig.\;\ref{outerexchange}.
To obtain a closed loop, step=12 needs to coincide with step=0.  The protocol for an exchange $\gamma_2 \leftrightarrow \gamma_3$ is visualized in Fig.\;\ref{innerexchange}. 
Numerically, we simulate these processes by solving the time dependent \mbox{BdG equation}, including the full Hamiltonian
\begin{align} \label{timedephamilton}
\mathcal{H}_\mathrm{BdG}(t) \equiv \mathcal{H}_\mathrm{BdG}(t=0) + \mu_\mathrm{gate} (t) \, \tau_3 \otimes \sigma_0 \ . 
\end{align}
In our system, state evolution is given by 
\begin{align}
\psi_i(t+\Delta t)= \mathrm{e}^{- \frac{\im}{\hbar} \mathcal{H}_\mathrm{BdG}(t) \Delta t} \psi_i(t), \nonumber
\end{align}
where $\Delta t \equiv t_\mathrm{step}/n$, and $n$ is the number of time discretization points\cite{Wu14,Sato14}. Initial values of this differential equation  are obtained by solving $\mathcal{H}_\mathrm{BdG}(t\!=\!0)$ for its eigensystem.
Moreover, the  energy of each state $\psi_i(t)$ is given by 
\begin{align} 
E_i(t)= \psi_i^\dagger (t) \mathcal{H}_\mathrm{BdG}(t) \psi_i (t) \ . \nonumber
\end{align}

\noindent
Based on the underlying PH symmetry, these energies always come in pairs $\pm E_i(t)$. 
We want to emphasize, that due to the finite energy of  HMF modes at $t=0$, their accumulated geometrical phase during any MF exchange process is accompanied by a dynamical one
\begin{align}
\phi_\mathrm{dyn}^i(t)=-\dfrac{1}{\hbar}  \int_0^t E_i(t') \mathrm{d}t' . \nonumber
\end{align}
All dynamical phases are real, PH anti-symmetric and
can be eliminated after complete MF braiding processes $(t=T)$, by gauging $\psi_i(T) \rightarrow \mathrm{e}^{- \im \phi^i_\mathrm{dyn}(T)} \psi_i (T)$.

\section{Exchange Statistics} \label{exchangesatsec}

\noindent
The basic idea of TQC algorithms is the realization of topologically protected quantum gates, using the geometrical exchange phases of non-abelian anyons. 
As explained in the previous section, we realize such processes by the braiding of four Majorana fermions $\gamma_{1,2,3,4}$. Theoretically, the pairing of zero energy MFs into the associated fermionic states is arbitrary and therefore represents a choice of basis \cite{ivanov}. However, beyond the zero energy limit, it is natural to combine two MFs to a nonlocal hybridized fermionic state (HMF mode) if they pairwise have a strong spatial overlap. Since this is the case in our finite size system, this dictates the definitions of our HMF modes $\psi_{1,2,3,4}$ in Tab.~\ref{exchangestatistics}\cite{Sato14}.  We want to remark that the two distinct topologically non-trivial regions of the tTt structure are not perfectly decoupled for $t=0$. As a consequence, the modes $\gamma_{2}$ and $\gamma_{3}$ weakly hybridize via the center T structure, which will be discussed in Sec.~\ref{Braidingstat}.

In particular, the finite hybridization energy of the HMF modes lifts the degeneracy of a perfect Majorana system, which has two important consequences: On the one hand, it allows us to uniquely identify and determine all HMF states and their corresponding energies in each exchange step. On the other hand it defines an upper bound for our exchange time $t_\mathrm{step}$. To realize a successful MF braiding process, $t_\mathrm{step}$ needs to be adiabatic with respect to the bulk gap, avoiding bulk excitations. At the same time, it needs to be diabatic with respect to the finite HMF hybridization gaps. Particularily, this ensures that the HMF modes can freely rotate within their non-degenerated  computational sub-space, which is a mandatory condition to obtain non-trivial exchange phases. While this requirement are absent for zero energy Majorana systems, it enforces our exchange protocols to be executed merely (quasi-)adiabatically \cite{Sato14,Bauer2018}.

\begin{figure}
\begin{minipage}{1.\columnwidth}
\includegraphics[width=.845\columnwidth]{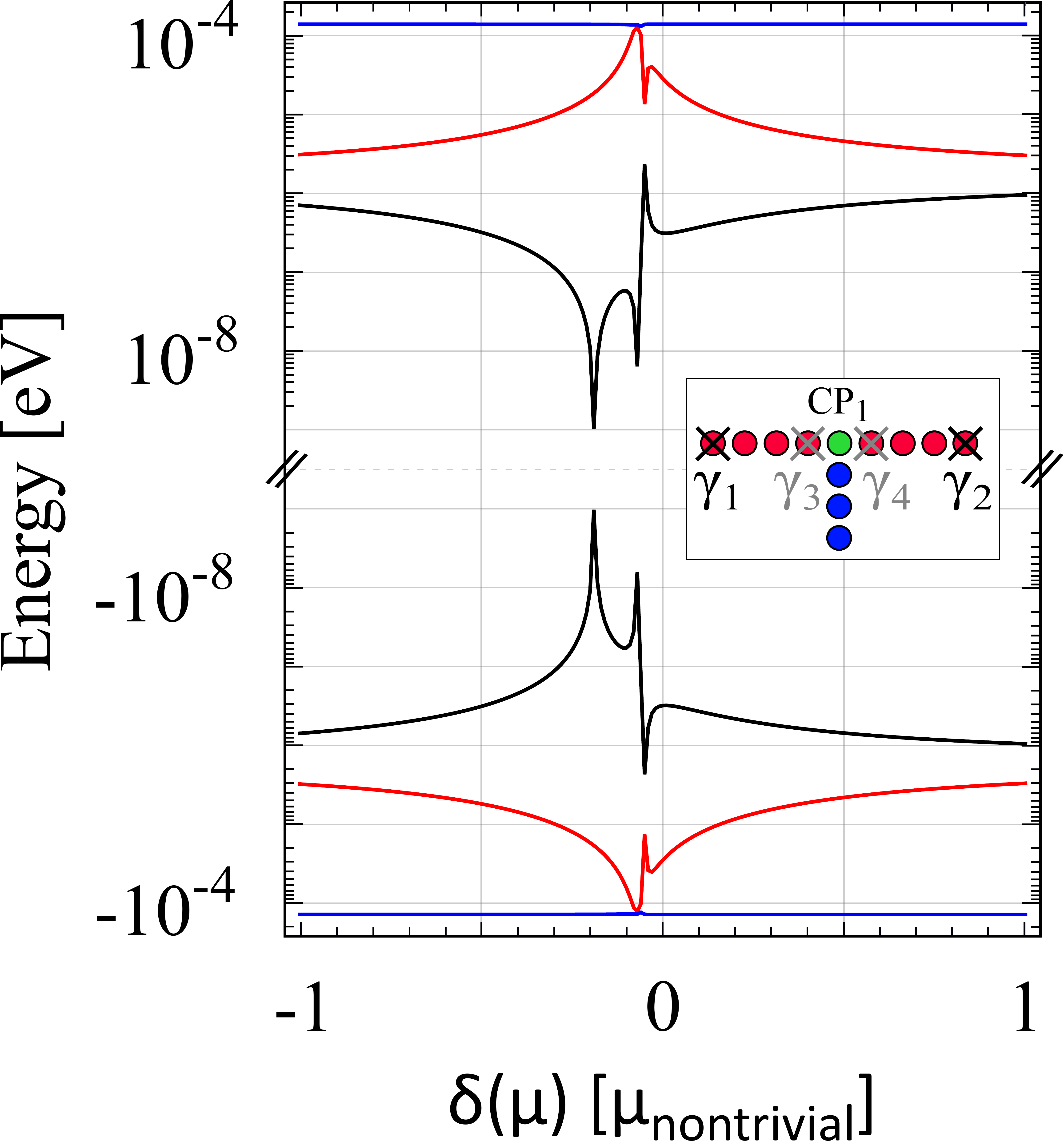}
\vspace{-1pt}
\caption{The three lowest gaps of a thin t-junction, formed by a non-trivial horizontal and a trivial vertical nanowire are shown as a function of the CP$_1$ on-site potential $\delta \mu$.  For \mbox{$\delta \mu \! =\! 0$}, one observes the formation of additional sub-gap states, a property which can be corrected by applying $\delta \mu_\mathrm{corr}$.}
\label{visualisationcooreection}
\end{minipage}
\vfil
\vspace{8pt}
\begin{minipage}{1.\columnwidth}
\includegraphics[width=.845\columnwidth]{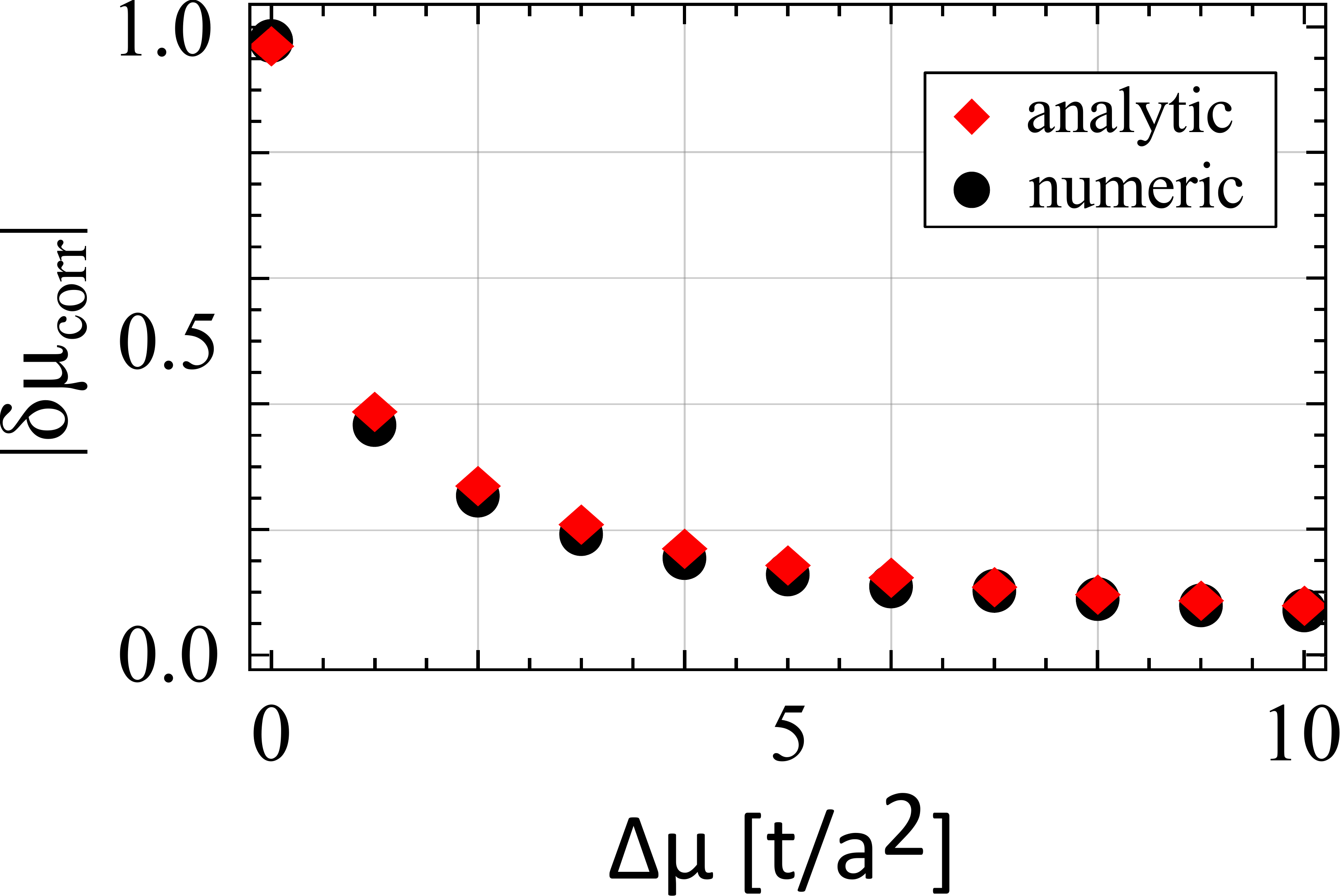} 
\vspace{-2pt}
\caption{Analytically and numerically determined values of $  \vert \delta \mu_\mathrm{corr} \vert $ as a function of $\Delta \mu $. 
For $\Delta \mu \! = \! 0$, one exactly has to correct the additional $t_\mathrm{hop}/a^2$  CP$_1$ hopping contribution.}
\label{calcgreenrashba}
\end{minipage}
\vfil
\vspace{7pt}
\begin{minipage}{1.\columnwidth}
\includegraphics[width=.845\columnwidth]{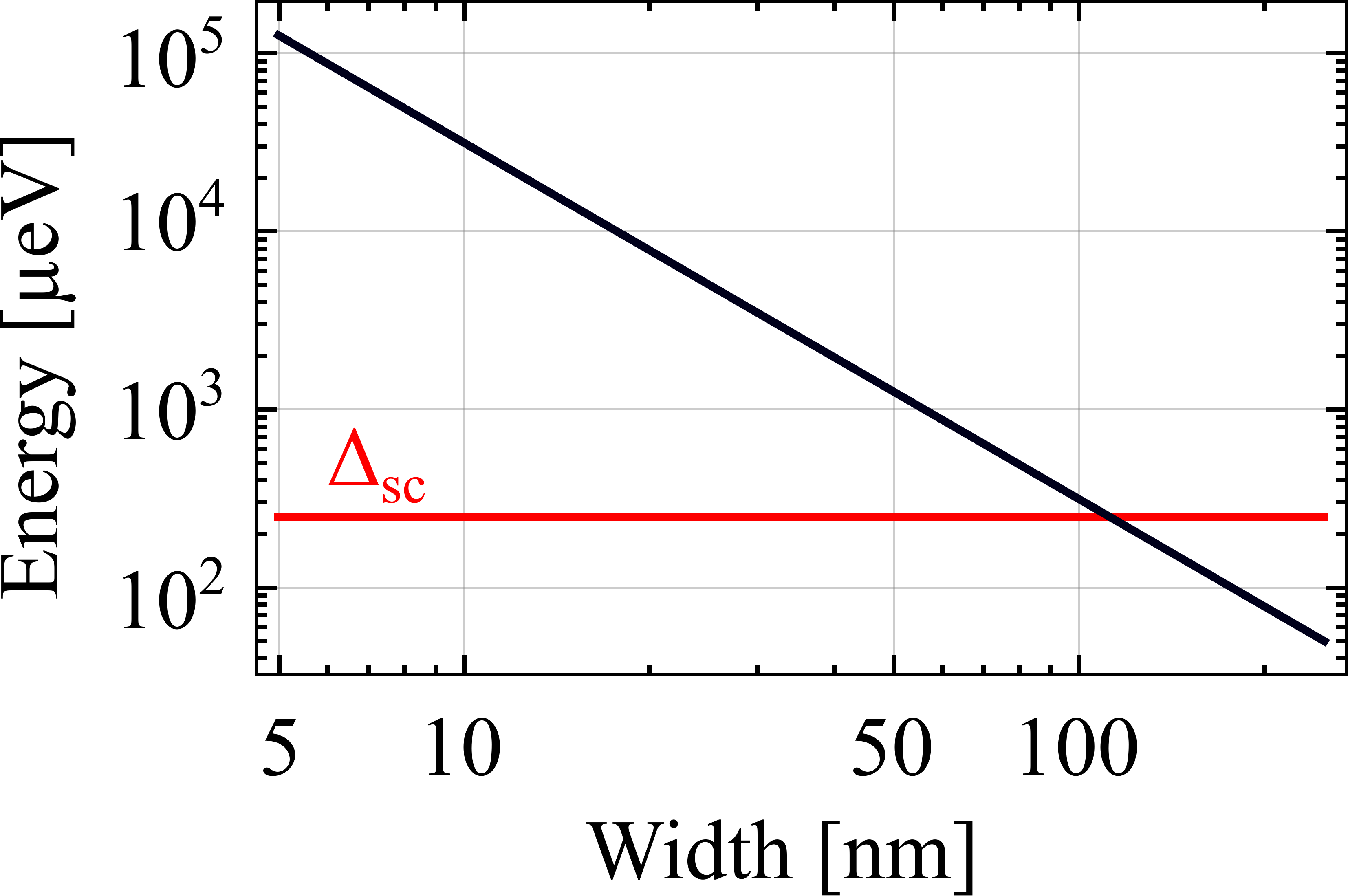}
\caption{Bound state energy $\Delta E_\mathrm{BS}(\Delta \mu \!=\! \delta \mu \!=\!0)$ in a InSb based t-junction as a function of the wire width $W$. $\vert \Delta_\mathrm{sc} \vert$ denotes the  $s$-wave gap. For $W \! \approx \! 90$nm, $\Delta E_\mathrm{BS}(\Delta \mu\!=\! \delta \mu \!=\! 0)$ starts to exceed $\vert \Delta_\mathrm{sc} \vert$, coming along with   CP$_1$ bound states.}
\label{additionalboundstateenergy1}
\end{minipage}

\end{figure}
In Tab.\;\ref{exchangestatistics}, we explicitly transfer the non-abelian braiding statistics of our four Majorana states $\gamma_{1,2,3,4}$ to the exchange statistics of the corresponding HMF modes $\psi_{1,2,3,4}$, where the operators $B_{l,l+1} \! \equiv \! \left( 1+ \gamma_l \gamma_{l+1}  \right) / \sqrt{2}$ ($l \! \in \! \lbrace 1,2,3 \rbrace$) mediate the particular braiding operation \cite{ivanov}. To check whether our simulated exchange processes provide the predicted exchange statistics, we define a time dependent quantity which calculates the overlap of states $\psi_\beta(t=0)$ and $\psi_\alpha(t)$ with $ \lbrace \alpha,\beta \rbrace \in \lbrace 1,2,3,4 \rbrace$:
\begin{align} \label{kappaTTT}
\left(\kappa_{ij} \right)_{\alpha,\beta}(t)= \left\langle \psi_\beta (t=0) \vert \left( B_{ij} \psi_\alpha  B_{ij}^{-1}\right) (t) \right\rangle  \ ,
\end{align}
where $\lbrace i,j \rbrace\!\in\!\lbrace 1,2,3,4 \rbrace$  characterizes the particular type of exchange. For successful MF exchange processes, the absolute values and geometrical phases of $(\kappa_{i,j})_{\alpha,\beta}(T)$  are explicitly listed in Tab.\;\ref{exchangestatistics}. In Sec.\,\ref{Braidingstat}, we will use these results to compare the simulated braiding processes
with the associated theoretical predictions.

\section{Branched Nanowire Networks}

\noindent
Before we present our Majorana braiding results, let us discuss an effect which also has been recently predicted in Refs.~\citen{Stanescu18}~and~\citen{Sarma18}, namely the formation of additional MF modes in networks made  of very thin nanowires. Originally, Majorana modes were predicted at the topological boundaries of strictly one-dimensional $p \,$-wave wires. 
If the wire width does not exceed the superconducting coherence length, such modes also occur beyond the single channel limit \cite{patricklee2010,Lutchyn}. Therefore, two-dimensional junctions formed by quasi-one-dimensional nanowires smaller then the underlying superconducting coherence length are appropriate for realizing MF exchange processes.   
However, as numerically has been shown by the authors of Refs.~\citen{Stanescu18}~and~\citen{Sarma18}, a naive realization of a T-junction by simply sticking together a non-trivial horizontal with a trivial vertical nanowire induces additional low-energy states besides the original MFs at the wire intersection point (cf. inset Fig.\;\ref{visualisationcooreection}). In general, these additional modes are caused by the non-uniform transverse confinement specifying the junction. At the  wire intersection point the local chemical potential is altered with respect to all remaining sites, due to the the additional hopping contribution, connecting the nanowire network. In the small wire limit, this implies the formation of fermionic bound states, which due to the Pauli principle effectively generate new topological boundaries. Beside the numerical analysis done in Refs.~\citen{Stanescu18}~and~\citen{Sarma18}, we will here analytically determine the parameter dependence of the sub-gap state formation, and predict a way to reverse this process by locally gating the wire intersection point.  This is mandatory since any additional Majorana sub-gap state formation prohibits the exchange of the computational MF modes due to  hybridization and fusing processes.

\noindent
Let us explain this more carefully, considering a single t-bar junction with wire intersection point CP$_1$, hopping parameter $t_\mathrm{hop}$ and lattice spacing $a$ \mbox{(cf. inset Fig.\;\ref{visualisationcooreection}).}
The vertical (horizontal) part of this structure is topologically trivial (non-trivial), which can be achieved by implementing $\mu_\mathrm{nontrivial}$ in the horizontal structure.
We define the difference between chemical potentials in the two nanowires by $ \Delta \mu \! \equiv \! \vert \mu_\mathrm{v} \! - \! \mu_\mathrm{h} \vert $ and introduce an on-site potential  \mbox{$V_\mathrm{CP1} \! \equiv \delta \mu    \cdot   t_\mathrm{hop} /a^2$} implemented at the wire intersection point. For a small lattice spacing, Fig.\;\ref{visualisationcooreection} schematically shows the three lowest energy gaps within such a t-junction as a function of $\delta \mu$. With increasing $\vert \delta \mu \vert$, 
one observes the formation of additional sub-gap states, 
indicating that the system hosts additional MF modes caused by the new topological boundaries in the junction. 
In particular, for small small $\vert \delta \mu \vert$ $\gamma_{1,2}$ and $\gamma_{3,4}$ form non-local fermionic states, whereas for large $\vert \delta \mu \vert$ $\gamma_{1,3}$ and $\gamma_{2,4}$ form the seperate HMF modes.In the intermediate $\vert \delta \mu \vert$ all four modes hybridize, causing the characteristic energy spectrum shown in Fig.~\ref{visualisationcooreection}.
Since already for $\delta \mu =0$ the sub-states are present, even the ungated t-junction shows fermionic bound states, a property which will be enhanced if we further reduce $a$. Notice, that for a certain value $\delta \mu_\mathrm{corr}$, the bulk energy gap can be reopened again, removing these bound states.  In what follows, we determine $\delta \mu_\mathrm{corr}$ analytically.

In a single, topologically non-trivial horizontal nanowire
the lowest PH symmetric bulk energy is given by $  \pm E_1^\mathrm{h}=  2 \, \ta$. 
If $E_1^\mathrm{t}(\Delta \mu,\delta \mu)$  denotes the corresponding energy in the full t-junction, the bound state energy is defined by $ \Delta E_\mathrm{BS}(\Delta \mu,\delta \mu) \equiv \vert E_1^\mathrm{h}-E_1^\mathrm{t}(\Delta \mu,\delta \mu) \vert $. This difference can be minimized by adjusting $\delta \mu$, allowing for a numerical scheme to determine $\delta \mu_\mathrm{corr}$. Analytically, one can calculate the corresponding strength by using the single particle Green's function in the disconnected, horizontal nanowire 
\begin{align}
G_\mathrm{h}^0(E) = \left( E-H_\mathrm{h} \right)^{-1} \ , \nonumber
\end{align}
where $H_\mathrm{h}$ is the associated Hamiltonian.
Connecting vertical and horizontal wire parts leads to an additional term in the horizontal Green's function
\begin{align} \label{Green's functionapproachformula}
G_\mathrm{h}^\mathrm{t}(E) & =\left(E-H_\mathrm{h}-[g_\mathrm{v}^{_\mathrm{CP1}}]^{-1}(E)\right)^{-1}  \\
 \left[g_\mathrm{v}^{_\mathrm{CP1}} \right]^{-1}(E) & = \sum \limits_{n, \mathrm{v}} \dfrac{\langle \mathrm{CP1} \vert H_\mathrm{hop} \vert n,\mathrm{v} \rangle \langle n,\mathrm{v}  \vert H_\mathrm{hop} \vert \mathrm{CP1}\rangle}{E-E_{n,\mathrm{v}}} \ . \nonumber
\end{align}
Here $\vert n,\mathrm{v} \rangle$ denotes a state $\vert n \rangle$ in the vertical wire and $H_\mathrm{hop}$ represents the hopping matrix, connecting the t-junction. Hence, $(g_\mathrm{v}^{_\mathrm{CP1}})^{-1}(E)$ describes the additional hopping from the horizontal into the vertical wire and back  ($\mathcal{O}(t_\mathrm{hop}^2)$). Hence, it resembles the inverse of the local, vertical Green's function. According to Eq.\;\eqref{Green's functionapproachformula}, 
We can exactly reverse the formation of bound states at site CP1, if we choose $V_{^{ \; \mathrm{CP1}}}^\mathrm{corr}= (g_\mathrm{v}^{_\mathrm{CP1}})^{-1}(E_1^\mathrm{h})$.

Fig.\;\ref{calcgreenrashba} shows numerically and analytically determined values of $\delta \mu_\mathrm{corr}$ as a function of $\Delta \mu$, which perfectly match each other. With increasing $ \Delta \mu $, hoppings from the horizontal into the vertical wire get more and more unfavorable, leading to a rapid decrease of $\delta \mu_\mathrm{corr}$. Notice, that for $\Dmu \! = \! 0$, we obtain $\delta \mu_\mathrm{corr} \! = \! 1 $, correcting the single, additional hopping term at CP1.\\

\noindent
As explained in the beginning of this section,  narrow, quasi-one dimensional channels are required to observe Majorana bound states in T-junctions. We proved that in such systems additional MF modes may occur at wire intersection points, requiring for a correction potential. Let us close this section by discussing the experimental relevance of this effect. For our set of parameters (cf. Sec.~\ref{ModelandSetup}),
the competing energy scales are given by\cite{Sols,LinChen,Schult}:
\begin{align}
\Delta  E_\mathrm{BS}(\Delta \mu \!= \! \delta \mu \!= \! 0)  & \approx \dfrac{1}{8} \frac{\left( \frac{\pi}{W} \right)^2 \hbar^2}{2 m^\star}   \nonumber \\
 E_\mathrm{SO} = \dfrac{\alpha^2 m^\star }{ \left( 2 \hbar \right)^2}  \approx 50 \mu \mathrm{eV} \quad  & \wedge \quad \vert \Delta_\mathrm{sc} \vert \approx 250 \mu \mathrm{eV} \nonumber 
\end{align}
Fig.\;\ref{additionalboundstateenergy1} shows $\Delta E_\mathrm{BS}(\Delta \mu \!=\! \delta \mu \!= \! 0)$ as a function of the wire width. A perceivable fermionic bound state formation is expected to be present as soon as $\Delta E_\mathrm{BS}(\Delta \mu \!=\! \delta \mu \!= \! 0) $ exceeds $ \vert \Delta_\mathrm{sc} \vert $, which happens for $W  \! \lesssim \! 100$nm. However, the associated formation of extra topological boundaries and correspondingly the formation of additional MF modes in the system  rather occurs for smaller values of $W$, since this requires a strong localization of the fermionic CP$_1$ bound states. 
Numerically, we observe this process for $W \! < \! 30\mathrm{nm}$, which corresponds to  $\Delta E_\mathrm{BS}(\Delta \mu \!=\! \delta \mu \!= \! 0) \! > \! 10 \, \vert \Delta_\mathrm{sc} \vert$. Nevertheless, we want to emphasize that due to $\Delta E_\mathrm{BS}(\Delta \mu \! \neq \! \delta \mu \!= \! 0) \ll \Delta E_\mathrm{BS}(\Delta \mu \!=\! \delta \mu \!= \! 0)$, this critical width decreases for $\Delta \mu \neq 0$, which is a mandatory requirement for all TQC algorithms.
Therefore, even though we expect additional HMF modes
in ungated, experimental TQC networks made of InSb nanowires of width $W\! \approx \!75\mathrm{nm}$ \cite{mourik}, we predict that such modes do not occur for our MF braiding operations due to the gating potential.

\section{Braiding Simulations} \label{Braidingstat}

\noindent
First, we analyze the braiding process $\gamma_1 \! \leftrightarrow \! \gamma_2$, using the exchange protocol, defined in Fig.~\ref{outerexchange}. We executed our simulations for various stub lengths $\LT=a\NT$ and $\Lt=a\Nt$, characterizing different  geometries of our tTt-setup (cf. setup Fig.\ref{outerexchange}), as well as for different gating times $\tstep$. Each of these parameters has a different effect on the braiding, which we separately discuss in the following.

\begin{figure*}[!t]
\centering
\includegraphics[width=.92\columnwidth]{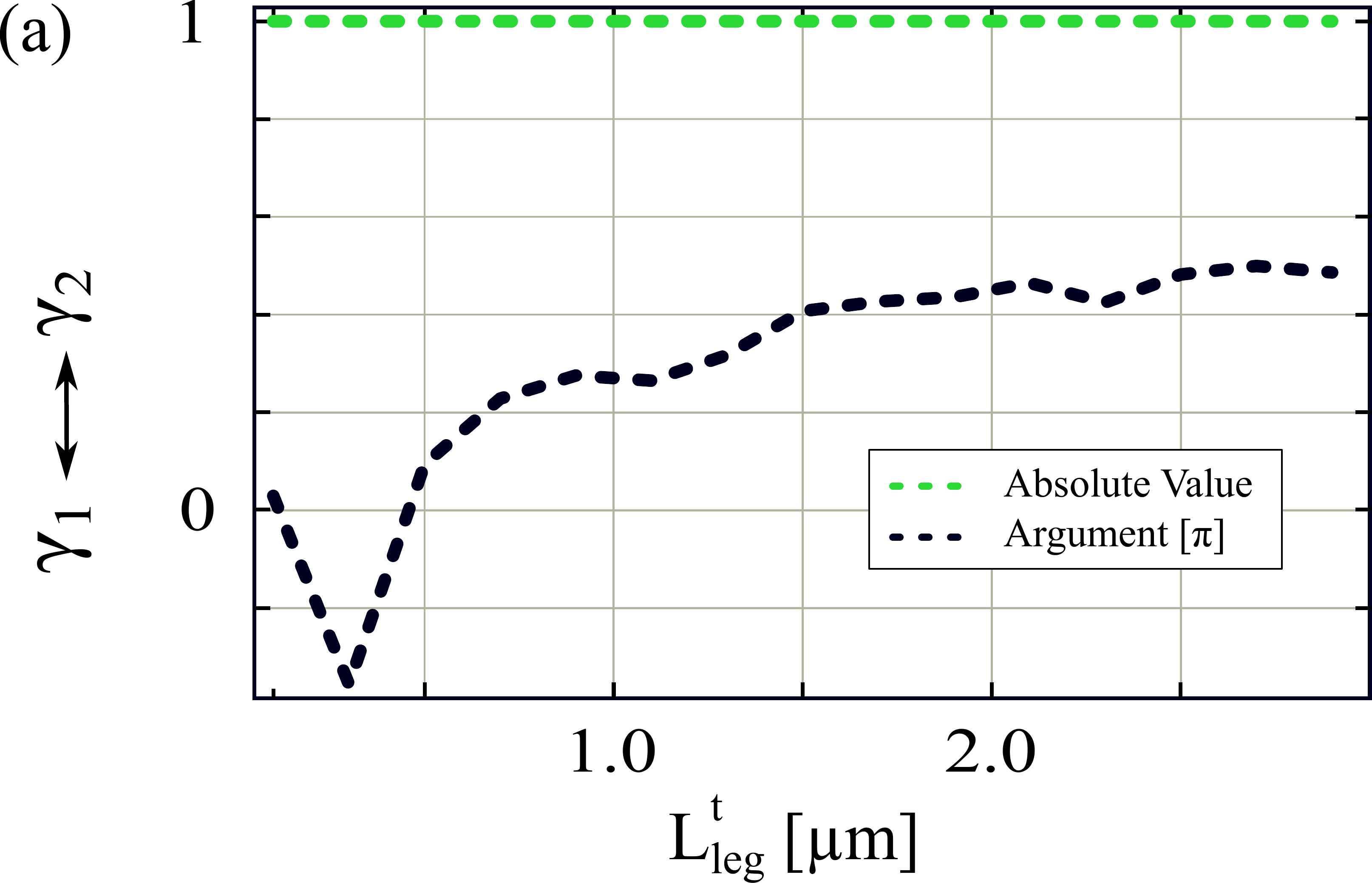}
\hfil
\includegraphics[width=.92\columnwidth]{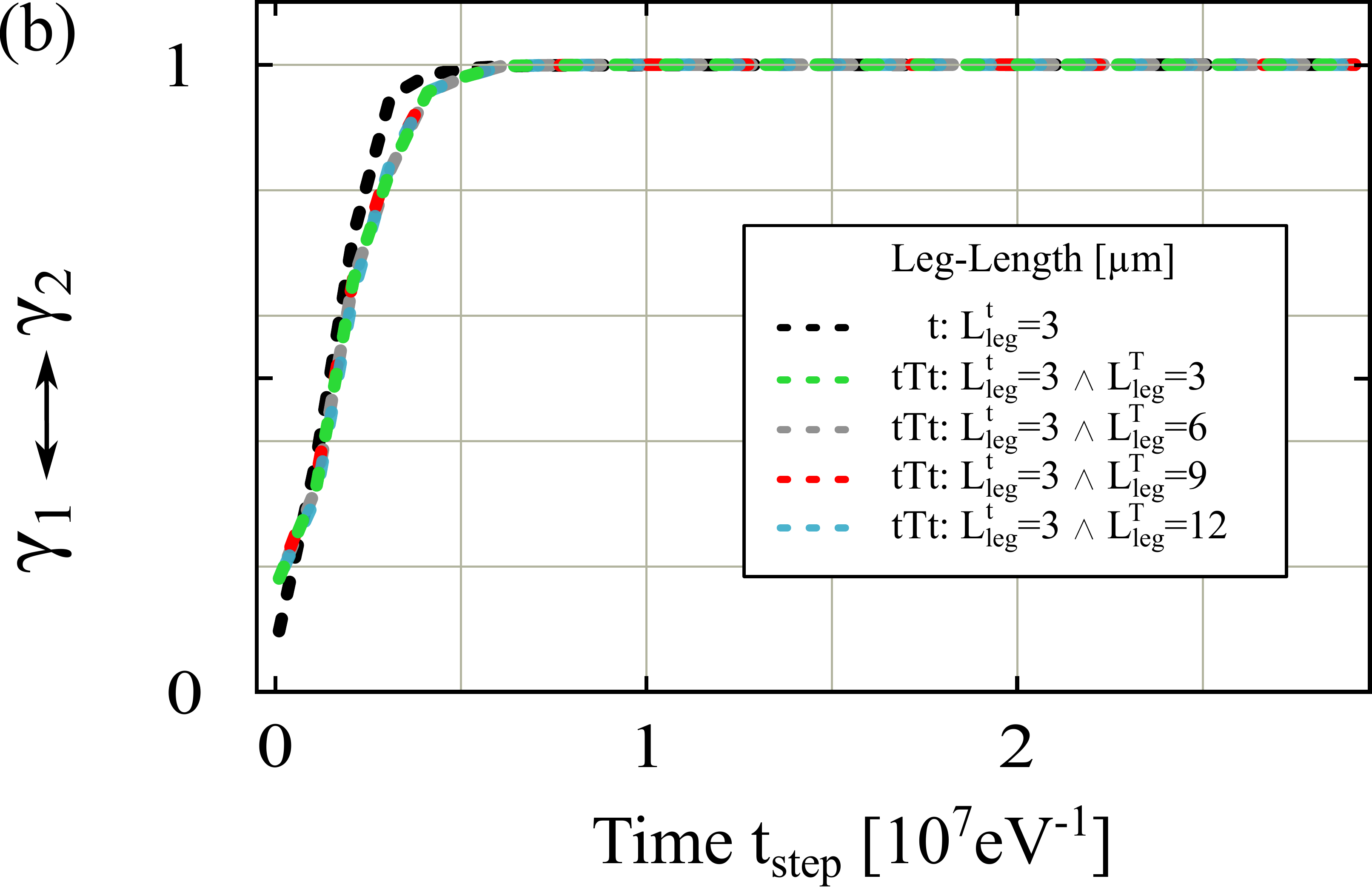}
\vfill
\includegraphics[width=.92\columnwidth]{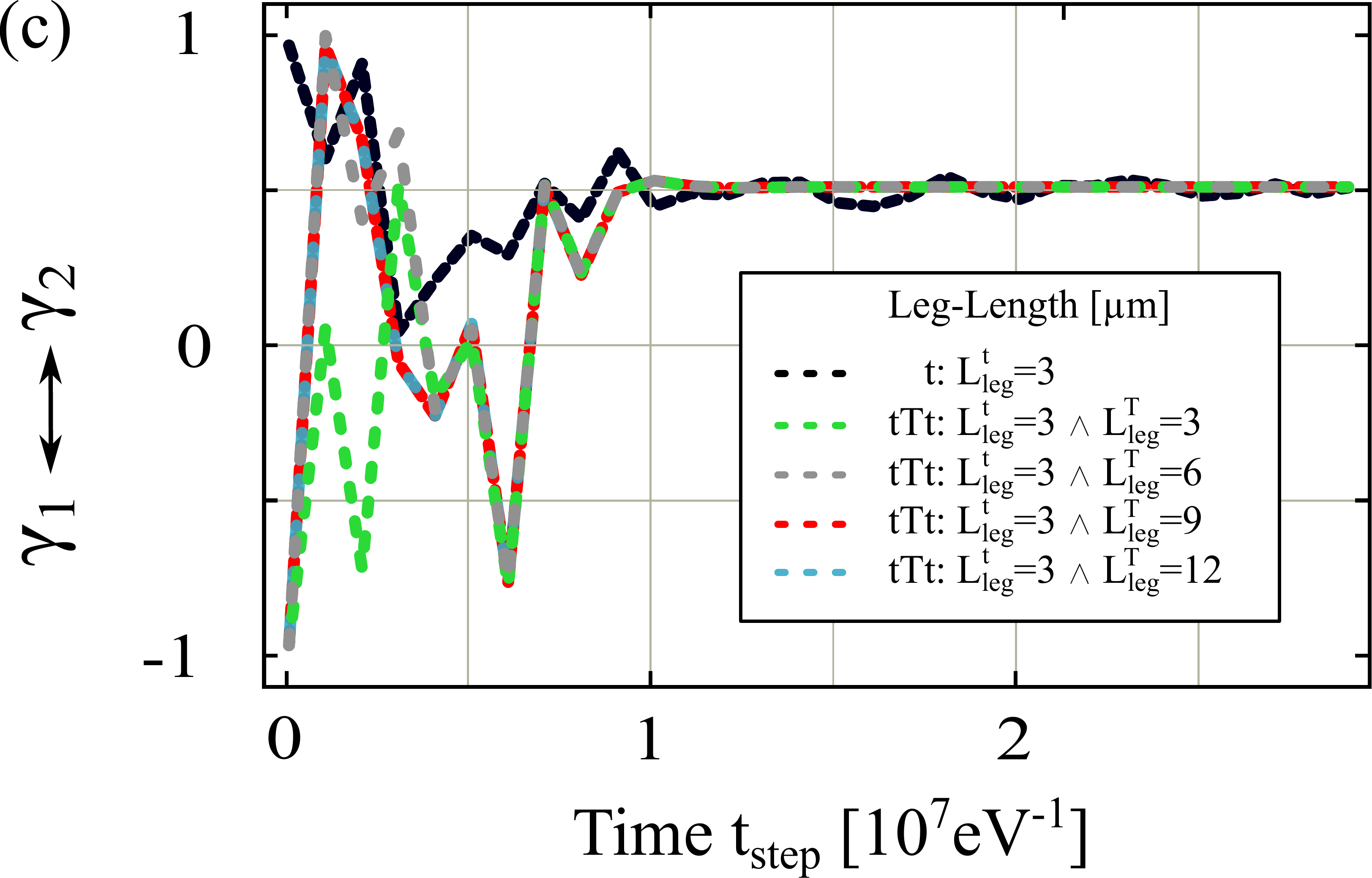}
\hfil
\includegraphics[width=.92\columnwidth]{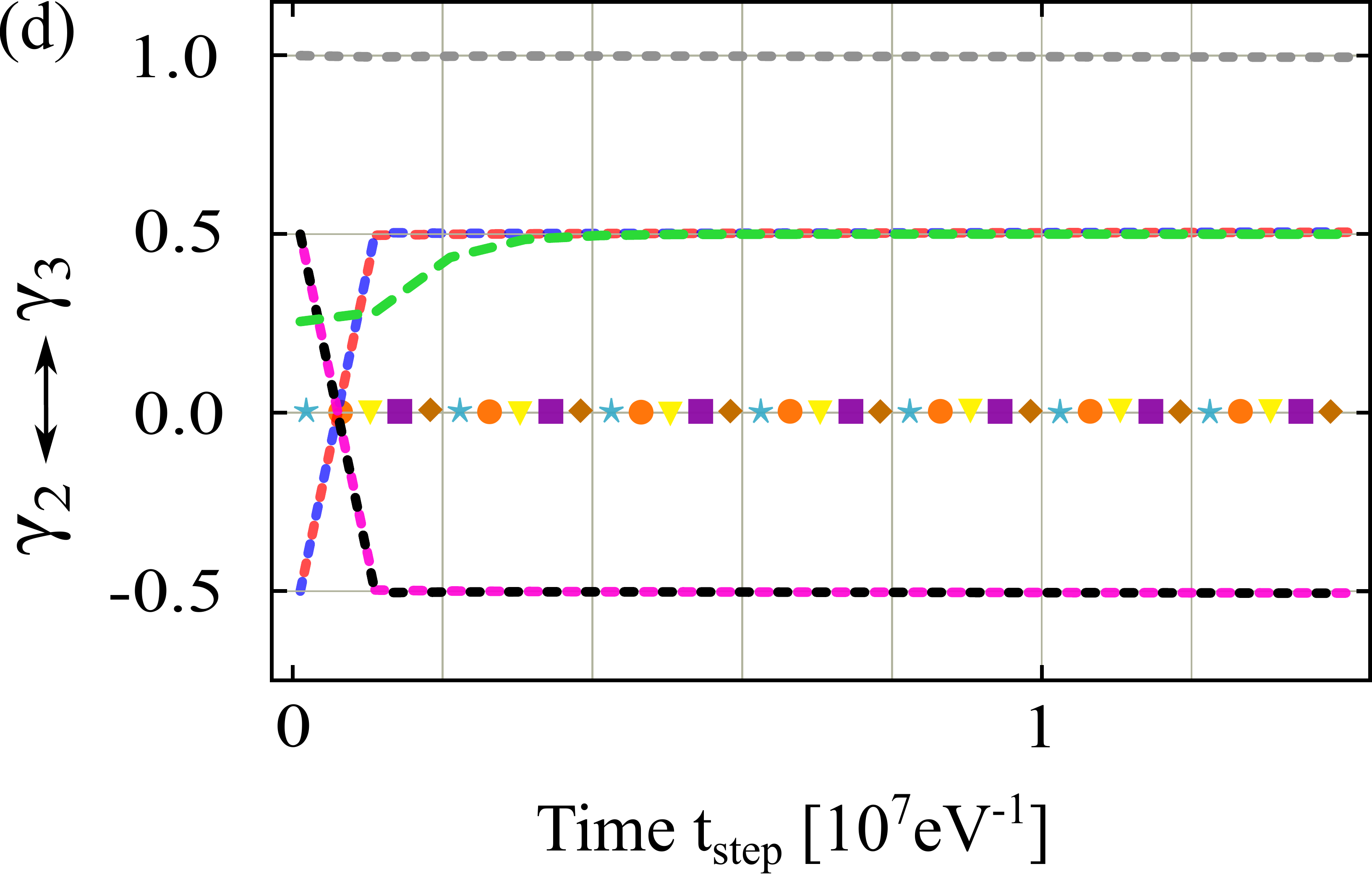}
\vfil
\caption{
\textbf{(a)} Absolute value and argument of $(\kappa_{12}(T))_{11}$ as a function of $\Lt$ for an (quasi-)adiabatic braiding $\gamma_1 \leftrightarrow \gamma_2$ with $L_\mathrm{Gate}=\Lt$ and $\LT= 6\mu$m. With increasing $\Lt$, $\mathrm{Arg}\left[ (\kappa_{12}(T))_{11} \right]$ approximates the theoretical prediction of $\pi/2$.
\textbf{(b)} $\vert \left(  \kappa_\mathrm{12}(T) \right)_\mathrm{11} \vert$ $\&$ 
\textbf{(c)} $ \mathrm{Arg} \left( \left( \kappa_\mathrm{12}(T) \right)_\mathrm{11} \right) $ $[\pi]$ as a function of $t_\mathrm{step}$ for several values $\LT \geq \Lt= L_\mathrm{Gate}$. For  all geometries and large $t_\mathrm{step}$ both quantities approximate their theoretical prediction. In the limit of small $t_\mathrm{step}$, non (quasi-)adiabatic effects lead to deviations from this limit.  
\textbf{(d)} $\overline{\mathrm{abs}}(T)$ $\&$ $\mathrm{Arg} [ \left( \kappa_\mathrm{23}(T) \right)_{\alpha,\beta} ][\pi]$ with $\lbrace \alpha,\beta \rbrace \in \lbrace 1,2,3,4 \rbrace$ are shown as a function of $t_\mathrm{step}$ for $\LT=2 \Lt= 2 L_\mathrm{gate} = 6 \mu \mathrm{m}$. The curves $  \mathrm{Arg} \left[ ( \kappa_\mathrm{12}(t_\mathrm{step}) )_{\alpha,\beta} \right][\pi]$, denoted in the form ($\alpha,\beta$,color), are given by (1,1,orange,$\bullet$), (2,2,brown,\begin{scriptsize} $\blacklozenge$\end{scriptsize}), (3,3,lightblue,$\star$), (4,4,yellow,\begin{scriptsize}$\blacktriangledown$\end{scriptsize}), (1,2,purple, \begin{tiny}$\blacksquare$\end{tiny}), (1,3,black), (2,3,red), (1,4,pink), (2,4,blue) and (3,4,gray). The average value $\overline{\mathrm{abs}}(T)$ is shown in green. In the (quasi-)adiabatic regime all quantities approximate their theoretical prediction.} 
\label{argandabs}
\end{figure*}

In Fig.\;\ref{argandabs}a, we illustrate $\left(\kappa_{12}(T)\right)_{11}$ as a function of $\Lt$, whereas $\tstep$ and $\LT$ are chosen appropriately.  
To ensure a successful braiding procedure, we need to satisfy  $\Delta_\mathrm{hyb}\left( \gamma_1, \gamma_2 \right) \ll \vert \Delta_\mathrm{sc} \vert$ throughout the entire exchange process, which can be achieved by keeping the  MF modes separated in space. For our set of parameters $\Lt > 2 \mu$m fulfills this condition, implying the predicted geometrical exchange phase of $\pi/2$ (cf. Tab.~\ref{exchangestatistics}). 
\noindent
Let us clarify the effect of $\LT$ on the exchange procedure $\gamma_1 \! \leftrightarrow \! \gamma_2$. The braiding statistics in Tab.\;\ref{exchangestatistics} was evaluated for the HMF basis vectors $\psi_{1,2,3,4}$. Explicitly, we assumed that the MF modes $\gamma_{1,2}$, as well as $\gamma_{3,4}$ hybridize to the associated HMF states. If the hybridization between $\gamma_{1,2}$ as well as between  $\gamma_{3,4}$ initially (finally) is  much stronger than the hybridization between $\gamma_{2,3}$, this is the proper description of our tTt structure at $t\!=\!0$ ($t\!=\!T$).
In general, this can be guaranteed either by choosing $\LT \gg \Lt$ or by initially (finally) applying a strong gating potential  $\mu_\mathrm{trivial} \! = \!  \mu_\mathrm{gate} \! + \! \mu_\mathrm{nontrivial}$, acting as a potential barrier in the central T structure\footnote{If this assumption is not satisfied this does not mean that we are not able to simulate topological braiding processes. However, for this scenario the extraction of topological phases becomes much more challenging, since the HMF modes are formed by a complex (arbitrary) composition of MF modes $\gamma_{1,2,3,4}$. Moreover, one needs to determine all hybridization amplitudes and derive the associated exchange statistics of HMF modes for each initial parameter configuration. Therefore, such a scenario is rather undesirable.}. We found that quantitatively this requirement corresponds to
\begin{align}  \label{geometric}
\dfrac{\Delta_\mathrm{hyb} \left( \gamma_1, \gamma_2 \right)}{\Delta_\mathrm{hyb} \left( \gamma_2, \gamma_3 \right)} &= \dfrac{\mathrm{e}^{-2\Lt}}{\mathrm{e}^{-2\LT} \times \left(\frac{\mu_\mathrm{trivial}}{\mu_\mathrm{nontrivial}} \right)^{-2 \LT}}  \gg 1  \nonumber
\\
\Rightarrow \quad \LT & \gg \Lt \left(1+\mathrm{ln} \left[ \dfrac{\mu_\mathrm{trivial}}{\mu_\mathrm{nontrivial}}  \right] \right)^{-1} \ . 
\end{align}
\noindent
Next, we specify requirements on $t_\mathrm{step}$ for a successful exchange process $\gamma_1 \leftrightarrow \gamma_2$ numerically as well as analytically. In Figs.~\ref{argandabs}b-c,  $ \vert \left( \kappa_\mathrm{12}(T) \right)_\mathrm{11} \vert$ and $\mathrm{Arg} \left[ \left( \kappa_\mathrm{12}(T) \right)_\mathrm{11} \right]$ are shown as a function of $t_\mathrm{step}$ for several values of $\LT \! \geq \! \Lt \!=\! L_\mathrm{gate}$. For large values of $t_\mathrm{step}$, all $\kappa_\mathrm{12}(T)$ matrix elements approximate their theoretical predictions (cf. Tab.~\ref{exchangestatistics}). In particular, we identify $t_\mathrm{step}^\mathrm{qad} \approx 6.5 \mathrm{n}$s to be the fastest, (quasi-)adiabatic gating time.
As introduced and discussed in Sec.~\ref{exchangestatistics}, the (quasi-)adiabatic time scale $t_\mathrm{step}^\mathrm{qad}$ defines the scale beyond which MF modes do not mix or scatter with bulk states, but are still able to freely rotate within their non-degenerated computational subspace. 
With $\mathrm{gate}=\Lt$, our  braiding protocol $\gamma_1 \leftrightarrow \gamma_2$ is related to six exchange steps. This leads to a minimal \mbox{(quasi-)}adiabatic braiding time of $ T_\mathrm{qad}(\gamma_1 \leftrightarrow \gamma_2) \approx 40 \ \mathrm{ns}$.
Let us analytically derive this lower boundary. In general, the adiabatic theorem of quantum mechanics predicts an adiabatic time evolution of a state $\vert n (t) \rangle$ with energy $E_n(t)$, if ($\forall \ m \neq n \ \wedge \ \forall \ t $) \cite{Puri,Griffiths}
\begin{align} \label{approx1}
\left \vert \left\langle m (t) \left\vert \dfrac{\mathrm{\partial} \hat H(t)}{\mathrm{\partial}t} \right\vert n (t) \right\rangle \right \vert \ll \hbar^{-1} \left\vert E_m (t) - E_n (t)\right\vert^2 \ .
\end{align}
Let us assume that $\vert n (t) \rangle$ defines a HMF mode, scattering with bulk modes $\vert m (t) \rangle$  in the non~(quasi-)adiabatic limit. The time dependent, TB scattering matrix $ \dot{H}(t)$ can be evaluated, using  Eq.~\eqref{timedephamilton}. For $\Ngate=1$, this matrix only  has one  nonzero entry related to lattice site $i$, if the gating potential is solely varied at this point in  the $j$-th exchange step ($t \in [t_{j-1};t_j]$):
\begin{align} \label{iielement}
\dfrac{\mathrm{\partial} H_{ii}(t)}{\mathrm{\partial}t} \! = \! \dfrac{\pi \mu_\mathrm{gate} }{  t_\mathrm{step}}  \sin \! \! \left[ \! \dfrac{\pi (t\!-\!t_{j\!-\!1})    }{2 t_\mathrm{step}} \! \right] \! \cos \! \! \left[\! \dfrac{\pi (t\!-\!t_{j\!-\!1})    }{2 t_\mathrm{step}} \! \right] .   
\end{align}
In general, Eq.~\eqref{approx1} needs to be valid for all exchange times. To determine an explicit \mbox{(quasi-)}adiabatic boundary for $t_\mathrm{step}$, we therefore need to derive the maximum value of $ \vert \langle m (t) \vert  [ \partial_t \hat H(t) ] \vert n (t) \rangle  \vert$.
With the maximum of Eq.~\eqref{iielement} and due to the fact that the HMF modes exactly peak at site $i \rightarrow i+1$ in the corresponding exchange step, we obtain
\begin{align} \label{approx2}
\mathrm{max} \! \left[ \left\langle \! m (t) \left\vert \dfrac{\mathrm{\partial} \hat H(t)}{\mathrm{\partial}t} \right\vert n (t) \! \right\rangle \right]\!  &   =  \dfrac{\pi}{2} \dfrac{ \mu_\mathrm{gate}}{t_\mathrm{step}}  \,
\mathrm{max} \left[  \leftidx{_i}{\langle} m (t) \vert n (t) \rangle_i \right]   \nonumber \\
 & =  \dfrac{\pi}{2 } \dfrac{\mu_\mathrm{gate}}{t_\mathrm{step}} \ . 
\end{align}
Therefore, Eqs.~\eqref{approx1}-\eqref{approx2} imply ($\left\vert E_m (t) \! - \! E_n (t) \right\vert  \approx  \vert \Delta_\mathrm{sc} \vert $):
\begin{align}\label{adiabaticeq}
\dfrac{\mu_\mathrm{gate}}{t_\mathrm{step}} \ll \hbar^{-1} \vert \Delta_\mathrm{sc} \vert^2 \ .
\end{align}
If we fix $\vert \Delta_\mathrm{sc} \vert$, Eq. \eqref{adiabaticeq} predicts a (quasi-)adiabatic exchange process either for  large $t_\mathrm{step}$, or rather for small $\mu_\mathrm{gate}$, which we confirmed numerically. For a fast exchange, one therefore needs to choose a small gating potential, which is still able to shift topological boundaries. Notice that Eq.~\eqref{adiabaticeq} satisfies the common and intuitive condition $t_\mathrm{step} \ll \vert \Delta_\mathrm{sc} \vert^{-1}$. However, for the first time, it also explicitly takes into account the rate with which the gating potential $\mu_\mathrm{gate}$ is ramped up/down. It is therefore much more accurate than all common qualitative estimations of (quasi-)adiabatic gating times in literature\cite{Sato14,Cheng14}.



\noindent
Let us analyze the exchange $\gamma_2 \! \leftrightarrow \! \gamma_3$, using the protocol defined in Fig.~\ref{innerexchange}. These simulations were executed as a function of $t_\mathrm{step}$ and $\LT$ with $L_\mathrm{gate}=\Lt=3\mu\mathrm{m}$.

To check the braiding statistics, given in Tab.~\ref{exchangestatistics}, we introduce a new, simplified quantity. For a successful exchange process $\gamma_2 \! \leftrightarrow \! \gamma_3$, any  matrix element of $\vert \kappa_\mathrm{23}(T)\vert$ is $\pi/2$. Thus, we define the average value
\vspace{.0cm}
\begin{align}
\overline{\mathrm{abs}}(T) \ \equiv 
\sum \limits_{ \lbrace \alpha,\beta \rbrace \in \lbrace 1,2,3,4 \rbrace} \dfrac{1}{16} \times \left\vert \left( \kappa_{23}(T) \right)_{\alpha,\beta} \right\vert \ , \nonumber
\end{align}
also counting $\pi/2$ for successful exchange processes of this kind.
Fig.\;\ref{argandabs}d, shows $\overline{\mathrm{abs}}(T)$ as well as $\mathrm{Arg} [ \left( \kappa_\mathrm{23}(T) \right)_{\alpha,\beta} ]$ as a function of $t_\mathrm{step}$ for $\LT = 2 \Lt$. One clearly determines (quasi-)adiabatic exchange processes for $t_\mathrm{step} \gtrsim 4$ns, perfectly agreeing with Eq.~\eqref{adiabaticeq}. We also checked that all single constituents of $\overline{\mathrm{abs}}(T)$ show the same feature. In comparison to $\gamma_1 \! \leftrightarrow \! \gamma_2$, the present exchange process comes along with a slightly smaller $t_\mathrm{step}^\mathrm{qad}$, resulting from slightly smaller HMF hybridization energies. 
In contrast to the first process, with $\mathrm{gate}=\Lt$ the current braiding is related to $14$ exchange steps, which leads to a minimal, (quasi-)adiabatic braiding time of $ T_\mathrm{qad}(\gamma_2 \! \leftrightarrow \! \gamma_3) \approx 55$ns. 
\\

\noindent
We close this section by discussing the upper, adiabatic boundary $t_\mathrm{step}^\mathrm{ad}$ for successful MF exchange processes. As introduced in Sec.~\ref{exchangesatsec}, this exchange time defines the scale above which all states, including HMF modes, completely evolve adiabatically.
For successful, (quasi-)adiabatic processes  with $t_\mathrm{step}^\mathrm{qad}  \ll   t   \ll  t_\mathrm{step}^\mathrm{ad}$, all HMF modes are allowed to rotate in the low energy computational subspace, while bulk excitations are effectively suppressed. In contrast, for $t_\mathrm{step}  >  t_\mathrm{step}^\mathrm{ad}$, these rotations are forbidden since all states purely evolve  adiabatically \cite{Sato14,Bauer2018}. Quantitatively, an exchange $\gamma_i \! \leftrightarrow \! \gamma_j$ is realized \mbox{(quasi-)adiabatically} if (cf. Eq.~\eqref{adiabaticeq})
\begin{align} \label{adiaboundary}
t_\mathrm{step} \ll \dfrac{\hbar \,  \mu_\mathrm{gate }}{\Delta_\mathrm{hyb}^{2} (\gamma_i, \gamma_j)} \ .
\end{align}
For our system $t_\mathrm{step}^\mathrm{ad} \! \approx \! 10^{5}\mathrm{s}$. Thus, $t_\mathrm{step} \! \ll \! t_\mathrm{step}^\mathrm{ad}$ is experimentally easy  accessible.

\section{Topological Quantum Computing}

\noindent
In this section, we will pedagogically derive how to realize universal TQC algorithms on our tTt structures. We start the discussion by defining a logical single qubit state in the even parity subspace, formed by the four MF modes $\gamma_{1,2,3,4}$  (cf. Tab.\ref{exchangestatistics})
\begin{align} \label{singleqbit}
\vert \tilde 0 \rangle \equiv \psi_1 \psi_3 \vert 00 \rangle = \vert 00 \rangle \ \ \wedge \ \ \vert \tilde 1 \rangle \equiv \psi_1^\dagger \psi_3^\dagger \vert 00 \rangle =  \vert 11 \rangle \ . 
\end{align}
To understand the following remarks, let us introduce some important mathematical group structures. For an $n$-qubit system, the Pauli group $\mathcal{P}_n$ is defined to be\cite{nielson}
\begin{align*}
\mathcal{P}_n \equiv \lbrace  \pm \im^{0,1} \left( \sigma_1 \otimes \cdot\cdot\cdot\otimes \sigma_n  \right)\vert \sigma_i \in \lbrace \sigma_0,\sigma_x,\sigma_y,\sigma_z \rbrace
 \rbrace \ ,
\end{align*}
where $\mathcal{P}_1$ is generated by $\sigma_x$, $\sigma_z$ and 
$\im \sigma_0$. Explicitly, this is characterized by $\mathcal{P}_1= \langle \sigma_x, \sigma_z,\im \sigma_0 \rangle$. Moreover, the group $\mathcal{P}_n=\mathcal{P}_1^{\otimes n}~\equiv~\mathcal{P}_1 \otimes~\cdot\cdot\cdot~\otimes~\mathcal{P}_1 $  
is a subgroup of the  $n$-qubit Clifford group $\mathcal{C}_n$,
which itself is generated by three characteristic operators: \cite{bravyikitaev} 
\begin{align*}
\mathcal{C}_n \equiv \langle H_i, K_i, \Lambda(\sigma_x)_{ij} \rangle \setminus U(1) \ .
\end{align*}
In the basis ($ \vert \tilde 0 \rangle, \vert \tilde 1 \rangle$), one defines:
\begin{small}
\begin{align*}
H \equiv \dfrac{1}{\sqrt{2}} \begin{pmatrix}
1&1\\
1&-1
\end{pmatrix} \ \wedge \ K \equiv \begin{pmatrix}
1&0\\
0& \im
\end{pmatrix}  \ \wedge \ \Lambda(\sigma_x) \equiv \begin{pmatrix}
I&0\\
0& \sigma_x
\end{pmatrix} ,
\end{align*}
\end{small}

\noindent
where $H$ represents the (single qubit $i$) Hadamard gate, $K$ denotes the (single qubit $i$) phase shift gate and $ \Lambda(\sigma_x)$ resembles the (two qubit $(i,j)$) \textbf{C}ontrolled \textbf{NOT} gate. Since $\mathcal{P}_1 \subset \mathcal{C}_1 $, Pauli gates can be constructed via	\cite{bravyikitaev}:
\begin{align} \nonumber
\sigma_z=K^2 \, , \, \sigma_x=HK^2H \, , \, \sigma_y = - \frac{\im}{2} \left[ \sigma_x, \sigma_y \right] \ .
\end{align}
Next, we introduce the  Gottesman-Knill theorem, stating that quantum computation schemes built on the following requirements are realizable in polynomial time on a probabilistic, classical computer \cite{Gottesman}:
\begin{enumerate}
\item[(I)]  Initialization of computational $\vert \tilde 0  \rangle$
\vspace{-.16cm}
\item[(II)] Realization of $\mathcal{C}_n$ operations
\vspace{-.16cm}
\item[(III)] Readout of $\mathcal{P}_n$ operators  
\vspace{-.16cm}
\item[(IV)] Classical control,  conditioned on (III).
\end{enumerate} 
We define $\mathcal{O}_\mathrm{ideal}$ to be the set of all algorithms based on this protocol \cite{bravyikitaev}. Since $\mathcal{C}_n$ alone does not represent a universal set of quantum gates, we extend $\mathcal{C}_n$ by another independent gate, ensuring this property \cite{bravialone}. For TQC algorithms it is convenient to extend $\mathcal{C}_n$ by the single qubit, $\pi/8$ phase gate \cite{hyart,karzig}
\begin{align*}
T \equiv \begin{pmatrix}
1&0\\
0& \mathrm{e}^{\im \pi/4}
\end{pmatrix} \ .
\end{align*}
\noindent
Up to overall phases, one can rewrite the MF braiding operators  $B_{ij}$ (cf. Tab.\;\ref{exchangestatistics}) in the even parity subspace via 
\begin{align*}
B_{1,2}=B_{3,4}= \mathrm{e}^{-\im \pi/4} \begin{pmatrix}
1&0\\
0 & \im 
\end{pmatrix} \   \wedge \ 
B_{2,3}= \dfrac{1}{\sqrt{2}} \begin{pmatrix}
1 &- \im \\
-\im & 1 \end{pmatrix}.
\end{align*}
\noindent
Thus, up to overall phases, we are able to construct the 
phase shift and Hadamard gate via consecutive braiding operations\cite{qubitsdeeper}:
\begin{align*}
K = B_{12}=B_{34} \quad \wedge \quad H = B_{1,2} B_{2,3} B_{1,2}   \ .
\end{align*} 
Hence, based on the Majorana braiding times in Sec.~\ref{Braidingstat}, we can realize the phase gate on the order of $\mathcal{O}(10\mathrm{ns})$ and the Hadamard gate on the order of  $\mathcal{O}(10^2\mathrm{ns})$. 
\noindent
To build the $\Lambda(\sigma_x)$ gate, we have to define a two qubit system, containing at least eight MF states $\gamma_{1,2, \dots ,8}$. In the even parity subspace, general two qubit states read
\begin{align*}
\vert \Psi \rangle = a \vert \tilde 0, \tilde 0 \rangle + b \vert \tilde 0, \tilde 1 \rangle + c  \vert \tilde 1, \tilde 0 \rangle + d  \vert \tilde 1, \tilde 1 \rangle \ ,
\end{align*}
with $(a,b,c,d) \in \mathbb{C}$.  Such a system can be defined on two adjacent tTt structures, where the first (second) qubit is characterized by $\gamma_{1,2,3,4}$ ($\gamma_{5,6,7,8}$). However, if MF braiding processes are the only present computational operations, an arbitrary two qubit state can always be written as a single qubit product state
\begin{align*}
\vert \Psi \rangle = \vert \Psi_1 \rangle \otimes \vert \Psi_2 \rangle \ ,
\end{align*}
known as the no-entanglement rule. 
Hence, the construction of entangled (Bell) states is not possible by MF braiding operations  alone \cite{bravialone}. 
This is a central problem of Majorana based TQC and justifies the requirement of projective measurements. We will prove this statement, using the stabilizer formalism \cite{Hastings17}.
Therefore, we define an abelian subgroup $S_{\vert \psi \rangle} \equiv \langle  g_1, \dots , g_n \rangle \subset P_n  $, generated by $n$-commuting and independent elements $g_i \in P_n $, such that $-\sigma^{\otimes n} \, \slashed \in \, S_{\vert \psi \rangle} $.  This group is called stabilizer and uniquely defines an $n$-qubit +1 eigenstate $\vert \psi \rangle$ via \cite{nielson,deVries}
\begin{align}\label{stabilizereg}
g_i \vert \psi \rangle=\vert \psi \rangle  \quad \forall \quad g_i \in S_{\vert \psi \rangle} \ \ \wedge  \ \ i \in \lbrace 1,2,...,n \rbrace \ .
\end{align}
For TQC codes, we define a vector space $V_S$ of dimension $2^k$, such that  $V_S$ is stabilized by $S=\langle g_1,...,g_{n-k}\rangle$\cite{nielson,Hastings17}. 
If one applies a unitary operation $U$ to $V_S$, one obtains 
\begin{align}
U \vert \psi \rangle = U g \vert \psi \rangle = UgU^\dagger U \vert \psi \rangle  \nonumber
\end{align}
for any $\vert \psi \rangle \in V_S$ and $g \in S$ \cite{nielson}. 
Hence, the vector space $V_S' \equiv U V_S$ is  stabilized by $S'  \equiv \lbrace UgU^\dagger \vert g \in S \rbrace$.
We are now in the position to prove the no-entanglement rule. Therefore, we consider an initial two qubit state  $ \vert \psi \rangle = \vert \tilde 0, \tilde 0 \rangle$, characterized by  (up to factors $\pm1,\pm \im$)\cite{bravialone}
\begin{align}
S_{\vert \psi \rangle}=\left( \gamma_1 \gamma_2,\gamma_3 \gamma_4,\gamma_5 \gamma_6,\gamma_7 \gamma_8 \right) \ . \nonumber
\end{align}
Applying a set of topological braiding gates, leads to 
\begin{align}
S_{\vert \psi \rangle '}=\left( \gamma_{p(1)} \gamma_{p(2)}, \gamma_{p(3)} \gamma_{p(4)},\gamma_{p(5)} \gamma_{p(6)},\gamma_{p(7)} \gamma_{p(8)} \right) \ , \nonumber
\end{align}
where $p(i) \in \lbrace 1,2, \dots ,8 \rbrace$. According to Eq.~\eqref{stabilizereg}, the two qubit state  
$\vert \psi \rangle '$ has to satisfy
\begin{align}
& \gamma_{p(1)} \gamma_{p(2)} \vert \psi \rangle' =  \vert \psi \rangle' \quad & \wedge \quad \gamma_{p(3)} \gamma_{p(4)} \vert \psi \rangle' =  \vert \psi \rangle' \nonumber \ \ \\ 
& \gamma_{p(5)} \gamma_{p(6)} \vert \psi \rangle' =  \vert \psi \rangle' \quad & \wedge \quad \gamma_{p(7)} \gamma_{p(8)} \vert \psi \rangle'=   \vert \psi \rangle' \nonumber \ .
\end{align}
Moreover, our even parity construction implies \cite{bravialone}
\begin{align} \nonumber
-\gamma_1 \gamma_2 \gamma_3 \gamma_4 \vert \psi \rangle '= - \gamma_5 \gamma_6 \gamma_7 \gamma_8  \vert  \psi \rangle '  \ .
\end{align}
To fulfill all these equations, one obtains for $1 \leq j \leq 4 \, :$
\begin{align}
  p(2j-1),p(2j) & \in \lbrace 1,2,3,4 \rbrace \nonumber \\
 \vee \quad p(2j-1),p(2j) &   \in \lbrace 5,6,7,8 \rbrace \ . \nonumber
\end{align}
Hence, each $g \in S_{\vert \psi \rangle '}$ is either  bilinear in $\gamma_{ 1,2,3,4 }$ or $\gamma_{ 5,6,7,8 }$, which implies the product structure of $\vert \psi \rangle ' \,$ \cite{bravialone}. 

To circumvent the no-entanglement rule, we will use the concept 
of so-called projective measurements, which will enable entanglement between topological (Majorana) qubits and therefore allow  the realization of CNOT gates. To understand how this comes about, let us consider that our system is defined by an $n$-qubit  state $\vert \psi \rangle$, stabilized by $S_{\vert \psi \rangle} = \langle g_1,...,g_n \rangle$. Further, we assume that one wants to measure an operator $g \in \mathcal{P}_n$. Without loss of generality we define $g$ to be a pure tensor product of Pauli matrices. After measuring $g$, the stabilizer $S_{\vert \psi \rangle}$  needs to be updated, leading to two different scenarios \cite{nielson}:
\begin{enumerate}
\item[(I)] $\forall i \in \lbrace 1,...,n \rbrace$: $[g,g_i]=0$.
\vspace{-.2cm}
\item[(II)] $\exists i \in \lbrace 1,...,n \rbrace$, such that   $\lbrace g,g_i \rbrace_+=0$.\\ Without loss of generality: $[ g,g_j]=0$ \\ for $j \in \lbrace 2,...,n \rbrace$ and $\lbrace g,g_1 \rbrace_+=0$.
\end{enumerate}
In case (I), either $g$ or $-g$ needs to be an element of $S_{\vert \psi \rangle}$. This is based on the fact that $g_i g \vert \psi \rangle = gg_i \vert \psi \rangle = g \vert \psi \rangle $ and $g^2=\sigma_0^{\otimes n} $ for all $i \in \lbrace 1,...,n \rbrace$, implying $g \vert \psi \rangle  = \pm \vert \psi \rangle$. Let us assume that $g \in S_{\vert \psi \rangle} $. In general, after measuring  $g$ with outcome $\lambda$, the $n$-qubit state $ \vert \psi \rangle $ needs to be multiplied by the projection operator
\begin{align}
\Pi_\lambda=\dfrac{1}{2} \left(\sigma_0^{\otimes n} + \lambda g \right). \nonumber
\end{align}
For $g \in S_{\vert \psi \rangle}$, $\lambda=1$ and $\Pi_1=\sigma_0^{\otimes n}$. Thus in scenario~(I), measuring $g$ does not effect $S_{\vert \psi \rangle}$\cite{nielson}. Such  processes can not generate any entanglement in the system. 
In scenario (II), the situation differs. This time, $\Pi_{\pm 1}$ dictates the  measurement probabilities 
\begin{align}
p(\pm 1)=\mathrm{Tr}\left( \dfrac{1 \pm g}{2} \vert \psi \rangle \langle \psi \vert \right) = \mathrm{Tr} \left( g_1 \dfrac{1 \mp g}{2} \vert \psi \rangle \langle \psi \vert \right) \ , \nonumber
\end{align}
where we used $g_1 \vert \psi \rangle = \vert \psi \rangle $ and $gg_1=-g_1 g$. With the cyclic property of the trace and with $g_1=g_1^\dagger$, one obtains $p(1)=p(-1)$. Due to $p(1)+p(-1)=1$, one finally gets $p(\pm1)=1/2$. After  measuring $g$ with outcome $\lambda=\pm1$, the projected states are given by \cite{nielson}
\begin{align}
\vert \psi^{\pm1} \rangle = \dfrac{\sigma_0^{\otimes n} \pm g}{\sqrt{2}} \vert \psi \rangle \quad \mathrm{with} \quad S_{\vert \psi^{\pm1}\rangle} = \langle \pm g,g_2,...,g_n \rangle \ . \nonumber
\end{align} 
Thus, in scenario (II), $g_1$ is replaced by $\pm g$ in $S_{\vert \psi^{\pm1}\rangle}$. This enables the addition of operator combinations $\gamma_{1,2,3,4} \gamma_{5,6,7,8}$ to the updated stabilizer, such that $S_{\vert \psi^{\pm1}\rangle}$ eventually defines an entangled (Bell) state. 

\begin{figure}
\begin{minipage}{1.\columnwidth}
\centering
\includegraphics[width=1.\columnwidth]{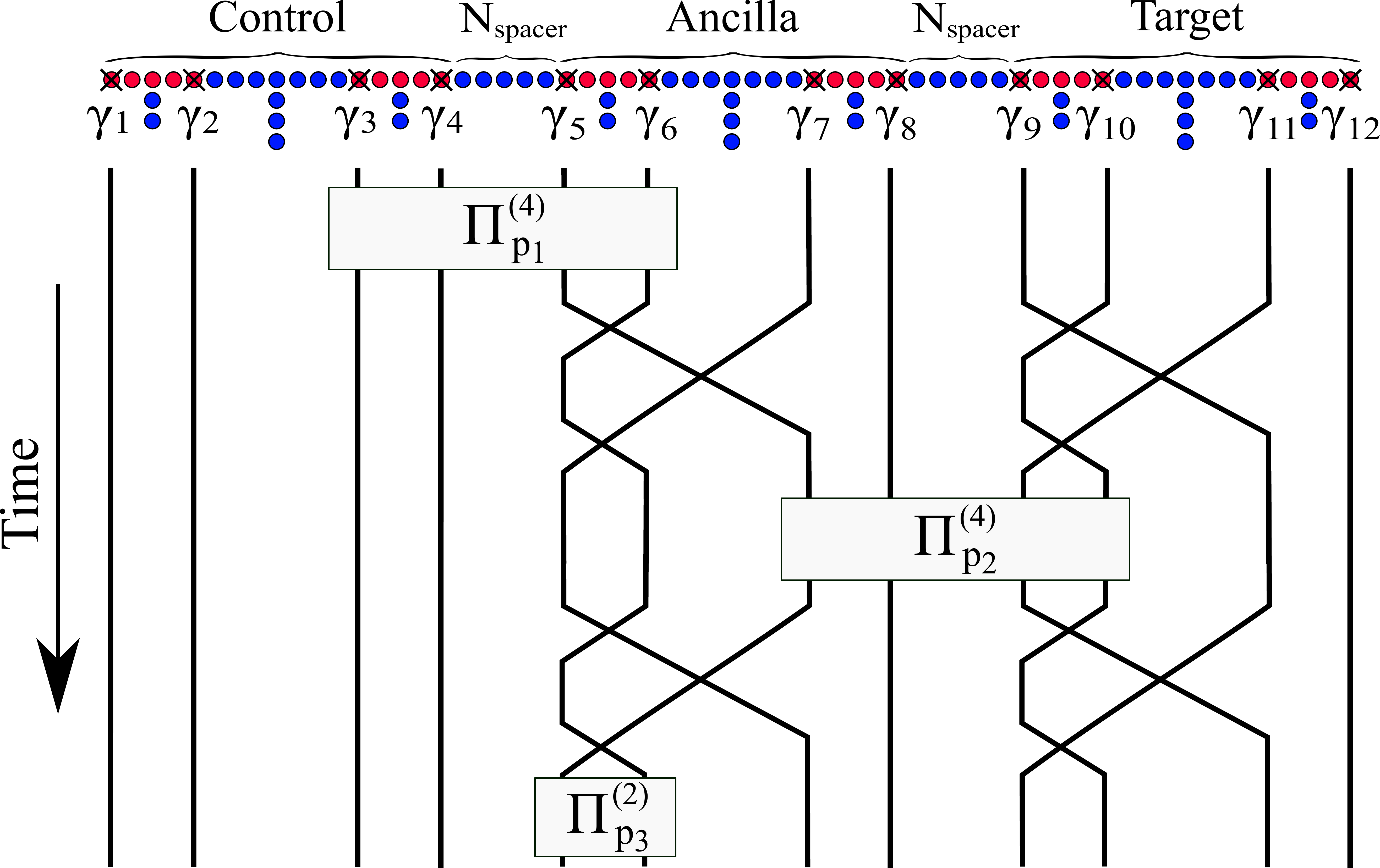} 
\vspace{-0pt}
\caption{Hardware setup of a CNOT gate. We show the world lines of all MF modes during this operation \cite{Zilberberg}. Between the two outer tTt building blocks, defining the logical qubits by their MF modes $\gamma_{1,2,3,4}$ and $\gamma_{9,10,11,12}$, one has to implement another tTt building block, hosting the ancillary MF modes $\gamma_{5,6,7,8}$. Blue sites indicate topologically trivial, whereas red ones encode topologically non-trivial regions. The single tTt junctions are separated by $N_\mathrm{spacer}$ lattice sites, satisfying Eq.~\eqref{geometric}. $\Pi_{1,2,3}$  represent non-destructive parity measurements with outcome $p_{1,2,3}$.}
\label{CNOTgaterealization}
\end{minipage}
\vfill
\vspace{15pt}
\begin{minipage}{1.\columnwidth}
\centering
\includegraphics[width=0.5\columnwidth]{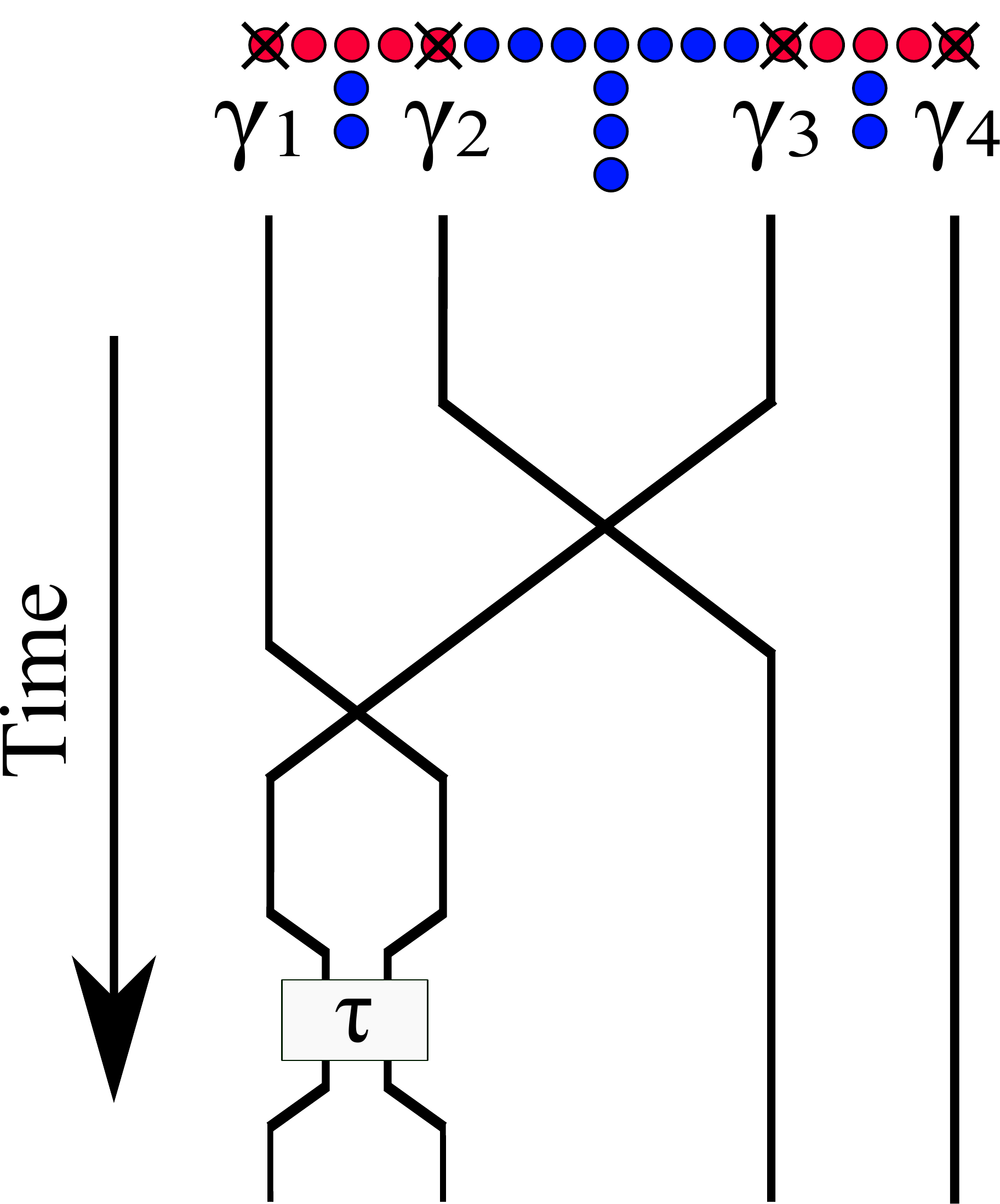} 
\vspace{10pt}
\caption{Preparation of a "noisy" copy of $\vert a_4 \rangle$, based on the short-range interaction of two MF modes during the time $\tau$ \cite{bravialone}.}
\label{prepa4}
\end{minipage}
\vfill
\vspace{25pt}
\begin{minipage}{1.\columnwidth}
\centering
\includegraphics[width=1.\columnwidth]{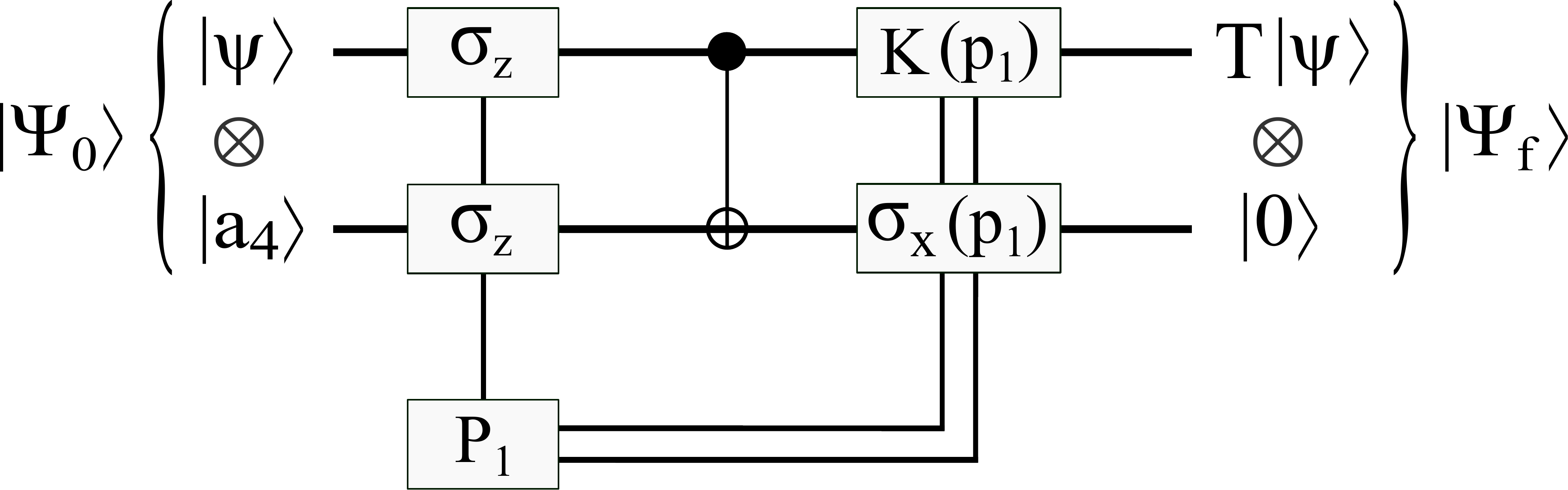} 
\caption{$\pi/8$ phase gate  with $\vert a_4 \rangle$ $\equiv$   $( \vert \bar 0 \rangle + \mathrm{e}^{\im \frac{\pi}{4}} \vert \bar 1\rangle )/\sqrt{2}$, $ K(p_1)\equiv \mathrm{e}^{-\im \frac{\pi}{8} \sigma_z(1-p_1)}$ and  $\sigma_x(p_1) \equiv \mathrm{e}^{\im \frac{\pi}{4} \sigma_x(1-p_1)}$. $p_1=\pm1$ represent measurement results of $P_1=\sigma_z \otimes \sigma_z$, which are transported via  classical channels  to  the subsequent  gates \cite{hyart}.}
\label{pi8gaterealization}
\end{minipage}

\end{figure}

Inspired by this idea, the first Majorana based CNOT gate was proposed by S. Bravyi and A. Kitaev in Refs.~[\citen{bravyikitaev,bravialone,Bravyikitaevfermonic}].
Their concrete gate realization relies on nondestructive parity measurements of ancillary MF states and long-range MF braiding operations.
For our tTt building blocks, we choose a slightly different CNOT gate structure, which was first suggested for $\nu=5/2$ Ising anyons in Ref.~\citen{Zilberberg} and has the advantage of strictly involving short-range MF braiding procedures. As a consequence, each MF mode stays in its initial tTt building block during all quantum operations, which is much easier to realize technically. Moreover, this setup is accompanied by smaller computation times in comparison to long-range based algorithms.  
In particular, Fig.\;\ref{CNOTgaterealization} illustrates
how to realize a CNOT gate on our nanowire based structure
and explicitly shows  the world lines of all involved MF modes. Explicitly, we insert a single tTt building block, hosting the four ancillary MF modes $\gamma_{5,6,7,8}$, between two tTt building blocks, which define the logical qubits. If \textbf{c},\textbf{t} and \textbf{a} encode the \textbf{c}ontrol, \textbf{t}arget and \textbf{a}ncillary qubit, our CNOT gate is given by
\begin{align} \label{CNOTgate}
\Lambda(\sigma_x) & \equiv  \Pi_{p_3}^{(2)}(5,6) \times H_\mathrm{a} \times H_\mathrm{t} \times \Pi_{p_2}^{(4)}(7,8,9,10) \\
& \times H_\mathrm{a} \times H_\mathrm{t} \times \Pi_{p_1}^{(4)}(3,4,5,6) \ .
\end{align}
Here $p_{1,2,3}\!=\!\pm1$ represent measurement results and $H_\mathrm{c,t,a}$ encode single qubit Hadamard gates. We use the following initial ancillary qubit configuration and the associated projectors of nondestructive measurements:
\begin{align*}
\left( \gamma_5+\im \gamma_6 \right) \vert \psi \rangle & = 0 & \Pi_{p_i}^{(2)}(p,q)& =\dfrac{1}{2} \left(1 -  \im p_i  \gamma_p \gamma_q \right)  \\
\left( \gamma_7+\im \gamma_8 \right) \vert \psi \rangle & = 0 & \Pi_{p_i}^{(4)}(p,q,r,s)& = \dfrac{1}{2} \left( \im + p_i \gamma_p \gamma_q \gamma_r \gamma_s \right)  \nonumber 
\end{align*}

\noindent
In particular, after initializing our qubits we act with a non-destructive parity measurement on the four MF modes $\gamma_{3,4,5,6}$. After that, we apply a Hadamard gate on the target and the ancilla qubit, respectively. Subsequently we non-destructively measure the parity of $\gamma_{7,8,9,10}$ followed by the application of another set of Hadamard gates acting on the ancilla and target qubit. Finally, we non-destructively measure the parity of the MF modes $\gamma_{5,6}$, which completes the protocol of our tTt-based CNOT gate.\\

\noindent
Last, let us discuss how to construct $\pi/8$ phase gates for tTt building blocks. Such operations can be realized by using ancillary $\vert a_4 \rangle \equiv ( \vert  \tilde 0 \rangle + \mathrm{e}^{\im \pi/4} \vert \tilde  1\rangle )/{\sqrt{2}}$ states \cite{Litinski172,bravialone}. While it is not possible to initialize these states topologically, we are able to construct so-called "noisy"  copies  of $\vert a_4 \rangle$ on our tTt structures, as  shown in Fig.\;\ref{prepa4}.  The initialization algorithm works as follows: We start with a single qubit state $\vert \tilde{0} \rangle$, defined by the two MF pairs $\gamma_{1,2}$ and $\gamma_{3,4}$. Then, via braiding processes,  we bring $\gamma_1$ and $\gamma_3 \rightarrow - \gamma_2$ sufficiently close to each other and afterwards let the system freely evolve for a certain time $\tau$. Eventually, we spatially separate  $\gamma_1$ and $\gamma_2$ again. Let us  analyze this construction scheme in detail. After the initial MF braiding operations, our system evolves to
\begin{align*}
\vert \Phi \rangle \equiv B_{1,2}^\dagger B_{2,3} \vert \tilde 0 \rangle =  \dfrac{1}{\sqrt{2}} \left( \vert \tilde 0 \rangle + \vert \tilde 1 \rangle \right) \ .
\end{align*}
During the time $\tau$, the short-range interaction between  MF states is described by the following Hamiltonian. which is constructed such that it commutes with all MF parity operators \cite{bravialone}:
\begin{align*}
H_\mathrm{int} \equiv  - \im \gamma_1 \gamma_2 \otimes X + \sigma_0 \otimes Y \ .
\end{align*} 
Here, $X$ and $Y$ are operators acting on the environment, which initially are  given by {\small $\vert \Psi_\mathrm{E} \rangle$}. Their explicit form is unimportant for
our line of reasoning, as will be clarified in the following.
Since $- \im \gamma_1 \gamma_2$ stabilizes $ \vert \tilde 0 \rangle $, our final state (including the environment)  is given by
\begin{small}
\begin{align*}
\vert \Phi \rangle \! \otimes \!  \vert \Psi_\mathrm{E} \rangle =\dfrac{1}{\sqrt{2}} \left( \vert \tilde 0 \rangle \! \otimes \!  \mathrm{e}^{\im (X+Y)\tau} \vert \Psi_\mathrm{E} \rangle + \vert  \tilde 1 \rangle \!  \otimes \! \mathrm{e}^{\im (-X+Y)\tau} \vert \Psi_\mathrm{E} \rangle \right)  .
\end{align*}
\end{small}

\vspace{-.2cm}
\noindent
Subsequently tracing over {\small $\vert \Psi_\mathrm{E} \rangle$} leads to a mixed state, described by the full density matrix \cite{bravialone}
\begin{small}
\begin{align*}
\rho \equiv \dfrac{1}{2} \begin{pmatrix}
1 & r \\
r^\star & 1
\end{pmatrix} \quad \mathrm{with} \quad r \equiv  \langle \Psi_\mathrm{E} \vert \mathrm{e}^{\im (X+Y)\tau} \mathrm{e}^{\im (X-Y)\tau} \vert  \Psi_\mathrm{E} \rangle \ .
\end{align*}
\end{small}

\noindent
By fine tuning $\tau$ we can achieve $r$=$\mathrm{e}^{\im \pi/4}$, which corresponds to $\rho$=$ \vert a_4 \rangle \langle a_4 \vert$. Since fine tuning is topologically not protected, we are just able to to create "noisy" copies of $\vert a_4 \rangle$ with a certain error probability $\epsilon$  \cite{sankar,bravialone}. Nevertheless, for $ \langle a_4 \vert \rho \vert a_4 \rangle = 1- \epsilon \gtrapprox 0.86$, these copies can be purified using 
a [[15,1,3]] Reed-Muller code, which is a special error correction code only including operations from $\mathcal{O}_\mathrm{ideal}$ and projective parity measurements. Therefore this code is perfectly realizable with our tTt building block structure  \cite{bravyikitaev}. For  $\epsilon \ll 1$, the final error probability after one round of purification  is given by $ \epsilon_\mathrm{out} \approx 35 \epsilon^3 + \mathcal{O}(\epsilon^4)$ \cite{bravyikitaev}.
    
As mentioned above, the ancillary $\vert a_4 \rangle$ state enables the construction of a $\pi/8$ phase gate\cite{bravialone}, which is visualized in Fig.\;\ref{pi8gaterealization}. For an arbitrary single qubit state
\begin{align*}
\vert \psi \rangle \equiv a \vert \tilde 0 \rangle + b   \vert \tilde 1 \rangle \quad \mathrm{with} \quad a,b \in \mathbb{C} \ \wedge \ \vert a \vert^2+\vert b \vert^2 =1 \  ,
\end{align*} 
we define the composite two qubit state 
\begin{align*}
\vert \Psi_0 \rangle \equiv \vert \psi \rangle \otimes \vert a_4 \rangle
\end{align*}
and measure its eigenvalues with respect to $P_1 \equiv  \sigma_z \otimes \sigma_z$. Both eigenvalues $p_1 = \pm1$   have probability $1/2$ and the  projected states are given by
\begin{align*}
\vert \Psi_1^+ \rangle & = a \vert \tilde 0, \tilde 0 \rangle + b  \mathrm{e}^{\im \frac{\pi}{4}} \vert \tilde 1, \tilde 1  \rangle \\
\vert \Psi_1^- \rangle & = a \mathrm{e}^{\im \frac{\pi}{4}} \vert \tilde 0, \tilde 1 \rangle + b   \vert \tilde 1, \tilde 0  \rangle \ . \nonumber
\end{align*}
If one subsequently  applies a CNOT gate, defining the first qubit as the control qubit, one gets
\begin{align*}
\vert \Psi_2^+ \rangle & = \left( a \vert \tilde 0 \rangle + b  \mathrm{e}^{\im \frac{\pi}{4}} \vert \tilde 1  \rangle \right) \otimes \vert \tilde 0 \rangle \\
\vert \Psi_2^- \rangle & = \left( a \mathrm{e}^{\im \frac{\pi}{4}} \vert \tilde 0 \rangle + b   \vert \tilde 1  \rangle \right) \otimes \vert \tilde 1 \rangle  \nonumber \ .
\end{align*}
Afterwards, if we measured $p_1=-1$, we apply a $K$ gate to the first and a $\sigma_x$ gate to the second qubit of $\vert \psi_2^- \rangle$, eventually leading to (for both results $p_1=\pm1$)
\begin{align*}
\vert \Psi_\mathrm{final} \rangle = \left(  a \vert \tilde 0 \rangle + b  \mathrm{e}^{\im \frac{\pi}{4}} \vert \tilde 1  \rangle \right) \otimes \vert \tilde  0 \rangle \ .
\end{align*}

\begin{sidewaysfigure}
\centering
\includegraphics[scale=0.235]{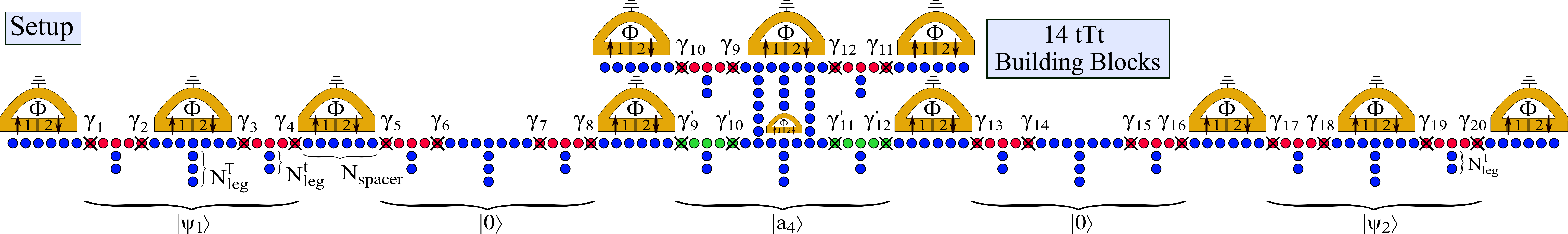}
    \caption{Hardware setup of a two qubit system, allowing for universal TQC algorithms. The two outer tTt building blocks define the logical qubits $\vert \psi_{1,2} \rangle$ via their MF modes $\gamma_{1,2,3,4}$ and $\gamma_{17,18,19,20}$. To enable a $\pi/8$ gate for each logical qubit, one has to generate $\vert a_4 \rangle$ in the middle tTt structure. This state is distilled in the top fifteen tTt building blocks via a [[15,1,3]] Reed-Muller code and finally transferred to the ancillary qubit structure, defined by $\gamma_{9,10,11,12}'$. CNOT operations between $\vert \psi_{1} \rangle $ and $ \vert \psi_{2} \rangle$ or between $ \vert \psi_{1,2} \rangle $ and $  \vert a_4 \rangle$, can be realized via the two ancillary tTt blocks, hosting the eight MF modes $\gamma_{5,6,7,8}$ and $\gamma_{13,14,15,16}$. They are initialized as stabilizer states $\vert \tilde 0 \rangle$. Blue sites indicate topologically trivial parameter regions, whereas red ones encode a topologically non-trivial ones. Green sites initially define topologically trivial junction parts, as well. To enable nondestructive, projective measurements, the device is connected to several superconducting flux qubits with islands $1$ and $2 $, which are shown in yellow. More explanations are given in the text.}
\label{completesetup2}
\end{sidewaysfigure}

\noindent
Hence, we realized a $\pi/8$ gate acting on $\vert \psi \rangle$ \cite{bravialone}. Similar to the CNOT gate construction in Fig.\;\ref{CNOTgaterealization}, the above algorithm can  be designed with tTt building blocks. 
In particular, one has to prepare  $\vert a_4 \rangle$ in an auxiliary tTt structure next to a tTt block, defining  $\vert \psi \rangle$.
To enable the intermediate CNOT operation, one needs to implement another auxiliary tTt building block between these two structures.

So far we did not clarify how to realize projective measurements as well as qubit readout schemes in our nanowire structure.  Particularly, one can achieve this by coupling all tTt building blocks to superconducting flux qubits with three Josephson junctions (\textbf{JJs})  \cite{hassler,treejunfc}. Their qubits are formed by clockwise and counterclockwise supercurrents, separated by an energy gap. Placing charge carriers next to the superconducting islands between the outer JJs closes this gap and therefore enables us to read off the implemented fermion parity electromagnetically. This allows the nondestructive, joined parity measurement of MF operator combinations, by bringing them close to these islands\cite{hassler,treejunfc}.\\

\noindent
In Fig.\;\ref{completesetup2}, we finally suggest our Majorana based two qubit system, allowing for universal TQC algorithms. Here, the logical qubits $\vert \psi_{1,2} \rangle$ are defined by $\gamma_{1,2,3,4}$ and $\gamma_{17,18,19,20}$. 
While all single qubit operations from $\mathcal{O}_\mathrm{ideal}$, such as the Hadamard gate H or the phase gate K, can be realized by the braiding of MF modes within the qubits they define, the inner tTt building blocks serve as ancillary qubits for $\pi/8$ and CNOT gate operations. In particular, the ancillary qubits, formed by  $\gamma_{5,6,7,8}$ and $\gamma_{13,14,15,16}$, are initialized in stabilizer states $\vert \tilde 0 \rangle$ and enable the CNOT gate operations by using the protocol given in Eq.\eqref{CNOTgate}. In the middle tTt building block, one can implement a purified $\vert a_4 \rangle$ state, allowing for $\pi/8$ gate operations. This state is distilled within the top fifteen tTt building blocks, using the error correction protocol, explained above. After one elementary purification round $\gamma_{9,10,11,12}$ define $\vert a_4 \rangle$ and are subsequently shifted to their final positions, indicated by $\gamma_{9,10,11,12}'$. Additionally, we use superconducting flux qubits with islands $1 \& 2$ for the realization of 
nondestructive projective parity measurements. We want to remark that such measurements can also be performed with superconducting charge qubits in a transmission line resonator \cite{transmons}. They promise a reduced sensitivity to charge noise, but are much larger in area ($300 \mu$m vs. $3 \mu$m) \cite{naturepics,transmons}. 
Last but not least, we emphasize that it is possible to horizontally stick together $m \in \mathbb{N}$ of these building blocks. This naturally defines a $2m$-qubit system.

\vspace{-0.2cm}

\section{Conclusion $\&$ Outlook}

\vspace{-0.2cm}

\noindent
In this work, we suggested a Majorana based multi qubit setup, which satisfies all DiVincenzo criteria \cite{nielson}. This structure is suitable for universal quantum computing, based on Majorana braiding operations, non-Clifford gate distillation protocols, as well as nondestructive, projective measurements. We explicitly simulated the exchange of four Majorana modes on a branched nanowire device, consisting of an inner T-shaped and two outer t-shaped structures.
To ensure our computational basis, one  has to make sure that the central T-structure is much larger than the two side ones. Moreover, we showed that in the adiabatic limit the Majorana gating time needs to be smaller than the inverse of the squared superconducting gap and scales linearly with the gating potential.  Further, we presented a formalism to correct the appearance of additional Majorana modes in very narrow, branched nanowire systems. We explicitly derived an associated correction potential, by using a Green's function approach.\\

\renewcommand{\arraystretch}{1.4}
\begin{table}[h]
\begin{center}
\begin{small} 
\begin{tabular}{|c||c|c|c|}
\hline
Quantum System & $ T_\mathrm{de}$ [s] & $T_\mathrm{op}$ [s] & $n_\mathrm{op} $  \\
\hline
\hline
Nuclear Spin & $10^{-2}$-$10^{-8}$ & $10^{-3}$-$10^{-6}$ & $10^{5}$-$10^{14}$\\
\hline
Electron Spin & $10^{-3}$ & $10^{-7}$ & $10^{4}$\\
\hline
Ion trap ($\mathrm{In}^+$) & $10^{-1}$ & $10^{-14}$ & $10^{13}$\\
\hline
Single $\mathrm{e}^-$-Au & $10^{-8}$ & $10^{-14}$ & $10^{6}$\\
\hline
Single $\mathrm{e}^-$-GaAs & $10^{-10}$ & $10^{-13}$ & $10^{3}$\\
\hline
Quantum Dot & $10^{-6}$ & $10^{-9}$ & $10^{3}$\\
\hline
Optical cavity & $10^{-5}$ & $10^{-14}$ & $10^{9}$\\
\hline
Microwave cavity  & $10^{0}$ & $10^{-4}$ & $10^{4}$\\
\hline
Majorana TQC  & $10^{-4}$-$10^{-5}$ & $10^{-8}$ & $10^3$-$10^{4}$\\
\hline
\end{tabular}
\caption{Decoherence $T_\mathrm{de}$ and operation $T_\mathrm{op}$ time scales of various  quantum computing concepts, as well as the amount  of possible quantum bit operations $n_\mathrm{op} \equiv T_\mathrm{de}/T_\mathrm{op}$ \cite{nielson}.}
\label{energyscalecpomparison}
\end{small} 
\end{center}
\vspace{-.3cm}
\end{table}
\noindent
From a broader perspective, there are many non-topological approaches to construct a quantum computer. The explicit decoherence  and operation  time scales $T_\mathrm{de}$ and $T_\mathrm{op}$, as well as the amount $n_\mathrm{op}$ of possible quantum bit operations during $T_\mathrm{de}$, are visualized within Tab.~\ref{energyscalecpomparison} for various qubit systems. 
We also embedded our TQC approach within this table. 
Recently, it was shown that the protection of Majorana based qubits against decoherence is heavily influenced by single electron tunneling from external sources, known as quasi-particle poisoning \cite{trauzettel,goldstein,loss}. 
In our system, the external source is given by the underlying superconductor.
Such processes change the fermion parity, generate bit flips and therefore make successful TQC algorithms impossible. Even worse, MF modes can hybridize with the mentioned states, decay and loose their anyonic features. 
Thus, the quasi-particle tunneling time scale sets an upper boundary for the coherence time in our device \cite{sankar}. 
Recent experiments predict Majorana coherence time scales which are longer than 10ms in proximitized semiconducting nanowires \cite{10ms}. 
Hence, if one compares the timescales in Tab.~\ref{energyscalecpomparison}, takes into account that our quantum algorithms are topologically protected and that all DiVincenzo criteria are satisfied, the present TQC approach is highly interesting for upcoming experimental realizations. 


\begin{acknowledgments}
\emph{Acknowledgments.} 
We thank Ronny Thomale and  Peter Schmitteckert for useful discussions. We acknowledge financial support through the 
Deutsche Forschungsgemeinschaft (DFG, German Research Foundation) – Project-ID 258499086 – SFB 1170, the ENB Graduate School on Topological Insulators and through the W\"urzburg-Dresden Cluster of Excellence on Complexity and Topology in Quantum Matter - ct.qmat \mbox{(EXC 2147, project-id 39085490)}. Additionally, this work was financially supported by the German research foundation DFG (Grant No.~HA 5893/3-1).
\end{acknowledgments}

\vspace{-.2cm}


\bibliographystyle{apsrev4-1}

\begin{thebibliography}{95}%
\makeatletter
\providecommand \@ifxundefined [1]{%
 \@ifx{#1\undefined}
}%
\providecommand \@ifnum [1]{%
 \ifnum #1\expandafter \@firstoftwo
 \else \expandafter \@secondoftwo
 \fi
}%
\providecommand \@ifx [1]{%
 \ifx #1\expandafter \@firstoftwo
 \else \expandafter \@secondoftwo
 \fi
}%
\providecommand \natexlab [1]{#1}%
\providecommand \enquote  [1]{``#1''}%
\providecommand \bibnamefont  [1]{#1}%
\providecommand \bibfnamefont [1]{#1}%
\providecommand \citenamefont [1]{#1}%
\providecommand \href@noop [0]{\@secondoftwo}%
\providecommand \href [0]{\begingroup \@sanitize@url \@href}%
\providecommand \@href[1]{\@@startlink{#1}\@@href}%
\providecommand \@@href[1]{\endgroup#1\@@endlink}%
\providecommand \@sanitize@url [0]{\catcode `\\12\catcode `\$12\catcode
  `\&12\catcode `\#12\catcode `\^12\catcode `\_12\catcode `\%12\relax}%
\providecommand \@@startlink[1]{}%
\providecommand \@@endlink[0]{}%
\providecommand \url  [0]{\begingroup\@sanitize@url \@url }%
\providecommand \@url [1]{\endgroup\@href {#1}{\urlprefix }}%
\providecommand \urlprefix  [0]{URL }%
\providecommand \Eprint [0]{\href }%
\providecommand \doibase [0]{http://dx.doi.org/}%
\providecommand \selectlanguage [0]{\@gobble}%
\providecommand \bibinfo  [0]{\@secondoftwo}%
\providecommand \bibfield  [0]{\@secondoftwo}%
\providecommand \translation [1]{[#1]}%
\providecommand \BibitemOpen [0]{}%
\providecommand \bibitemStop [0]{}%
\providecommand \bibitemNoStop [0]{.\EOS\space}%
\providecommand \EOS [0]{\spacefactor3000\relax}%
\providecommand \BibitemShut  [1]{\csname bibitem#1\endcsname}%
\let\auto@bib@innerbib\@empty
\bibitem [{\citenamefont {{Kitaev}}(2003)}]{Kitaev20032}%
  \BibitemOpen
  \bibfield  {author} {\bibinfo {author} {\bibfnamefont {A.~Y.}\ \bibnamefont
  {{Kitaev}}},\ }\href@noop {} {\bibfield  {journal} {\bibinfo  {journal}
  {Annals of Physics}\ }\textbf {\bibinfo {volume} {303}},\ \bibinfo {pages} {2
  } (\bibinfo {year} {2003})}\BibitemShut {NoStop}%
\bibitem [{\citenamefont {{Das Sarma}}\ \emph {et~al.}(2015)\citenamefont {{Das
  Sarma}}, \citenamefont {{Freedman}},\ and\ \citenamefont {{Nayak}}}]{sankar}%
  \BibitemOpen
  \bibfield  {author} {\bibinfo {author} {\bibfnamefont {S.}~\bibnamefont {{Das
  Sarma}}}, \bibinfo {author} {\bibfnamefont {M.}~\bibnamefont {{Freedman}}}, \
  and\ \bibinfo {author} {\bibfnamefont {C.}~\bibnamefont {{Nayak}}},\
  }\href@noop {} {\bibfield  {journal} {\bibinfo  {journal} {NPJ Quantum
  Information}\ }\textbf {\bibinfo {volume} {1}},\ \bibinfo {pages} {15001}
  (\bibinfo {year} {2015})}\BibitemShut {NoStop}%
\bibitem [{\citenamefont {Fu}\ and\ \citenamefont {Kane}(2008)}]{Fu08}%
  \BibitemOpen
  \bibfield  {author} {\bibinfo {author} {\bibfnamefont {L.}~\bibnamefont
  {Fu}}\ and\ \bibinfo {author} {\bibfnamefont {C.~L.}\ \bibnamefont {Kane}},\
  }\href {\doibase 10.1103/PhysRevLett.100.096407} {\bibfield  {journal}
  {\bibinfo  {journal} {Phys. Rev. Lett.}\ }\textbf {\bibinfo {volume} {100}},\
  \bibinfo {pages} {096407} (\bibinfo {year} {2008})}\BibitemShut {NoStop}%
\bibitem [{\citenamefont {Fu}\ and\ \citenamefont {Kane}(2009)}]{Fu09}%
  \BibitemOpen
  \bibfield  {author} {\bibinfo {author} {\bibfnamefont {L.}~\bibnamefont
  {Fu}}\ and\ \bibinfo {author} {\bibfnamefont {C.~L.}\ \bibnamefont {Kane}},\
  }\href {\doibase 10.1103/PhysRevB.79.161408} {\bibfield  {journal} {\bibinfo
  {journal} {Phys. Rev. B}\ }\textbf {\bibinfo {volume} {79}},\ \bibinfo
  {pages} {161408} (\bibinfo {year} {2009})}\BibitemShut {NoStop}%
\bibitem [{\citenamefont {Cook}\ and\ \citenamefont {Franz}(2011)}]{Cook11}%
  \BibitemOpen
  \bibfield  {author} {\bibinfo {author} {\bibfnamefont {A.}~\bibnamefont
  {Cook}}\ and\ \bibinfo {author} {\bibfnamefont {M.}~\bibnamefont {Franz}},\
  }\href {\doibase 10.1103/PhysRevB.84.201105} {\bibfield  {journal} {\bibinfo
  {journal} {Phys. Rev. B}\ }\textbf {\bibinfo {volume} {84}},\ \bibinfo
  {pages} {201105} (\bibinfo {year} {2011})}\BibitemShut {NoStop}%
\bibitem [{\citenamefont {Choy}\ \emph {et~al.}(2011)\citenamefont {Choy},
  \citenamefont {Edge}, \citenamefont {Akhmerov},\ and\ \citenamefont
  {Beenakker}}]{Choy11}%
  \BibitemOpen
  \bibfield  {author} {\bibinfo {author} {\bibfnamefont {T.-P.}\ \bibnamefont
  {Choy}}, \bibinfo {author} {\bibfnamefont {J.~M.}\ \bibnamefont {Edge}},
  \bibinfo {author} {\bibfnamefont {A.~R.}\ \bibnamefont {Akhmerov}}, \ and\
  \bibinfo {author} {\bibfnamefont {C.~W.~J.}\ \bibnamefont {Beenakker}},\
  }\href {\doibase 10.1103/PhysRevB.84.195442} {\bibfield  {journal} {\bibinfo
  {journal} {Phys. Rev. B}\ }\textbf {\bibinfo {volume} {84}},\ \bibinfo
  {pages} {195442} (\bibinfo {year} {2011})}\BibitemShut {NoStop}%
\bibitem [{\citenamefont {{Kitaev}}(2006)}]{Kitaev06}%
  \BibitemOpen
  \bibfield  {author} {\bibinfo {author} {\bibfnamefont {A.}~\bibnamefont
  {{Kitaev}}},\ }\href@noop {} {\bibfield  {journal} {\bibinfo  {journal}
  {Annals of Physics}\ }\textbf {\bibinfo {volume} {321}},\ \bibinfo {pages}
  {2} (\bibinfo {year} {2006})}\BibitemShut {NoStop}%
\bibitem [{\citenamefont {Greiter}\ and\ \citenamefont
  {Thomale}(2009)}]{Greiter09}%
  \BibitemOpen
  \bibfield  {author} {\bibinfo {author} {\bibfnamefont {M.}~\bibnamefont
  {Greiter}}\ and\ \bibinfo {author} {\bibfnamefont {R.}~\bibnamefont
  {Thomale}},\ }\href {\doibase 10.1103/PhysRevLett.102.207203} {\bibfield
  {journal} {\bibinfo  {journal} {Phys. Rev. Lett.}\ }\textbf {\bibinfo
  {volume} {102}},\ \bibinfo {pages} {207203} (\bibinfo {year}
  {2009})}\BibitemShut {NoStop}%
\bibitem [{\citenamefont {Moore}\ and\ \citenamefont {Read}(1991)}]{Moore91}%
  \BibitemOpen
  \bibfield  {author} {\bibinfo {author} {\bibfnamefont {G.}~\bibnamefont
  {Moore}}\ and\ \bibinfo {author} {\bibfnamefont {N.}~\bibnamefont {Read}},\
  }\href {\doibase https://doi.org/10.1016/0550-3213(91)90407-O} {\bibfield
  {journal} {\bibinfo  {journal} {Nuclear Physics B}\ }\textbf {\bibinfo
  {volume} {360}},\ \bibinfo {pages} {362 } (\bibinfo {year}
  {1991})}\BibitemShut {NoStop}%
\bibitem [{\citenamefont {{Read}}\ and\ \citenamefont
  {{Green}}(2000)}]{readgreen}%
  \BibitemOpen
  \bibfield  {author} {\bibinfo {author} {\bibfnamefont {N.}~\bibnamefont
  {{Read}}}\ and\ \bibinfo {author} {\bibfnamefont {D.}~\bibnamefont
  {{Green}}},\ }\href@noop {} {\bibfield  {journal} {\bibinfo  {journal} {Phys.
  Rev. B}\ }\textbf {\bibinfo {volume} {61}},\ \bibinfo {pages} {10267}
  (\bibinfo {year} {2000})}\BibitemShut {NoStop}%
\bibitem [{\citenamefont {{Kitaev}}(2001)}]{kitaevoriginalmajoranapaper}%
  \BibitemOpen
  \bibfield  {author} {\bibinfo {author} {\bibfnamefont {A.~Y.}\ \bibnamefont
  {{Kitaev}}},\ }\href {http://stacks.iop.org/1063-7869/44/i=10S/a=S29}
  {\bibfield  {journal} {\bibinfo  {journal} {Physics-Uspekhi}\ }\textbf
  {\bibinfo {volume} {44}},\ \bibinfo {pages} {131} (\bibinfo {year}
  {2001})}\BibitemShut {NoStop}%
\bibitem [{\citenamefont {{Ivanov}}(2001)}]{ivanov}%
  \BibitemOpen
  \bibfield  {author} {\bibinfo {author} {\bibfnamefont {D.~A.}\ \bibnamefont
  {{Ivanov}}},\ }\href@noop {} {\bibfield  {journal} {\bibinfo  {journal}
  {Phys. Rev. Lett.}\ }\textbf {\bibinfo {volume} {86}},\ \bibinfo {pages}
  {268} (\bibinfo {year} {2001})}\BibitemShut {NoStop}%
\bibitem [{\citenamefont {Sau}\ \emph {et~al.}(2010)\citenamefont {Sau},
  \citenamefont {Lutchyn}, \citenamefont {Tewari},\ and\ \citenamefont
  {Das~Sarma}}]{sau10}%
  \BibitemOpen
  \bibfield  {author} {\bibinfo {author} {\bibfnamefont {J.~D.}\ \bibnamefont
  {Sau}}, \bibinfo {author} {\bibfnamefont {R.~M.}\ \bibnamefont {Lutchyn}},
  \bibinfo {author} {\bibfnamefont {S.}~\bibnamefont {Tewari}}, \ and\ \bibinfo
  {author} {\bibfnamefont {S.}~\bibnamefont {Das~Sarma}},\ }\href {\doibase
  10.1103/PhysRevLett.104.040502} {\bibfield  {journal} {\bibinfo  {journal}
  {Phys. Rev. Lett.}\ }\textbf {\bibinfo {volume} {104}},\ \bibinfo {pages}
  {040502} (\bibinfo {year} {2010})}\BibitemShut {NoStop}%
\bibitem [{\citenamefont {{Alicea}}(2010)}]{aliceaprojection}%
  \BibitemOpen
  \bibfield  {author} {\bibinfo {author} {\bibfnamefont {J.}~\bibnamefont
  {{Alicea}}},\ }\href@noop {} {\bibfield  {journal} {\bibinfo  {journal}
  {Phys. Rev. B}\ }\textbf {\bibinfo {volume} {81}},\ \bibinfo {pages} {125318}
  (\bibinfo {year} {2010})}\BibitemShut {NoStop}%
\bibitem [{\citenamefont {Lutchyn}\ \emph {et~al.}(2010)\citenamefont
  {Lutchyn}, \citenamefont {Sau},\ and\ \citenamefont {Das~Sarma}}]{Lutchyn10}%
  \BibitemOpen
  \bibfield  {author} {\bibinfo {author} {\bibfnamefont {R.~M.}\ \bibnamefont
  {Lutchyn}}, \bibinfo {author} {\bibfnamefont {J.~D.}\ \bibnamefont {Sau}}, \
  and\ \bibinfo {author} {\bibfnamefont {S.}~\bibnamefont {Das~Sarma}},\ }\href
  {\doibase 10.1103/PhysRevLett.105.077001} {\bibfield  {journal} {\bibinfo
  {journal} {Phys. Rev. Lett.}\ }\textbf {\bibinfo {volume} {105}},\ \bibinfo
  {pages} {077001} (\bibinfo {year} {2010})}\BibitemShut {NoStop}%
\bibitem [{\citenamefont {Oreg}\ \emph {et~al.}(2010)\citenamefont {Oreg},
  \citenamefont {Refael},\ and\ \citenamefont {von Oppen}}]{Oreg10}%
  \BibitemOpen
  \bibfield  {author} {\bibinfo {author} {\bibfnamefont {Y.}~\bibnamefont
  {Oreg}}, \bibinfo {author} {\bibfnamefont {G.}~\bibnamefont {Refael}}, \ and\
  \bibinfo {author} {\bibfnamefont {F.}~\bibnamefont {von Oppen}},\ }\href
  {\doibase 10.1103/PhysRevLett.105.177002} {\bibfield  {journal} {\bibinfo
  {journal} {Phys. Rev. Lett.}\ }\textbf {\bibinfo {volume} {105}},\ \bibinfo
  {pages} {177002} (\bibinfo {year} {2010})}\BibitemShut {NoStop}%
\bibitem [{\citenamefont {{Elliott}}\ and\ \citenamefont
  {{Franz}}(2015)}]{majoranavorkommen}%
  \BibitemOpen
  \bibfield  {author} {\bibinfo {author} {\bibfnamefont {S.~R.}\ \bibnamefont
  {{Elliott}}}\ and\ \bibinfo {author} {\bibfnamefont {M.}~\bibnamefont
  {{Franz}}},\ }\href {\doibase 10.1103/RevModPhys.87.137} {\bibfield
  {journal} {\bibinfo  {journal} {Rev. Mod. Phys.}\ }\textbf {\bibinfo {volume}
  {87}},\ \bibinfo {pages} {137} (\bibinfo {year} {2015})}\BibitemShut
  {NoStop}%
\bibitem [{\citenamefont {{Aasen}}\ \emph {et~al.}(2016)\citenamefont
  {{Aasen}}, \citenamefont {{Hell}}, \citenamefont {{Mishmash}}, \citenamefont
  {{Higginbotham}}, \citenamefont {{Danon}}, \citenamefont {{Leijnse}},
  \citenamefont {{Jespersen}}, \citenamefont {{Folk}}, \citenamefont
  {{Marcus}}, \citenamefont {{Flensberg}},\ and\ \citenamefont
  {{Alicea}}}]{milestones}%
  \BibitemOpen
  \bibfield  {author} {\bibinfo {author} {\bibfnamefont {D.}~\bibnamefont
  {{Aasen}}}, \bibinfo {author} {\bibfnamefont {M.}~\bibnamefont {{Hell}}},
  \bibinfo {author} {\bibfnamefont {R.~V.}\ \bibnamefont {{Mishmash}}},
  \bibinfo {author} {\bibfnamefont {A.}~\bibnamefont {{Higginbotham}}},
  \bibinfo {author} {\bibfnamefont {J.}~\bibnamefont {{Danon}}}, \bibinfo
  {author} {\bibfnamefont {M.}~\bibnamefont {{Leijnse}}}, \bibinfo {author}
  {\bibfnamefont {T.~S.}\ \bibnamefont {{Jespersen}}}, \bibinfo {author}
  {\bibfnamefont {J.~A.}\ \bibnamefont {{Folk}}}, \bibinfo {author}
  {\bibfnamefont {C.~M.}\ \bibnamefont {{Marcus}}}, \bibinfo {author}
  {\bibfnamefont {K.}~\bibnamefont {{Flensberg}}}, \ and\ \bibinfo {author}
  {\bibfnamefont {J.}~\bibnamefont {{Alicea}}},\ }\href@noop {} {\bibfield
  {journal} {\bibinfo  {journal} {Phys. Rev. X}\ }\textbf {\bibinfo {volume}
  {6}},\ \bibinfo {pages} {031016} (\bibinfo {year} {2016})}\BibitemShut
  {NoStop}%
\bibitem [{\citenamefont {{He}}\ \emph {et~al.}(2017)\citenamefont {{He}},
  \citenamefont {{Pan}}, \citenamefont {{Stern}}, \citenamefont {{Burks}},
  \citenamefont {{Che}}, \citenamefont {{Yin}}, \citenamefont {{Wang}},
  \citenamefont {{Lian}}, \citenamefont {{Zhou}}, \citenamefont {{Choi}},
  \citenamefont {{Murata}}, \citenamefont {{Kou}}, \citenamefont {{Chen}},
  \citenamefont {{Nie}}, \citenamefont {{Shao}}, \citenamefont {{Fan}},
  \citenamefont {{Zhang}}, \citenamefont {{Liu}}, \citenamefont {{Xia}},\ and\
  \citenamefont {{Wang}}}]{He17}%
  \BibitemOpen
  \bibfield  {author} {\bibinfo {author} {\bibfnamefont {Q.~L.}\ \bibnamefont
  {{He}}}, \bibinfo {author} {\bibfnamefont {L.}~\bibnamefont {{Pan}}},
  \bibinfo {author} {\bibfnamefont {A.~L.}\ \bibnamefont {{Stern}}}, \bibinfo
  {author} {\bibfnamefont {E.~C.}\ \bibnamefont {{Burks}}}, \bibinfo {author}
  {\bibfnamefont {X.}~\bibnamefont {{Che}}}, \bibinfo {author} {\bibfnamefont
  {G.}~\bibnamefont {{Yin}}}, \bibinfo {author} {\bibfnamefont
  {J.}~\bibnamefont {{Wang}}}, \bibinfo {author} {\bibfnamefont
  {B.}~\bibnamefont {{Lian}}}, \bibinfo {author} {\bibfnamefont
  {Q.}~\bibnamefont {{Zhou}}}, \bibinfo {author} {\bibfnamefont {E.~S.}\
  \bibnamefont {{Choi}}}, \bibinfo {author} {\bibfnamefont {K.}~\bibnamefont
  {{Murata}}}, \bibinfo {author} {\bibfnamefont {X.}~\bibnamefont {{Kou}}},
  \bibinfo {author} {\bibfnamefont {Z.}~\bibnamefont {{Chen}}}, \bibinfo
  {author} {\bibfnamefont {T.}~\bibnamefont {{Nie}}}, \bibinfo {author}
  {\bibfnamefont {Q.}~\bibnamefont {{Shao}}}, \bibinfo {author} {\bibfnamefont
  {Y.}~\bibnamefont {{Fan}}}, \bibinfo {author} {\bibfnamefont {S.-C.}\
  \bibnamefont {{Zhang}}}, \bibinfo {author} {\bibfnamefont {K.}~\bibnamefont
  {{Liu}}}, \bibinfo {author} {\bibfnamefont {J.}~\bibnamefont {{Xia}}}, \ and\
  \bibinfo {author} {\bibfnamefont {K.~L.}\ \bibnamefont {{Wang}}},\
  }\href@noop {} {\bibfield  {journal} {\bibinfo  {journal} {Science}\ }\textbf
  {\bibinfo {volume} {357}},\ \bibinfo {pages} {294} (\bibinfo {year}
  {2017})}\BibitemShut {NoStop}%
\bibitem [{\citenamefont {{Mourik}}\ \emph {et~al.}(2012)\citenamefont
  {{Mourik}}, \citenamefont {{Zuo}}, \citenamefont {{Frolov}}, \citenamefont
  {{Plissard}}, \citenamefont {{Bakkers}},\ and\ \citenamefont
  {{Kouwenhoven}}}]{mourik}%
  \BibitemOpen
  \bibfield  {author} {\bibinfo {author} {\bibfnamefont {V.}~\bibnamefont
  {{Mourik}}}, \bibinfo {author} {\bibfnamefont {K.}~\bibnamefont {{Zuo}}},
  \bibinfo {author} {\bibfnamefont {S.~M.}\ \bibnamefont {{Frolov}}}, \bibinfo
  {author} {\bibfnamefont {S.~R.}\ \bibnamefont {{Plissard}}}, \bibinfo
  {author} {\bibfnamefont {E.~P.~A.~M.}\ \bibnamefont {{Bakkers}}}, \ and\
  \bibinfo {author} {\bibfnamefont {L.~P.}\ \bibnamefont {{Kouwenhoven}}},\
  }\href@noop {} {\bibfield  {journal} {\bibinfo  {journal} {Science}\ }\textbf
  {\bibinfo {volume} {336}},\ \bibinfo {pages} {1003} (\bibinfo {year}
  {2012})}\BibitemShut {NoStop}%
\bibitem [{\citenamefont {{Rokhinson}}\ \emph {et~al.}(2012)\citenamefont
  {{Rokhinson}}, \citenamefont {{Liu}},\ and\ \citenamefont
  {{Furdyna}}}]{Rokhinson12}%
  \BibitemOpen
  \bibfield  {author} {\bibinfo {author} {\bibfnamefont {L.~P.}\ \bibnamefont
  {{Rokhinson}}}, \bibinfo {author} {\bibfnamefont {X.}~\bibnamefont {{Liu}}},
  \ and\ \bibinfo {author} {\bibfnamefont {J.~K.}\ \bibnamefont {{Furdyna}}},\
  }\href@noop {} {\bibfield  {journal} {\bibinfo  {journal} {Nature Physics}\
  }\textbf {\bibinfo {volume} {8}},\ \bibinfo {pages} {795} (\bibinfo {year}
  {2012})}\BibitemShut {NoStop}%
\bibitem [{\citenamefont {{Deng}}\ \emph {et~al.}(2012)\citenamefont {{Deng}},
  \citenamefont {{Yu}}, \citenamefont {{Huang}}, \citenamefont {{Larsson}},
  \citenamefont {{Caroff}},\ and\ \citenamefont {{Xu}}}]{Deng12}%
  \BibitemOpen
  \bibfield  {author} {\bibinfo {author} {\bibfnamefont {M.~T.}\ \bibnamefont
  {{Deng}}}, \bibinfo {author} {\bibfnamefont {C.~L.}\ \bibnamefont {{Yu}}},
  \bibinfo {author} {\bibfnamefont {G.~Y.}\ \bibnamefont {{Huang}}}, \bibinfo
  {author} {\bibfnamefont {M.}~\bibnamefont {{Larsson}}}, \bibinfo {author}
  {\bibfnamefont {P.}~\bibnamefont {{Caroff}}}, \ and\ \bibinfo {author}
  {\bibfnamefont {H.~Q.}\ \bibnamefont {{Xu}}},\ }\href {\doibase
  10.1021/nl303758w} {\bibfield  {journal} {\bibinfo  {journal} {Nano Letters}\
  }\textbf {\bibinfo {volume} {12}},\ \bibinfo {pages} {6414} (\bibinfo {year}
  {2012})}\BibitemShut {NoStop}%
\bibitem [{\citenamefont {{Das}}\ \emph {et~al.}(2012)\citenamefont {{Das}},
  \citenamefont {{Ronen}}, \citenamefont {{Most}}, \citenamefont {{Oreg}},
  \citenamefont {{Heiblum}},\ and\ \citenamefont {{Shtrikman}}}]{Heiblum}%
  \BibitemOpen
  \bibfield  {author} {\bibinfo {author} {\bibfnamefont {A.}~\bibnamefont
  {{Das}}}, \bibinfo {author} {\bibfnamefont {Y.}~\bibnamefont {{Ronen}}},
  \bibinfo {author} {\bibfnamefont {Y.}~\bibnamefont {{Most}}}, \bibinfo
  {author} {\bibfnamefont {Y.}~\bibnamefont {{Oreg}}}, \bibinfo {author}
  {\bibfnamefont {M.}~\bibnamefont {{Heiblum}}}, \ and\ \bibinfo {author}
  {\bibfnamefont {H.}~\bibnamefont {{Shtrikman}}},\ }\href@noop {} {\bibfield
  {journal} {\bibinfo  {journal} {Nature Physics}\ }\textbf {\bibinfo {volume}
  {8}},\ \bibinfo {pages} {887} (\bibinfo {year} {2012})}\BibitemShut {NoStop}%
\bibitem [{\citenamefont {{Churchill}}\ \emph {et~al.}(2013)\citenamefont
  {{Churchill}}, \citenamefont {{Fatemi}}, \citenamefont {{Grove-Rasmussen}},
  \citenamefont {{Deng}}, \citenamefont {{Caroff}}, \citenamefont {{Xu}},\ and\
  \citenamefont {{Marcus}}}]{Churchill13}%
  \BibitemOpen
  \bibfield  {author} {\bibinfo {author} {\bibfnamefont {H.~O.~H.}\
  \bibnamefont {{Churchill}}}, \bibinfo {author} {\bibfnamefont
  {V.}~\bibnamefont {{Fatemi}}}, \bibinfo {author} {\bibfnamefont
  {K.}~\bibnamefont {{Grove-Rasmussen}}}, \bibinfo {author} {\bibfnamefont
  {M.~T.}\ \bibnamefont {{Deng}}}, \bibinfo {author} {\bibfnamefont
  {P.}~\bibnamefont {{Caroff}}}, \bibinfo {author} {\bibfnamefont {H.~Q.}\
  \bibnamefont {{Xu}}}, \ and\ \bibinfo {author} {\bibfnamefont {C.~M.}\
  \bibnamefont {{Marcus}}},\ }\href@noop {} {\bibfield  {journal} {\bibinfo
  {journal} {Phys. Rev. B}\ }\textbf {\bibinfo {volume} {87}},\ \bibinfo
  {pages} {241401} (\bibinfo {year} {2013})}\BibitemShut {NoStop}%
\bibitem [{\citenamefont {{Nadj-Perge}}\ \emph {et~al.}(2014)\citenamefont
  {{Nadj-Perge}}, \citenamefont {{Drozdov}}, \citenamefont {{Li}},
  \citenamefont {{Chen}}, \citenamefont {{Jeon}}, \citenamefont {{Seo}},
  \citenamefont {{MacDonald}}, \citenamefont {{Bernevig}},\ and\ \citenamefont
  {{Yazdani}}}]{Nadj-Perge14}%
  \BibitemOpen
  \bibfield  {author} {\bibinfo {author} {\bibfnamefont {S.}~\bibnamefont
  {{Nadj-Perge}}}, \bibinfo {author} {\bibfnamefont {I.~K.}\ \bibnamefont
  {{Drozdov}}}, \bibinfo {author} {\bibfnamefont {J.}~\bibnamefont {{Li}}},
  \bibinfo {author} {\bibfnamefont {H.}~\bibnamefont {{Chen}}}, \bibinfo
  {author} {\bibfnamefont {S.}~\bibnamefont {{Jeon}}}, \bibinfo {author}
  {\bibfnamefont {J.}~\bibnamefont {{Seo}}}, \bibinfo {author} {\bibfnamefont
  {A.~H.}\ \bibnamefont {{MacDonald}}}, \bibinfo {author} {\bibfnamefont
  {B.~A.}\ \bibnamefont {{Bernevig}}}, \ and\ \bibinfo {author} {\bibfnamefont
  {A.}~\bibnamefont {{Yazdani}}},\ }\href@noop {} {\bibfield  {journal}
  {\bibinfo  {journal} {Science}\ }\textbf {\bibinfo {volume} {346}},\ \bibinfo
  {pages} {602} (\bibinfo {year} {2014})}\BibitemShut {NoStop}%
\bibitem [{\citenamefont {{Albrecht}}\ \emph {et~al.}(2016)\citenamefont
  {{Albrecht}}, \citenamefont {{Higginbotham}}, \citenamefont {{Madsen}},
  \citenamefont {{Kuemmeth}}, \citenamefont {{Jespersen}}, \citenamefont
  {{Nygard}}, \citenamefont {{Krogstrup}},\ and\ \citenamefont
  {{Marcus}}}]{cmarcus}%
  \BibitemOpen
  \bibfield  {author} {\bibinfo {author} {\bibfnamefont {S.~M.}\ \bibnamefont
  {{Albrecht}}}, \bibinfo {author} {\bibfnamefont {A.~P.}\ \bibnamefont
  {{Higginbotham}}}, \bibinfo {author} {\bibfnamefont {M.}~\bibnamefont
  {{Madsen}}}, \bibinfo {author} {\bibfnamefont {F.}~\bibnamefont
  {{Kuemmeth}}}, \bibinfo {author} {\bibfnamefont {T.~S.}\ \bibnamefont
  {{Jespersen}}}, \bibinfo {author} {\bibfnamefont {J.}~\bibnamefont
  {{Nygard}}}, \bibinfo {author} {\bibfnamefont {P.}~\bibnamefont
  {{Krogstrup}}}, \ and\ \bibinfo {author} {\bibfnamefont {C.~M.}\ \bibnamefont
  {{Marcus}}},\ }\href@noop {} {\bibfield  {journal} {\bibinfo  {journal}
  {Nature Physics}\ }\textbf {\bibinfo {volume} {531}},\ \bibinfo {pages} {206}
  (\bibinfo {year} {2016})}\BibitemShut {NoStop}%
\bibitem [{\citenamefont {{Pawlak}}\ \emph {et~al.}(2016)\citenamefont
  {{Pawlak}}, \citenamefont {{Kisiel}}, \citenamefont {{Klinovaja}},
  \citenamefont {{Meier}}, \citenamefont {{Kawai}}, \citenamefont {{Glatzel}},
  \citenamefont {{Loss}},\ and\ \citenamefont {{Meyer}}}]{Pawlak16}%
  \BibitemOpen
  \bibfield  {author} {\bibinfo {author} {\bibfnamefont {R.}~\bibnamefont
  {{Pawlak}}}, \bibinfo {author} {\bibfnamefont {M.}~\bibnamefont {{Kisiel}}},
  \bibinfo {author} {\bibfnamefont {J.}~\bibnamefont {{Klinovaja}}}, \bibinfo
  {author} {\bibfnamefont {T.}~\bibnamefont {{Meier}}}, \bibinfo {author}
  {\bibfnamefont {S.}~\bibnamefont {{Kawai}}}, \bibinfo {author} {\bibfnamefont
  {T.}~\bibnamefont {{Glatzel}}}, \bibinfo {author} {\bibfnamefont
  {D.}~\bibnamefont {{Loss}}}, \ and\ \bibinfo {author} {\bibfnamefont
  {E.}~\bibnamefont {{Meyer}}},\ }\href@noop {} {\bibfield  {journal} {\bibinfo
   {journal} {NPJ Quantum Information}\ }\textbf {\bibinfo {volume} {2}},\
  \bibinfo {pages} {16035} (\bibinfo {year} {2016})}\BibitemShut {NoStop}%
\bibitem [{\citenamefont {Deng}\ \emph {et~al.}(2016)\citenamefont {Deng},
  \citenamefont {Vaitiekenas}, \citenamefont {Hansen}, \citenamefont {Danon},
  \citenamefont {Leijnse}, \citenamefont {Flensberg}, \citenamefont {Nyg{\r
  a}rd}, \citenamefont {Krogstrup},\ and\ \citenamefont {Marcus}}]{Deng16}%
  \BibitemOpen
  \bibfield  {author} {\bibinfo {author} {\bibfnamefont {M.~T.}\ \bibnamefont
  {Deng}}, \bibinfo {author} {\bibfnamefont {S.}~\bibnamefont {Vaitiekenas}},
  \bibinfo {author} {\bibfnamefont {E.~B.}\ \bibnamefont {Hansen}}, \bibinfo
  {author} {\bibfnamefont {J.}~\bibnamefont {Danon}}, \bibinfo {author}
  {\bibfnamefont {M.}~\bibnamefont {Leijnse}}, \bibinfo {author} {\bibfnamefont
  {K.}~\bibnamefont {Flensberg}}, \bibinfo {author} {\bibfnamefont
  {J.}~\bibnamefont {Nyg{\r a}rd}}, \bibinfo {author} {\bibfnamefont
  {P.}~\bibnamefont {Krogstrup}}, \ and\ \bibinfo {author} {\bibfnamefont
  {C.~M.}\ \bibnamefont {Marcus}},\ }\href@noop {} {\bibfield  {journal}
  {\bibinfo  {journal} {Science}\ }\textbf {\bibinfo {volume} {354}},\ \bibinfo
  {pages} {1557} (\bibinfo {year} {2016})}\BibitemShut {NoStop}%
\bibitem [{\citenamefont {{G{\"u}l}}\ \emph {et~al.}(2018)\citenamefont
  {{G{\"u}l}}, \citenamefont {{Zhang}}, \citenamefont {{Bommer}}, \citenamefont
  {{de Moor}}, \citenamefont {{Car}}, \citenamefont {{Plissard}}, \citenamefont
  {{Bakkers}}, \citenamefont {{Geresdi}}, \citenamefont {{Watanabe}},
  \citenamefont {{Taniguchi}},\ and\ \citenamefont {{Kouwenhoven}}}]{Zhang16}%
  \BibitemOpen
  \bibfield  {author} {\bibinfo {author} {\bibfnamefont {{\~A}.-n.}\
  \bibnamefont {{G{\"u}l}}}, \bibinfo {author} {\bibfnamefont {H.}~\bibnamefont
  {{Zhang}}}, \bibinfo {author} {\bibfnamefont {J.~D.~S.}\ \bibnamefont
  {{Bommer}}}, \bibinfo {author} {\bibfnamefont {M.~W.~A.}\ \bibnamefont {{de
  Moor}}}, \bibinfo {author} {\bibfnamefont {D.}~\bibnamefont {{Car}}},
  \bibinfo {author} {\bibfnamefont {S.~R.}\ \bibnamefont {{Plissard}}},
  \bibinfo {author} {\bibfnamefont {E.~P.~A.~M.}\ \bibnamefont {{Bakkers}}},
  \bibinfo {author} {\bibfnamefont {A.}~\bibnamefont {{Geresdi}}}, \bibinfo
  {author} {\bibfnamefont {K.}~\bibnamefont {{Watanabe}}}, \bibinfo {author}
  {\bibfnamefont {T.}~\bibnamefont {{Taniguchi}}}, \ and\ \bibinfo {author}
  {\bibfnamefont {L.~P.}\ \bibnamefont {{Kouwenhoven}}},\ }\href@noop {}
  {\bibfield  {journal} {\bibinfo  {journal} {Nature Nanotechnology}\ }\textbf
  {\bibinfo {volume} {13}},\ \bibinfo {pages} {192} (\bibinfo {year}
  {2018})}\BibitemShut {NoStop}%
\bibitem [{\citenamefont {{Feldman}}\ \emph {et~al.}(2017)\citenamefont
  {{Feldman}}, \citenamefont {{Randeria}}, \citenamefont {{Li}}, \citenamefont
  {{Jeon}}, \citenamefont {{Xie}}, \citenamefont {{Wang}}, \citenamefont
  {{Drozdov}}, \citenamefont {{Andrei Bernevig}},\ and\ \citenamefont
  {{Yazdani}}}]{Feldman17}%
  \BibitemOpen
  \bibfield  {author} {\bibinfo {author} {\bibfnamefont {B.~E.}\ \bibnamefont
  {{Feldman}}}, \bibinfo {author} {\bibfnamefont {M.~T.}\ \bibnamefont
  {{Randeria}}}, \bibinfo {author} {\bibfnamefont {J.}~\bibnamefont {{Li}}},
  \bibinfo {author} {\bibfnamefont {S.}~\bibnamefont {{Jeon}}}, \bibinfo
  {author} {\bibfnamefont {Y.}~\bibnamefont {{Xie}}}, \bibinfo {author}
  {\bibfnamefont {Z.}~\bibnamefont {{Wang}}}, \bibinfo {author} {\bibfnamefont
  {I.~K.}\ \bibnamefont {{Drozdov}}}, \bibinfo {author} {\bibfnamefont
  {B.}~\bibnamefont {{Andrei Bernevig}}}, \ and\ \bibinfo {author}
  {\bibfnamefont {A.}~\bibnamefont {{Yazdani}}},\ }\href@noop {} {\bibfield
  {journal} {\bibinfo  {journal} {Nature Physics}\ }\textbf {\bibinfo {volume}
  {13}},\ \bibinfo {pages} {286} (\bibinfo {year} {2017})}\BibitemShut
  {NoStop}%
\bibitem [{\citenamefont {{Suominen}}\ \emph {et~al.}(2017)\citenamefont
  {{Suominen}}, \citenamefont {{Kjaergaard}}, \citenamefont {{Hamilton}},
  \citenamefont {{Shabani}}, \citenamefont {{Palmstr{\o}m}}, \citenamefont
  {{Marcus}},\ and\ \citenamefont {{Nichele}}}]{Suominen17}%
  \BibitemOpen
  \bibfield  {author} {\bibinfo {author} {\bibfnamefont {H.~J.}\ \bibnamefont
  {{Suominen}}}, \bibinfo {author} {\bibfnamefont {M.}~\bibnamefont
  {{Kjaergaard}}}, \bibinfo {author} {\bibfnamefont {A.~R.}\ \bibnamefont
  {{Hamilton}}}, \bibinfo {author} {\bibfnamefont {J.}~\bibnamefont
  {{Shabani}}}, \bibinfo {author} {\bibfnamefont {C.~J.}\ \bibnamefont
  {{Palmstr{\o}m}}}, \bibinfo {author} {\bibfnamefont {C.~M.}\ \bibnamefont
  {{Marcus}}}, \ and\ \bibinfo {author} {\bibfnamefont {F.}~\bibnamefont
  {{Nichele}}},\ }\href@noop {} {\bibfield  {journal} {\bibinfo  {journal}
  {\prl}\ }\textbf {\bibinfo {volume} {119}},\ \bibinfo {pages} {176805}
  (\bibinfo {year} {2017})}\BibitemShut {NoStop}%
\bibitem [{\citenamefont {Hell}\ \emph {et~al.}(2017)\citenamefont {Hell},
  \citenamefont {Flensberg},\ and\ \citenamefont {Leijnse}}]{Hell17}%
  \BibitemOpen
  \bibfield  {author} {\bibinfo {author} {\bibfnamefont {M.}~\bibnamefont
  {Hell}}, \bibinfo {author} {\bibfnamefont {K.}~\bibnamefont {Flensberg}}, \
  and\ \bibinfo {author} {\bibfnamefont {M.}~\bibnamefont {Leijnse}},\ }\href
  {\doibase 10.1103/PhysRevB.96.035444} {\bibfield  {journal} {\bibinfo
  {journal} {Phys. Rev. B}\ }\textbf {\bibinfo {volume} {96}},\ \bibinfo
  {pages} {035444} (\bibinfo {year} {2017})}\BibitemShut {NoStop}%
\bibitem [{\citenamefont {{Zhang}}\ \emph {et~al.}(2017)\citenamefont
  {{Zhang}}, \citenamefont {{G{\"u}l}}, \citenamefont {{Conesa-Boj}},
  \citenamefont {{Nowak}}, \citenamefont {{Wimmer}}, \citenamefont {{Zuo}},
  \citenamefont {{Mourik}}, \citenamefont {{de Vries}}, \citenamefont {{van
  Veen}}, \citenamefont {{de Moor}}, \citenamefont {{Bommer}}, \citenamefont
  {{van Woerkom}}, \citenamefont {{Car}}, \citenamefont {{Plissard}},
  \citenamefont {{Bakkers}}, \citenamefont {{Quintero-P{\'e}rez}},
  \citenamefont {{Cassidy}}, \citenamefont {{Koelling}}, \citenamefont
  {{Goswami}}, \citenamefont {{Watanabe}}, \citenamefont {{Taniguchi}},\ and\
  \citenamefont {{Kouwenhoven}}}]{Zhang17}%
  \BibitemOpen
  \bibfield  {author} {\bibinfo {author} {\bibfnamefont {H.}~\bibnamefont
  {{Zhang}}}, \bibinfo {author} {\bibfnamefont {{\"O}.}~\bibnamefont
  {{G{\"u}l}}}, \bibinfo {author} {\bibfnamefont {S.}~\bibnamefont
  {{Conesa-Boj}}}, \bibinfo {author} {\bibfnamefont {M.~P.}\ \bibnamefont
  {{Nowak}}}, \bibinfo {author} {\bibfnamefont {M.}~\bibnamefont {{Wimmer}}},
  \bibinfo {author} {\bibfnamefont {K.}~\bibnamefont {{Zuo}}}, \bibinfo
  {author} {\bibfnamefont {V.}~\bibnamefont {{Mourik}}}, \bibinfo {author}
  {\bibfnamefont {F.~K.}\ \bibnamefont {{de Vries}}}, \bibinfo {author}
  {\bibfnamefont {J.}~\bibnamefont {{van Veen}}}, \bibinfo {author}
  {\bibfnamefont {M.~W.~A.}\ \bibnamefont {{de Moor}}}, \bibinfo {author}
  {\bibfnamefont {J.~D.~S.}\ \bibnamefont {{Bommer}}}, \bibinfo {author}
  {\bibfnamefont {D.~J.}\ \bibnamefont {{van Woerkom}}}, \bibinfo {author}
  {\bibfnamefont {D.}~\bibnamefont {{Car}}}, \bibinfo {author} {\bibfnamefont
  {S.~R.}\ \bibnamefont {{Plissard}}}, \bibinfo {author} {\bibfnamefont
  {E.~P.~A.~M.}\ \bibnamefont {{Bakkers}}}, \bibinfo {author} {\bibfnamefont
  {M.}~\bibnamefont {{Quintero-P{\'e}rez}}}, \bibinfo {author} {\bibfnamefont
  {M.~C.}\ \bibnamefont {{Cassidy}}}, \bibinfo {author} {\bibfnamefont
  {S.}~\bibnamefont {{Koelling}}}, \bibinfo {author} {\bibfnamefont
  {S.}~\bibnamefont {{Goswami}}}, \bibinfo {author} {\bibfnamefont
  {K.}~\bibnamefont {{Watanabe}}}, \bibinfo {author} {\bibfnamefont
  {T.}~\bibnamefont {{Taniguchi}}}, \ and\ \bibinfo {author} {\bibfnamefont
  {L.~P.}\ \bibnamefont {{Kouwenhoven}}},\ }\href@noop {} {\bibfield  {journal}
  {\bibinfo  {journal} {Nature Communications}\ }\textbf {\bibinfo {volume}
  {8}},\ \bibinfo {pages} {16025} (\bibinfo {year} {2017})}\BibitemShut
  {NoStop}%
\bibitem [{\citenamefont {{Nichele}}\ \emph {et~al.}(2017)\citenamefont
  {{Nichele}}, \citenamefont {{Drachmann}}, \citenamefont {{Whiticar}},
  \citenamefont {{O'Farrell}}, \citenamefont {{Suominen}}, \citenamefont
  {{Fornieri}}, \citenamefont {{Wang}}, \citenamefont {{Gardner}},
  \citenamefont {{Thomas}}, \citenamefont {{Hatke}}, \citenamefont
  {{Krogstrup}}, \citenamefont {{Manfra}}, \citenamefont {{Flensberg}},\ and\
  \citenamefont {{Marcus}}}]{Nichele17}%
  \BibitemOpen
  \bibfield  {author} {\bibinfo {author} {\bibfnamefont {F.}~\bibnamefont
  {{Nichele}}}, \bibinfo {author} {\bibfnamefont {A.~C.~C.}\ \bibnamefont
  {{Drachmann}}}, \bibinfo {author} {\bibfnamefont {A.~M.}\ \bibnamefont
  {{Whiticar}}}, \bibinfo {author} {\bibfnamefont {E.~C.~T.}\ \bibnamefont
  {{O'Farrell}}}, \bibinfo {author} {\bibfnamefont {H.~J.}\ \bibnamefont
  {{Suominen}}}, \bibinfo {author} {\bibfnamefont {A.}~\bibnamefont
  {{Fornieri}}}, \bibinfo {author} {\bibfnamefont {T.}~\bibnamefont {{Wang}}},
  \bibinfo {author} {\bibfnamefont {G.~C.}\ \bibnamefont {{Gardner}}}, \bibinfo
  {author} {\bibfnamefont {C.}~\bibnamefont {{Thomas}}}, \bibinfo {author}
  {\bibfnamefont {A.~T.}\ \bibnamefont {{Hatke}}}, \bibinfo {author}
  {\bibfnamefont {P.}~\bibnamefont {{Krogstrup}}}, \bibinfo {author}
  {\bibfnamefont {M.~J.}\ \bibnamefont {{Manfra}}}, \bibinfo {author}
  {\bibfnamefont {K.}~\bibnamefont {{Flensberg}}}, \ and\ \bibinfo {author}
  {\bibfnamefont {C.~M.}\ \bibnamefont {{Marcus}}},\ }\href@noop {} {\bibfield
  {journal} {\bibinfo  {journal} {\prl}\ }\textbf {\bibinfo {volume} {119}},\
  \bibinfo {pages} {136803} (\bibinfo {year} {2017})}\BibitemShut {NoStop}%
\bibitem [{\citenamefont {{Zhang}}\ \emph {et~al.}(2018)\citenamefont
  {{Zhang}}, \citenamefont {{Liu}}, \citenamefont {{Gazibegovic}},
  \citenamefont {{Xu}}, \citenamefont {{Logan}}, \citenamefont {{Wang}},
  \citenamefont {{van Loo}}, \citenamefont {{Bommer}}, \citenamefont {{de
  Moor}}, \citenamefont {{Car}}, \citenamefont {{Op Het Veld}}, \citenamefont
  {{van Veldhoven}}, \citenamefont {{Koelling}}, \citenamefont {{Verheijen}},
  \citenamefont {{Pendharkar}}, \citenamefont {{Pennachio}}, \citenamefont
  {{Shojaei}}, \citenamefont {{Lee}}, \citenamefont {{Palmstr{\o}m}},
  \citenamefont {{Bakkers}}, \citenamefont {{Sarma}},\ and\ \citenamefont
  {{Kouwenhoven}}}]{Zhang18}%
  \BibitemOpen
  \bibfield  {author} {\bibinfo {author} {\bibfnamefont {H.}~\bibnamefont
  {{Zhang}}}, \bibinfo {author} {\bibfnamefont {C.-X.}\ \bibnamefont {{Liu}}},
  \bibinfo {author} {\bibfnamefont {S.}~\bibnamefont {{Gazibegovic}}}, \bibinfo
  {author} {\bibfnamefont {D.}~\bibnamefont {{Xu}}}, \bibinfo {author}
  {\bibfnamefont {J.~A.}\ \bibnamefont {{Logan}}}, \bibinfo {author}
  {\bibfnamefont {G.}~\bibnamefont {{Wang}}}, \bibinfo {author} {\bibfnamefont
  {N.}~\bibnamefont {{van Loo}}}, \bibinfo {author} {\bibfnamefont {J.~D.~S.}\
  \bibnamefont {{Bommer}}}, \bibinfo {author} {\bibfnamefont {M.~W.~A.}\
  \bibnamefont {{de Moor}}}, \bibinfo {author} {\bibfnamefont {D.}~\bibnamefont
  {{Car}}}, \bibinfo {author} {\bibfnamefont {R.~L.~M.}\ \bibnamefont {{Op Het
  Veld}}}, \bibinfo {author} {\bibfnamefont {P.~J.}\ \bibnamefont {{van
  Veldhoven}}}, \bibinfo {author} {\bibfnamefont {S.}~\bibnamefont
  {{Koelling}}}, \bibinfo {author} {\bibfnamefont {M.~A.}\ \bibnamefont
  {{Verheijen}}}, \bibinfo {author} {\bibfnamefont {M.}~\bibnamefont
  {{Pendharkar}}}, \bibinfo {author} {\bibfnamefont {D.~J.}\ \bibnamefont
  {{Pennachio}}}, \bibinfo {author} {\bibfnamefont {B.}~\bibnamefont
  {{Shojaei}}}, \bibinfo {author} {\bibfnamefont {J.~S.}\ \bibnamefont
  {{Lee}}}, \bibinfo {author} {\bibfnamefont {C.~J.}\ \bibnamefont
  {{Palmstr{\o}m}}}, \bibinfo {author} {\bibfnamefont {E.~P.~A.~M.}\
  \bibnamefont {{Bakkers}}}, \bibinfo {author} {\bibfnamefont {S.~D.}\
  \bibnamefont {{Sarma}}}, \ and\ \bibinfo {author} {\bibfnamefont {L.~P.}\
  \bibnamefont {{Kouwenhoven}}},\ }\href@noop {} {\bibfield  {journal}
  {\bibinfo  {journal} {\nat}\ }\textbf {\bibinfo {volume} {556}},\ \bibinfo
  {pages} {74} (\bibinfo {year} {2018})}\BibitemShut {NoStop}%
\bibitem [{\citenamefont {{Grivnin}}\ \emph {et~al.}(2018)\citenamefont
  {{Grivnin}}, \citenamefont {{Bor}}, \citenamefont {{Heiblum}}, \citenamefont
  {{Oreg}},\ and\ \citenamefont {{Shtrikman}}}]{Grivnin18}%
  \BibitemOpen
  \bibfield  {author} {\bibinfo {author} {\bibfnamefont {A.}~\bibnamefont
  {{Grivnin}}}, \bibinfo {author} {\bibfnamefont {E.}~\bibnamefont {{Bor}}},
  \bibinfo {author} {\bibfnamefont {M.}~\bibnamefont {{Heiblum}}}, \bibinfo
  {author} {\bibfnamefont {Y.}~\bibnamefont {{Oreg}}}, \ and\ \bibinfo {author}
  {\bibfnamefont {H.}~\bibnamefont {{Shtrikman}}},\ }\href@noop {} {\bibfield
  {journal} {\bibinfo  {journal} {arXiv: 1807.06632}\ } (\bibinfo {year}
  {2018})}\BibitemShut {NoStop}%
\bibitem [{\citenamefont {Vaitiek\ifmmode~\dot{e}\else \.{e}\fi{}nas}\ \emph
  {et~al.}(2018)\citenamefont {Vaitiek\ifmmode~\dot{e}\else \.{e}\fi{}nas},
  \citenamefont {Deng}, \citenamefont {Nyg\aa{}rd}, \citenamefont {Krogstrup},\
  and\ \citenamefont {Marcus}}]{Marcus18}%
  \BibitemOpen
  \bibfield  {author} {\bibinfo {author} {\bibfnamefont {S.}~\bibnamefont
  {Vaitiek\ifmmode~\dot{e}\else \.{e}\fi{}nas}}, \bibinfo {author}
  {\bibfnamefont {M.-T.}\ \bibnamefont {Deng}}, \bibinfo {author}
  {\bibfnamefont {J.}~\bibnamefont {Nyg\aa{}rd}}, \bibinfo {author}
  {\bibfnamefont {P.}~\bibnamefont {Krogstrup}}, \ and\ \bibinfo {author}
  {\bibfnamefont {C.~M.}\ \bibnamefont {Marcus}},\ }\href {\doibase
  10.1103/PhysRevLett.121.037703} {\bibfield  {journal} {\bibinfo  {journal}
  {Phys. Rev. Lett.}\ }\textbf {\bibinfo {volume} {121}},\ \bibinfo {pages}
  {037703} (\bibinfo {year} {2018})}\BibitemShut {NoStop}%
\bibitem [{\citenamefont {{Chiu}}\ \emph {et~al.}(2015)\citenamefont {{Chiu}},
  \citenamefont {{Vazifeh}},\ and\ \citenamefont
  {{Franz}}}]{onedimensionalexchange}%
  \BibitemOpen
  \bibfield  {author} {\bibinfo {author} {\bibfnamefont {C.~K.}\ \bibnamefont
  {{Chiu}}}, \bibinfo {author} {\bibfnamefont {M.~M.}\ \bibnamefont
  {{Vazifeh}}}, \ and\ \bibinfo {author} {\bibfnamefont {M.}~\bibnamefont
  {{Franz}}},\ }\href@noop {} {\bibfield  {journal} {\bibinfo  {journal}
  {Europhysics Letters}\ }\textbf {\bibinfo {volume} {110}},\ \bibinfo {pages}
  {10001} (\bibinfo {year} {2015})}\BibitemShut {NoStop}%
\bibitem [{\citenamefont {Vijay}\ and\ \citenamefont {Fu}(2016)}]{Vijay16}%
  \BibitemOpen
  \bibfield  {author} {\bibinfo {author} {\bibfnamefont {S.}~\bibnamefont
  {Vijay}}\ and\ \bibinfo {author} {\bibfnamefont {L.}~\bibnamefont {Fu}},\
  }\href {\doibase 10.1103/PhysRevB.94.235446} {\bibfield  {journal} {\bibinfo
  {journal} {Phys. Rev. B}\ }\textbf {\bibinfo {volume} {94}},\ \bibinfo
  {pages} {235446} (\bibinfo {year} {2016})}\BibitemShut {NoStop}%
\bibitem [{\citenamefont {Plugge}\ \emph {et~al.}(2016)\citenamefont {Plugge},
  \citenamefont {Landau}, \citenamefont {Sela}, \citenamefont {Altland},
  \citenamefont {Flensberg},\ and\ \citenamefont {Egger}}]{Plugge16}%
  \BibitemOpen
  \bibfield  {author} {\bibinfo {author} {\bibfnamefont {S.}~\bibnamefont
  {Plugge}}, \bibinfo {author} {\bibfnamefont {L.~A.}\ \bibnamefont {Landau}},
  \bibinfo {author} {\bibfnamefont {E.}~\bibnamefont {Sela}}, \bibinfo {author}
  {\bibfnamefont {A.}~\bibnamefont {Altland}}, \bibinfo {author} {\bibfnamefont
  {K.}~\bibnamefont {Flensberg}}, \ and\ \bibinfo {author} {\bibfnamefont
  {R.}~\bibnamefont {Egger}},\ }\href {\doibase 10.1103/PhysRevB.94.174514}
  {\bibfield  {journal} {\bibinfo  {journal} {Phys. Rev. B}\ }\textbf {\bibinfo
  {volume} {94}},\ \bibinfo {pages} {174514} (\bibinfo {year}
  {2016})}\BibitemShut {NoStop}%
\bibitem [{\citenamefont {{Landau}}\ \emph {et~al.}(2016)\citenamefont
  {{Landau}}, \citenamefont {{Plugge}}, \citenamefont {{Sela}}, \citenamefont
  {{Altland}}, \citenamefont {{Albrecht}},\ and\ \citenamefont
  {{Egger}}}]{Landau16}%
  \BibitemOpen
  \bibfield  {author} {\bibinfo {author} {\bibfnamefont {L.~A.}\ \bibnamefont
  {{Landau}}}, \bibinfo {author} {\bibfnamefont {S.}~\bibnamefont {{Plugge}}},
  \bibinfo {author} {\bibfnamefont {E.}~\bibnamefont {{Sela}}}, \bibinfo
  {author} {\bibfnamefont {A.}~\bibnamefont {{Altland}}}, \bibinfo {author}
  {\bibfnamefont {S.~M.}\ \bibnamefont {{Albrecht}}}, \ and\ \bibinfo {author}
  {\bibfnamefont {R.}~\bibnamefont {{Egger}}},\ }\href@noop {} {\bibfield
  {journal} {\bibinfo  {journal} {\prl}\ }\textbf {\bibinfo {volume} {116}},\
  \bibinfo {pages} {050501} (\bibinfo {year} {2016})}\BibitemShut {NoStop}%
\bibitem [{\citenamefont {{Wiedenmann}}\ \emph {et~al.}(2016)\citenamefont
  {{Wiedenmann}}, \citenamefont {{Bocquillon}}, \citenamefont {{Deacon}},
  \citenamefont {{Hartinger}}, \citenamefont {{Herrmann}}, \citenamefont
  {{Klapwijk}}, \citenamefont {{Maier}}, \citenamefont {{Ames}}, \citenamefont
  {{Br{\"u}ne}}, \citenamefont {{Gould}}, \citenamefont {{Oiwa}}, \citenamefont
  {{Ishibashi}}, \citenamefont {{Tarucha}}, \citenamefont {{Buhmann}},\ and\
  \citenamefont {{Molenkamp}}}]{Molenkamp16}%
  \BibitemOpen
  \bibfield  {author} {\bibinfo {author} {\bibfnamefont {J.}~\bibnamefont
  {{Wiedenmann}}}, \bibinfo {author} {\bibfnamefont {E.}~\bibnamefont
  {{Bocquillon}}}, \bibinfo {author} {\bibfnamefont {R.~S.}\ \bibnamefont
  {{Deacon}}}, \bibinfo {author} {\bibfnamefont {S.}~\bibnamefont
  {{Hartinger}}}, \bibinfo {author} {\bibfnamefont {O.}~\bibnamefont
  {{Herrmann}}}, \bibinfo {author} {\bibfnamefont {T.~M.}\ \bibnamefont
  {{Klapwijk}}}, \bibinfo {author} {\bibfnamefont {L.}~\bibnamefont {{Maier}}},
  \bibinfo {author} {\bibfnamefont {C.}~\bibnamefont {{Ames}}}, \bibinfo
  {author} {\bibfnamefont {C.}~\bibnamefont {{Br{\"u}ne}}}, \bibinfo {author}
  {\bibfnamefont {C.}~\bibnamefont {{Gould}}}, \bibinfo {author} {\bibfnamefont
  {A.}~\bibnamefont {{Oiwa}}}, \bibinfo {author} {\bibfnamefont
  {K.}~\bibnamefont {{Ishibashi}}}, \bibinfo {author} {\bibfnamefont
  {S.}~\bibnamefont {{Tarucha}}}, \bibinfo {author} {\bibfnamefont
  {H.}~\bibnamefont {{Buhmann}}}, \ and\ \bibinfo {author} {\bibfnamefont
  {L.~W.}\ \bibnamefont {{Molenkamp}}},\ }\href {\doibase 10.1038/ncomms10303}
  {\bibfield  {journal} {\bibinfo  {journal} {Nature Communications}\ }\textbf
  {\bibinfo {volume} {7}},\ \bibinfo {eid} {10303} (\bibinfo {year}
  {2016})}\BibitemShut {NoStop}%
\bibitem [{\citenamefont {Plugge}\ \emph {et~al.}(2017)\citenamefont {Plugge},
  \citenamefont {Rasmussen}, \citenamefont {Egger},\ and\ \citenamefont
  {Flensberg}}]{Plugge17}%
  \BibitemOpen
  \bibfield  {author} {\bibinfo {author} {\bibfnamefont {S.}~\bibnamefont
  {Plugge}}, \bibinfo {author} {\bibfnamefont {A.}~\bibnamefont {Rasmussen}},
  \bibinfo {author} {\bibfnamefont {R.}~\bibnamefont {Egger}}, \ and\ \bibinfo
  {author} {\bibfnamefont {K.}~\bibnamefont {Flensberg}},\ }\href@noop {}
  {\bibfield  {journal} {\bibinfo  {journal} {New Journal of Physics}\ }\textbf
  {\bibinfo {volume} {19}},\ \bibinfo {pages} {012001} (\bibinfo {year}
  {2017})}\BibitemShut {NoStop}%
\bibitem [{\citenamefont {Karzig}\ \emph {et~al.}(2017)\citenamefont {Karzig},
  \citenamefont {Knapp}, \citenamefont {Lutchyn}, \citenamefont {Bonderson},
  \citenamefont {Hastings}, \citenamefont {Nayak}, \citenamefont {Alicea},
  \citenamefont {Flensberg}, \citenamefont {Plugge}, \citenamefont {Oreg},
  \citenamefont {Marcus},\ and\ \citenamefont {Freedman}}]{Karzig17}%
  \BibitemOpen
  \bibfield  {author} {\bibinfo {author} {\bibfnamefont {T.}~\bibnamefont
  {Karzig}}, \bibinfo {author} {\bibfnamefont {C.}~\bibnamefont {Knapp}},
  \bibinfo {author} {\bibfnamefont {R.~M.}\ \bibnamefont {Lutchyn}}, \bibinfo
  {author} {\bibfnamefont {P.}~\bibnamefont {Bonderson}}, \bibinfo {author}
  {\bibfnamefont {M.~B.}\ \bibnamefont {Hastings}}, \bibinfo {author}
  {\bibfnamefont {C.}~\bibnamefont {Nayak}}, \bibinfo {author} {\bibfnamefont
  {J.}~\bibnamefont {Alicea}}, \bibinfo {author} {\bibfnamefont
  {K.}~\bibnamefont {Flensberg}}, \bibinfo {author} {\bibfnamefont
  {S.}~\bibnamefont {Plugge}}, \bibinfo {author} {\bibfnamefont
  {Y.}~\bibnamefont {Oreg}}, \bibinfo {author} {\bibfnamefont {C.~M.}\
  \bibnamefont {Marcus}}, \ and\ \bibinfo {author} {\bibfnamefont {M.~H.}\
  \bibnamefont {Freedman}},\ }\href {\doibase 10.1103/PhysRevB.95.235305}
  {\bibfield  {journal} {\bibinfo  {journal} {Phys. Rev. B}\ }\textbf {\bibinfo
  {volume} {95}},\ \bibinfo {pages} {235305} (\bibinfo {year}
  {2017})}\BibitemShut {NoStop}%
\bibitem [{\citenamefont {Litinski}\ and\ \citenamefont {von
  Oppen}(2017)}]{Litinski17}%
  \BibitemOpen
  \bibfield  {author} {\bibinfo {author} {\bibfnamefont {D.}~\bibnamefont
  {Litinski}}\ and\ \bibinfo {author} {\bibfnamefont {F.}~\bibnamefont {von
  Oppen}},\ }\href {\doibase 10.1103/PhysRevB.96.205413} {\bibfield  {journal}
  {\bibinfo  {journal} {Phys. Rev. B}\ }\textbf {\bibinfo {volume} {96}},\
  \bibinfo {pages} {205413} (\bibinfo {year} {2017})}\BibitemShut {NoStop}%
\bibitem [{\citenamefont {{Litinski}}\ \emph {et~al.}(2017)\citenamefont
  {{Litinski}}, \citenamefont {{Kesselring}}, \citenamefont {{Eisert}},\ and\
  \citenamefont {{von Oppen}}}]{Litinski172}%
  \BibitemOpen
  \bibfield  {author} {\bibinfo {author} {\bibfnamefont {D.}~\bibnamefont
  {{Litinski}}}, \bibinfo {author} {\bibfnamefont {M.~S.}\ \bibnamefont
  {{Kesselring}}}, \bibinfo {author} {\bibfnamefont {J.}~\bibnamefont
  {{Eisert}}}, \ and\ \bibinfo {author} {\bibfnamefont {F.}~\bibnamefont {{von
  Oppen}}},\ }\href@noop {} {\bibfield  {journal} {\bibinfo  {journal} {Phys.
  Rev. X}\ }\textbf {\bibinfo {volume} {7}},\ \bibinfo {pages} {031048}
  (\bibinfo {year} {2017})}\BibitemShut {NoStop}%
\bibitem [{\citenamefont {Dom\'{\i}nguez}\ \emph {et~al.}(2017)\citenamefont
  {Dom\'{\i}nguez}, \citenamefont {Kashuba}, \citenamefont {Bocquillon},
  \citenamefont {Wiedenmann}, \citenamefont {Deacon}, \citenamefont {Klapwijk},
  \citenamefont {Platero}, \citenamefont {Molenkamp}, \citenamefont
  {Trauzettel},\ and\ \citenamefont {Hankiewicz}}]{Hankiewicz17}%
  \BibitemOpen
  \bibfield  {author} {\bibinfo {author} {\bibfnamefont {F.}~\bibnamefont
  {Dom\'{\i}nguez}}, \bibinfo {author} {\bibfnamefont {O.}~\bibnamefont
  {Kashuba}}, \bibinfo {author} {\bibfnamefont {E.}~\bibnamefont {Bocquillon}},
  \bibinfo {author} {\bibfnamefont {J.}~\bibnamefont {Wiedenmann}}, \bibinfo
  {author} {\bibfnamefont {R.~S.}\ \bibnamefont {Deacon}}, \bibinfo {author}
  {\bibfnamefont {T.~M.}\ \bibnamefont {Klapwijk}}, \bibinfo {author}
  {\bibfnamefont {G.}~\bibnamefont {Platero}}, \bibinfo {author} {\bibfnamefont
  {L.~W.}\ \bibnamefont {Molenkamp}}, \bibinfo {author} {\bibfnamefont
  {B.}~\bibnamefont {Trauzettel}}, \ and\ \bibinfo {author} {\bibfnamefont
  {E.~M.}\ \bibnamefont {Hankiewicz}},\ }\href {\doibase
  10.1103/PhysRevB.95.195430} {\bibfield  {journal} {\bibinfo  {journal} {Phys.
  Rev. B}\ }\textbf {\bibinfo {volume} {95}},\ \bibinfo {pages} {195430}
  (\bibinfo {year} {2017})}\BibitemShut {NoStop}%
\bibitem [{\citenamefont {{Bocquillon}}\ \emph {et~al.}(2017)\citenamefont
  {{Bocquillon}}, \citenamefont {{Deacon}}, \citenamefont {{Wiedenmann}},
  \citenamefont {{Leubner}}, \citenamefont {{Klapwijk}}, \citenamefont
  {{Br{\"u}ne}}, \citenamefont {{Ishibashi}}, \citenamefont {{Buhmann}},\ and\
  \citenamefont {{Molenkamp}}}]{Molenkamp17}%
  \BibitemOpen
  \bibfield  {author} {\bibinfo {author} {\bibfnamefont {E.}~\bibnamefont
  {{Bocquillon}}}, \bibinfo {author} {\bibfnamefont {R.~S.}\ \bibnamefont
  {{Deacon}}}, \bibinfo {author} {\bibfnamefont {J.}~\bibnamefont
  {{Wiedenmann}}}, \bibinfo {author} {\bibfnamefont {P.}~\bibnamefont
  {{Leubner}}}, \bibinfo {author} {\bibfnamefont {T.~M.}\ \bibnamefont
  {{Klapwijk}}}, \bibinfo {author} {\bibfnamefont {C.}~\bibnamefont
  {{Br{\"u}ne}}}, \bibinfo {author} {\bibfnamefont {K.}~\bibnamefont
  {{Ishibashi}}}, \bibinfo {author} {\bibfnamefont {H.}~\bibnamefont
  {{Buhmann}}}, \ and\ \bibinfo {author} {\bibfnamefont {L.~W.}\ \bibnamefont
  {{Molenkamp}}},\ }\href {\doibase 10.1038/nnano.2016.159} {\bibfield
  {journal} {\bibinfo  {journal} {Nature Nanotechnology}\ }\textbf {\bibinfo
  {volume} {12}},\ \bibinfo {pages} {137} (\bibinfo {year} {2017})}\BibitemShut
  {NoStop}%
\bibitem [{\citenamefont {{Beenakker}}(2013)}]{beenaker}%
  \BibitemOpen
  \bibfield  {author} {\bibinfo {author} {\bibfnamefont {C.~W.~J.}\
  \bibnamefont {{Beenakker}}},\ }\href@noop {} {\bibfield  {journal} {\bibinfo
  {journal} {Annu. Rev. Con. Mat. Phys.}\ }\textbf {\bibinfo {volume} {4}},\
  \bibinfo {pages} {113} (\bibinfo {year} {2013})}\BibitemShut {NoStop}%
\bibitem [{\citenamefont {{Alicea}}(2012)}]{aliceareview}%
  \BibitemOpen
  \bibfield  {author} {\bibinfo {author} {\bibfnamefont {J.}~\bibnamefont
  {{Alicea}}},\ }\href@noop {} {\bibfield  {journal} {\bibinfo  {journal}
  {Reports on Progress in Physics}\ }\textbf {\bibinfo {volume} {75}},\
  \bibinfo {pages} {076501} (\bibinfo {year} {2012})}\BibitemShut {NoStop}%
\bibitem [{\citenamefont {{Leijnse}}\ and\ \citenamefont
  {{Flensberg}}(2012)}]{flensberg}%
  \BibitemOpen
  \bibfield  {author} {\bibinfo {author} {\bibfnamefont {M.}~\bibnamefont
  {{Leijnse}}}\ and\ \bibinfo {author} {\bibfnamefont {K.}~\bibnamefont
  {{Flensberg}}},\ }\href@noop {} {\bibfield  {journal} {\bibinfo  {journal}
  {Semiconductor Science and Technology}\ }\textbf {\bibinfo {volume} {27}},\
  \bibinfo {pages} {124003} (\bibinfo {year} {2012})}\BibitemShut {NoStop}%
\bibitem [{\citenamefont {{Stanescu}}\ and\ \citenamefont
  {{Tewari}}(2013)}]{tudor}%
  \BibitemOpen
  \bibfield  {author} {\bibinfo {author} {\bibfnamefont {T.~D.}\ \bibnamefont
  {{Stanescu}}}\ and\ \bibinfo {author} {\bibfnamefont {S.}~\bibnamefont
  {{Tewari}}},\ }\href@noop {} {\bibfield  {journal} {\bibinfo  {journal} {J.
  Phys.: Condens. Matter}\ }\textbf {\bibinfo {volume} {25}},\ \bibinfo {pages}
  {233201} (\bibinfo {year} {2013})}\BibitemShut {NoStop}%
\bibitem [{\citenamefont {{Pedrocchi}}\ and\ \citenamefont
  {{DiVincenzo}}(2015)}]{DiVincenzo15}%
  \BibitemOpen
  \bibfield  {author} {\bibinfo {author} {\bibfnamefont {F.~L.}\ \bibnamefont
  {{Pedrocchi}}}\ and\ \bibinfo {author} {\bibfnamefont {D.~P.}\ \bibnamefont
  {{DiVincenzo}}},\ }\href@noop {} {\bibfield  {journal} {\bibinfo  {journal}
  {\prl}\ }\textbf {\bibinfo {volume} {115}},\ \bibinfo {pages} {120402}
  (\bibinfo {year} {2015})}\BibitemShut {NoStop}%
\bibitem [{\citenamefont {Pedrocchi}\ \emph {et~al.}(2015)\citenamefont
  {Pedrocchi}, \citenamefont {Bonesteel},\ and\ \citenamefont
  {DiVincenzo}}]{DiVincenzo15long}%
  \BibitemOpen
  \bibfield  {author} {\bibinfo {author} {\bibfnamefont {F.~L.}\ \bibnamefont
  {Pedrocchi}}, \bibinfo {author} {\bibfnamefont {N.~E.}\ \bibnamefont
  {Bonesteel}}, \ and\ \bibinfo {author} {\bibfnamefont {D.~P.}\ \bibnamefont
  {DiVincenzo}},\ }\href {\doibase 10.1103/PhysRevB.92.115441} {\bibfield
  {journal} {\bibinfo  {journal} {Phys. Rev. B}\ }\textbf {\bibinfo {volume}
  {92}},\ \bibinfo {pages} {115441} (\bibinfo {year} {2015})}\BibitemShut
  {NoStop}%
\bibitem [{\citenamefont {{Stanescu}}\ and\ \citenamefont {{Das
  Sarma}}(2018)}]{Stanescu18}%
  \BibitemOpen
  \bibfield  {author} {\bibinfo {author} {\bibfnamefont {T.~D.}\ \bibnamefont
  {{Stanescu}}}\ and\ \bibinfo {author} {\bibfnamefont {S.}~\bibnamefont {{Das
  Sarma}}},\ }\href@noop {} {\bibfield  {journal} {\bibinfo  {journal} {Phys.
  Rev. B}\ }\textbf {\bibinfo {volume} {97}},\ \bibinfo {pages} {045410}
  (\bibinfo {year} {2018})}\BibitemShut {NoStop}%
\bibitem [{\citenamefont {{Huang}}\ \emph {et~al.}(2018)\citenamefont
  {{Huang}}, \citenamefont {{Sau}}, \citenamefont {{Stanescu}},\ and\
  \citenamefont {{Das Sarma}}}]{Sarma18}%
  \BibitemOpen
  \bibfield  {author} {\bibinfo {author} {\bibfnamefont {Y.}~\bibnamefont
  {{Huang}}}, \bibinfo {author} {\bibfnamefont {J.~D.}\ \bibnamefont {{Sau}}},
  \bibinfo {author} {\bibfnamefont {T.~D.}\ \bibnamefont {{Stanescu}}}, \ and\
  \bibinfo {author} {\bibfnamefont {S.}~\bibnamefont {{Das Sarma}}},\
  }\href@noop {} {\bibfield  {journal} {\bibinfo  {journal} {\prb}\ }\textbf
  {\bibinfo {volume} {98}},\ \bibinfo {eid} {224512} (\bibinfo {year}
  {2018})}\BibitemShut {NoStop}%
\bibitem [{\citenamefont {{Scheurer}}\ and\ \citenamefont
  {{Shnirman}}(2013)}]{Scheurer13}%
  \BibitemOpen
  \bibfield  {author} {\bibinfo {author} {\bibfnamefont {M.~S.}\ \bibnamefont
  {{Scheurer}}}\ and\ \bibinfo {author} {\bibfnamefont {A.}~\bibnamefont
  {{Shnirman}}},\ }\href@noop {} {\bibfield  {journal} {\bibinfo  {journal}
  {Phys. Rev. B}\ }\textbf {\bibinfo {volume} {88}},\ \bibinfo {pages} {064515}
  (\bibinfo {year} {2013})}\BibitemShut {NoStop}%
\bibitem [{\citenamefont {Wu}\ \emph {et~al.}(2014)\citenamefont {Wu},
  \citenamefont {Liang},\ and\ \citenamefont {Hu}}]{Wu14}%
  \BibitemOpen
  \bibfield  {author} {\bibinfo {author} {\bibfnamefont {L.-H.}\ \bibnamefont
  {Wu}}, \bibinfo {author} {\bibfnamefont {Q.-F.}\ \bibnamefont {Liang}}, \
  and\ \bibinfo {author} {\bibfnamefont {X.}~\bibnamefont {Hu}},\ }\href
  {\doibase 10.1088/1468-6996/15/6/064402} {\bibfield  {journal} {\bibinfo
  {journal} {Science and Technology of Advanced Materials}\ }\textbf {\bibinfo
  {volume} {15}},\ \bibinfo {pages} {064402} (\bibinfo {year}
  {2014})}\BibitemShut {NoStop}%
\bibitem [{\citenamefont {Cheng}\ \emph {et~al.}(2016)\citenamefont {Cheng},
  \citenamefont {He},\ and\ \citenamefont {Kou}}]{Cheng14}%
  \BibitemOpen
  \bibfield  {author} {\bibinfo {author} {\bibfnamefont {Q.-B.}\ \bibnamefont
  {Cheng}}, \bibinfo {author} {\bibfnamefont {J.}~\bibnamefont {He}}, \ and\
  \bibinfo {author} {\bibfnamefont {S.-P.}\ \bibnamefont {Kou}},\ }\href
  {\doibase https://doi.org/10.1016/j.physleta.2015.11.030} {\bibfield
  {journal} {\bibinfo  {journal} {Physics Letters A}\ }\textbf {\bibinfo
  {volume} {380}},\ \bibinfo {pages} {779 } (\bibinfo {year}
  {2016})}\BibitemShut {NoStop}%
\bibitem [{\citenamefont {{Karzig}}\ \emph {et~al.}(2015)\citenamefont
  {{Karzig}}, \citenamefont {{Pientka}}, \citenamefont {{Refael}},\ and\
  \citenamefont {{von Oppen}}}]{Karzig15}%
  \BibitemOpen
  \bibfield  {author} {\bibinfo {author} {\bibfnamefont {T.}~\bibnamefont
  {{Karzig}}}, \bibinfo {author} {\bibfnamefont {F.}~\bibnamefont {{Pientka}}},
  \bibinfo {author} {\bibfnamefont {G.}~\bibnamefont {{Refael}}}, \ and\
  \bibinfo {author} {\bibfnamefont {F.}~\bibnamefont {{von Oppen}}},\
  }\href@noop {} {\bibfield  {journal} {\bibinfo  {journal} {Phys. Rev. B}\
  }\textbf {\bibinfo {volume} {91}},\ \bibinfo {pages} {201102} (\bibinfo
  {year} {2015})}\BibitemShut {NoStop}%
\bibitem [{\citenamefont {{Sozinho Amorim}}\ \emph {et~al.}(2015)\citenamefont
  {{Sozinho Amorim}}, \citenamefont {{Ebihara}}, \citenamefont {{Yamakage}},
  \citenamefont {{Tanaka}},\ and\ \citenamefont {{Sato}}}]{otherbraidingex}%
  \BibitemOpen
  \bibfield  {author} {\bibinfo {author} {\bibfnamefont {C.}~\bibnamefont
  {{Sozinho Amorim}}}, \bibinfo {author} {\bibfnamefont {K.}~\bibnamefont
  {{Ebihara}}}, \bibinfo {author} {\bibfnamefont {A.}~\bibnamefont
  {{Yamakage}}}, \bibinfo {author} {\bibfnamefont {Y.}~\bibnamefont
  {{Tanaka}}}, \ and\ \bibinfo {author} {\bibfnamefont {M.}~\bibnamefont
  {{Sato}}},\ }\href {\doibase 10.1103/PhysRevB.91.174305} {\bibfield
  {journal} {\bibinfo  {journal} {Phys. Rev. B}\ }\textbf {\bibinfo {volume}
  {91}},\ \bibinfo {pages} {174305} (\bibinfo {year} {2015})}\BibitemShut
  {NoStop}%
\bibitem [{\citenamefont {Knapp}\ \emph {et~al.}(2016)\citenamefont {Knapp},
  \citenamefont {Zaletel}, \citenamefont {Liu}, \citenamefont {Cheng},
  \citenamefont {Bonderson},\ and\ \citenamefont {Nayak}}]{Knapp16}%
  \BibitemOpen
  \bibfield  {author} {\bibinfo {author} {\bibfnamefont {C.}~\bibnamefont
  {Knapp}}, \bibinfo {author} {\bibfnamefont {M.}~\bibnamefont {Zaletel}},
  \bibinfo {author} {\bibfnamefont {D.~E.}\ \bibnamefont {Liu}}, \bibinfo
  {author} {\bibfnamefont {M.}~\bibnamefont {Cheng}}, \bibinfo {author}
  {\bibfnamefont {P.}~\bibnamefont {Bonderson}}, \ and\ \bibinfo {author}
  {\bibfnamefont {C.}~\bibnamefont {Nayak}},\ }\href@noop {} {\bibfield
  {journal} {\bibinfo  {journal} {Phys. Rev.}\ }\textbf {\bibinfo {volume}
  {X6}},\ \bibinfo {pages} {041003} (\bibinfo {year} {2016})}\BibitemShut
  {NoStop}%
\bibitem [{\citenamefont {{Hell}}\ \emph {et~al.}(2016)\citenamefont {{Hell}},
  \citenamefont {{Danon}}, \citenamefont {{Flensberg}},\ and\ \citenamefont
  {{Leijnse}}}]{Hell162}%
  \BibitemOpen
  \bibfield  {author} {\bibinfo {author} {\bibfnamefont {M.}~\bibnamefont
  {{Hell}}}, \bibinfo {author} {\bibfnamefont {J.}~\bibnamefont {{Danon}}},
  \bibinfo {author} {\bibfnamefont {K.}~\bibnamefont {{Flensberg}}}, \ and\
  \bibinfo {author} {\bibfnamefont {M.}~\bibnamefont {{Leijnse}}},\ }\href@noop
  {} {\bibfield  {journal} {\bibinfo  {journal} {Phys. Rev. B}\ }\textbf
  {\bibinfo {volume} {94}},\ \bibinfo {pages} {035424} (\bibinfo {year}
  {2016})}\BibitemShut {NoStop}%
\bibitem [{\citenamefont {{Rahmani}}\ \emph {et~al.}(2017)\citenamefont
  {{Rahmani}}, \citenamefont {{Seradjeh}},\ and\ \citenamefont
  {{Franz}}}]{Rahmani17}%
  \BibitemOpen
  \bibfield  {author} {\bibinfo {author} {\bibfnamefont {A.}~\bibnamefont
  {{Rahmani}}}, \bibinfo {author} {\bibfnamefont {B.}~\bibnamefont
  {{Seradjeh}}}, \ and\ \bibinfo {author} {\bibfnamefont {M.}~\bibnamefont
  {{Franz}}},\ }\href@noop {} {\bibfield  {journal} {\bibinfo  {journal} {Phys.
  Rev. B}\ }\textbf {\bibinfo {volume} {96}},\ \bibinfo {pages} {075158}
  (\bibinfo {year} {2017})}\BibitemShut {NoStop}%
\bibitem [{\citenamefont {{Sekania}}\ \emph {et~al.}(2017)\citenamefont
  {{Sekania}}, \citenamefont {{Plugge}}, \citenamefont {{Greiter}},
  \citenamefont {{Thomale}},\ and\ \citenamefont
  {{Schmitteckert}}}]{Schmitteckert17}%
  \BibitemOpen
  \bibfield  {author} {\bibinfo {author} {\bibfnamefont {M.}~\bibnamefont
  {{Sekania}}}, \bibinfo {author} {\bibfnamefont {S.}~\bibnamefont {{Plugge}}},
  \bibinfo {author} {\bibfnamefont {M.}~\bibnamefont {{Greiter}}}, \bibinfo
  {author} {\bibfnamefont {R.}~\bibnamefont {{Thomale}}}, \ and\ \bibinfo
  {author} {\bibfnamefont {P.}~\bibnamefont {{Schmitteckert}}},\ }\href@noop {}
  {\bibfield  {journal} {\bibinfo  {journal} {Phys. Rev. B}\ }\textbf {\bibinfo
  {volume} {96}},\ \bibinfo {pages} {094307} (\bibinfo {year}
  {2017})}\BibitemShut {NoStop}%
\bibitem [{\citenamefont {{Bauer}}\ \emph {et~al.}(2018)\citenamefont
  {{Bauer}}, \citenamefont {{Karzig}}, \citenamefont {{Mishmash}},
  \citenamefont {{Antipov}},\ and\ \citenamefont {{Alicea}}}]{Bauer2018}%
  \BibitemOpen
  \bibfield  {author} {\bibinfo {author} {\bibfnamefont {B.}~\bibnamefont
  {{Bauer}}}, \bibinfo {author} {\bibfnamefont {T.}~\bibnamefont {{Karzig}}},
  \bibinfo {author} {\bibfnamefont {R.}~\bibnamefont {{Mishmash}}}, \bibinfo
  {author} {\bibfnamefont {A.}~\bibnamefont {{Antipov}}}, \ and\ \bibinfo
  {author} {\bibfnamefont {J.}~\bibnamefont {{Alicea}}},\ }\href@noop {}
  {\bibfield  {journal} {\bibinfo  {journal} {SciPost Physics}\ }\textbf
  {\bibinfo {volume} {5}},\ \bibinfo {pages} {004} (\bibinfo {year}
  {2018})}\BibitemShut {NoStop}%
\bibitem [{\citenamefont {{Alicea}}\ \emph {et~al.}(2011)\citenamefont
  {{Alicea}}, \citenamefont {{Oreg}}, \citenamefont {{Refael}}, \citenamefont
  {{von Oppen}},\ and\ \citenamefont {{Fisher}}}]{aliceanature}%
  \BibitemOpen
  \bibfield  {author} {\bibinfo {author} {\bibfnamefont {J.}~\bibnamefont
  {{Alicea}}}, \bibinfo {author} {\bibfnamefont {Y.}~\bibnamefont {{Oreg}}},
  \bibinfo {author} {\bibfnamefont {G.}~\bibnamefont {{Refael}}}, \bibinfo
  {author} {\bibfnamefont {F.}~\bibnamefont {{von Oppen}}}, \ and\ \bibinfo
  {author} {\bibfnamefont {M.~P.~A.}\ \bibnamefont {{Fisher}}},\ }\href@noop {}
  {\bibfield  {journal} {\bibinfo  {journal} {Nature Physics}\ }\textbf
  {\bibinfo {volume} {7}},\ \bibinfo {pages} {412} (\bibinfo {year}
  {2011})}\BibitemShut {NoStop}%
\bibitem [{\citenamefont {Sozinho~Amorim}\ \emph {et~al.}(2015)\citenamefont
  {Sozinho~Amorim}, \citenamefont {Ebihara}, \citenamefont {Yamakage},
  \citenamefont {Tanaka},\ and\ \citenamefont {Sato}}]{Sato14}%
  \BibitemOpen
  \bibfield  {author} {\bibinfo {author} {\bibfnamefont {C.}~\bibnamefont
  {Sozinho~Amorim}}, \bibinfo {author} {\bibfnamefont {K.}~\bibnamefont
  {Ebihara}}, \bibinfo {author} {\bibfnamefont {A.}~\bibnamefont {Yamakage}},
  \bibinfo {author} {\bibfnamefont {Y.}~\bibnamefont {Tanaka}}, \ and\ \bibinfo
  {author} {\bibfnamefont {M.}~\bibnamefont {Sato}},\ }\href@noop {} {\bibfield
   {journal} {\bibinfo  {journal} {Phys. Rev.}\ }\textbf {\bibinfo {volume}
  {B91}},\ \bibinfo {pages} {174305} (\bibinfo {year} {2015})}\BibitemShut
  {NoStop}%
\bibitem [{\citenamefont {{Potter}}\ and\ \citenamefont
  {{Lee}}(2010)}]{patricklee2010}%
  \BibitemOpen
  \bibfield  {author} {\bibinfo {author} {\bibfnamefont {A.~C.}\ \bibnamefont
  {{Potter}}}\ and\ \bibinfo {author} {\bibfnamefont {P.~A.}\ \bibnamefont
  {{Lee}}},\ }\href@noop {} {\bibfield  {journal} {\bibinfo  {journal} {Phys.
  Rev. Lett.}\ }\textbf {\bibinfo {volume} {105}},\ \bibinfo {pages} {227003}
  (\bibinfo {year} {2010})}\BibitemShut {NoStop}%
\bibitem [{\citenamefont {{Lutchyn}}\ \emph {et~al.}(2011)\citenamefont
  {{Lutchyn}}, \citenamefont {{Stanescu}},\ and\ \citenamefont {{Das
  Sarma}}}]{Lutchyn}%
  \BibitemOpen
  \bibfield  {author} {\bibinfo {author} {\bibfnamefont {R.~M.}\ \bibnamefont
  {{Lutchyn}}}, \bibinfo {author} {\bibfnamefont {T.~D.}\ \bibnamefont
  {{Stanescu}}}, \ and\ \bibinfo {author} {\bibfnamefont {S.}~\bibnamefont
  {{Das Sarma}}},\ }\href@noop {} {\bibfield  {journal} {\bibinfo  {journal}
  {Phys. Rev. Lett.}\ }\textbf {\bibinfo {volume} {106}},\ \bibinfo {pages}
  {127001} (\bibinfo {year} {2011})}\BibitemShut {NoStop}%
\bibitem [{\citenamefont {{Sols}}\ \emph {et~al.}(1989)\citenamefont {{Sols}},
  \citenamefont {{Macucci}}, \citenamefont {{Ravaioli}},\ and\ \citenamefont
  {{Hess}}}]{Sols}%
  \BibitemOpen
  \bibfield  {author} {\bibinfo {author} {\bibfnamefont {F.}~\bibnamefont
  {{Sols}}}, \bibinfo {author} {\bibfnamefont {M.}~\bibnamefont {{Macucci}}},
  \bibinfo {author} {\bibfnamefont {U.}~\bibnamefont {{Ravaioli}}}, \ and\
  \bibinfo {author} {\bibfnamefont {K.}~\bibnamefont {{Hess}}},\ }\href
  {\doibase 10.1063/1.344032} {\bibfield  {journal} {\bibinfo  {journal}
  {Journal of Applied Physics}\ }\textbf {\bibinfo {volume} {66}},\ \bibinfo
  {pages} {3892} (\bibinfo {year} {1989})}\BibitemShut {NoStop}%
\bibitem [{\citenamefont {{Lin}}\ \emph {et~al.}(2002)\citenamefont {{Lin}},
  \citenamefont {{Chen}},\ and\ \citenamefont {{Chuu}}}]{LinChen}%
  \BibitemOpen
  \bibfield  {author} {\bibinfo {author} {\bibfnamefont {Y.~K.}\ \bibnamefont
  {{Lin}}}, \bibinfo {author} {\bibfnamefont {Y.~N.}\ \bibnamefont {{Chen}}}, \
  and\ \bibinfo {author} {\bibfnamefont {D.~S.}\ \bibnamefont {{Chuu}}},\
  }\href {\doibase 10.1063/1.1446233} {\bibfield  {journal} {\bibinfo
  {journal} {Journal of Applied Physics}\ }\textbf {\bibinfo {volume} {91}},\
  \bibinfo {pages} {3054} (\bibinfo {year} {2002})}\BibitemShut {NoStop}%
\bibitem [{\citenamefont {{Schult}}\ \emph {et~al.}(1989)\citenamefont
  {{Schult}}, \citenamefont {{Ravenhall}},\ and\ \citenamefont
  {{Wyld}}}]{Schult}%
  \BibitemOpen
  \bibfield  {author} {\bibinfo {author} {\bibfnamefont {R.~L.}\ \bibnamefont
  {{Schult}}}, \bibinfo {author} {\bibfnamefont {D.~G.}\ \bibnamefont
  {{Ravenhall}}}, \ and\ \bibinfo {author} {\bibfnamefont {H.~W.}\ \bibnamefont
  {{Wyld}}},\ }\href@noop {} {\bibfield  {journal} {\bibinfo  {journal} {Phys.
  Rev. B}\ }\textbf {\bibinfo {volume} {39}},\ \bibinfo {pages} {5476}
  (\bibinfo {year} {1989})}\BibitemShut {NoStop}%
\bibitem [{Note1()}]{Note1}%
  \BibitemOpen
  \bibinfo {note} {If this assumption is not satisfied this does not mean that
  we are not able to simulate topological braiding processes. However, for this
  scenario the extraction of topological phases becomes much more challenging,
  since the HMF modes are formed by a complex (arbitrary) composition of MF
  modes $\gamma _{1,2,3,4}$. Moreover, one needs to determine all hybridization
  amplitudes and derive the associated exchange statistics of HMF modes for
  each initial parameter configuration. Therefore, such a scenario is rather
  undesirable.}\BibitemShut {Stop}%
\bibitem [{\citenamefont {{Puri}}(2017)}]{Puri}%
  \BibitemOpen
  \bibfield  {author} {\bibinfo {author} {\bibfnamefont {R.~R.}\ \bibnamefont
  {{Puri}}},\ }\href@noop {} {\emph {\bibinfo {title} {Non-Relativistic Quantum
  Mechanics}}}\ (\bibinfo  {publisher} {Cambridge University Press},\ \bibinfo
  {year} {2017})\BibitemShut {NoStop}%
\bibitem [{\citenamefont {{Griffiths}}(2016)}]{Griffiths}%
  \BibitemOpen
  \bibfield  {author} {\bibinfo {author} {\bibfnamefont {D.~J.}\ \bibnamefont
  {{Griffiths}}},\ }\href@noop {} {\emph {\bibinfo {title} {Introduction to
  Quantum Mechanics}}}\ (\bibinfo  {publisher} {Cambridge University Press},\
  \bibinfo {year} {2016})\BibitemShut {NoStop}%
\bibitem [{\citenamefont {{Nielsen}}\ and\ \citenamefont
  {{Chuang}}(2011)}]{nielson}%
  \BibitemOpen
  \bibfield  {author} {\bibinfo {author} {\bibfnamefont {M.~A.}\ \bibnamefont
  {{Nielsen}}}\ and\ \bibinfo {author} {\bibfnamefont {I.~L.}\ \bibnamefont
  {{Chuang}}},\ }\href@noop {} {\emph {\bibinfo {title} {{Quantum Computation
  and Quantum Information}}}}\ (\bibinfo  {publisher} {Cambridge University
  Press},\ \bibinfo {year} {2011})\BibitemShut {NoStop}%
\bibitem [{\citenamefont {{Bravyi}}\ and\ \citenamefont
  {{Kitaev}}(2005)}]{bravyikitaev}%
  \BibitemOpen
  \bibfield  {author} {\bibinfo {author} {\bibfnamefont {S.}~\bibnamefont
  {{Bravyi}}}\ and\ \bibinfo {author} {\bibfnamefont {A.~Y.}\ \bibnamefont
  {{Kitaev}}},\ }\href@noop {} {\bibfield  {journal} {\bibinfo  {journal}
  {Phys. Rev. A}\ }\textbf {\bibinfo {volume} {71}},\ \bibinfo {pages} {022316}
  (\bibinfo {year} {2005})}\BibitemShut {NoStop}%
\bibitem [{\citenamefont {{Gottesman}}(1998)}]{Gottesman}%
  \BibitemOpen
  \bibfield  {author} {\bibinfo {author} {\bibfnamefont {D.}~\bibnamefont
  {{Gottesman}}},\ }\href@noop {} {\bibfield  {journal} {\bibinfo  {journal}
  {arXiv:9807006}\ } (\bibinfo {year} {1998})}\BibitemShut {NoStop}%
\bibitem [{\citenamefont {{Bravyi}}(2006)}]{bravialone}%
  \BibitemOpen
  \bibfield  {author} {\bibinfo {author} {\bibfnamefont {S.}~\bibnamefont
  {{Bravyi}}},\ }\href@noop {} {\bibfield  {journal} {\bibinfo  {journal}
  {Phys. Rev. A}\ }\textbf {\bibinfo {volume} {73}},\ \bibinfo {pages} {042313}
  (\bibinfo {year} {2006})}\BibitemShut {NoStop}%
\bibitem [{\citenamefont {{Hyart}}\ \emph {et~al.}(2013)\citenamefont
  {{Hyart}}, \citenamefont {{van Heck}}, \citenamefont {{Fulga}}, \citenamefont
  {{Burrello}}, \citenamefont {{Akhmerov}},\ and\ \citenamefont
  {{Beenakker}}}]{hyart}%
  \BibitemOpen
  \bibfield  {author} {\bibinfo {author} {\bibfnamefont {T.}~\bibnamefont
  {{Hyart}}}, \bibinfo {author} {\bibfnamefont {B.}~\bibnamefont {{van Heck}}},
  \bibinfo {author} {\bibfnamefont {I.~C.}\ \bibnamefont {{Fulga}}}, \bibinfo
  {author} {\bibfnamefont {M.}~\bibnamefont {{Burrello}}}, \bibinfo {author}
  {\bibfnamefont {A.~R.}\ \bibnamefont {{Akhmerov}}}, \ and\ \bibinfo {author}
  {\bibfnamefont {C.~W.~J.}\ \bibnamefont {{Beenakker}}},\ }\href@noop {}
  {\bibfield  {journal} {\bibinfo  {journal} {Phys. Rev. B}\ }\textbf {\bibinfo
  {volume} {88}},\ \bibinfo {pages} {035121} (\bibinfo {year}
  {2013})}\BibitemShut {NoStop}%
\bibitem [{\citenamefont {Karzig}\ \emph {et~al.}(2016)\citenamefont {Karzig},
  \citenamefont {Oreg}, \citenamefont {Refael},\ and\ \citenamefont
  {Freedman}}]{karzig}%
  \BibitemOpen
  \bibfield  {author} {\bibinfo {author} {\bibfnamefont {T.}~\bibnamefont
  {Karzig}}, \bibinfo {author} {\bibfnamefont {Y.}~\bibnamefont {Oreg}},
  \bibinfo {author} {\bibfnamefont {G.}~\bibnamefont {Refael}}, \ and\ \bibinfo
  {author} {\bibfnamefont {M.~H.}\ \bibnamefont {Freedman}},\ }\href {\doibase
  10.1103/PhysRevX.6.031019} {\bibfield  {journal} {\bibinfo  {journal} {Phys.
  Rev. X}\ }\textbf {\bibinfo {volume} {6}},\ \bibinfo {pages} {031019}
  (\bibinfo {year} {2016})}\BibitemShut {NoStop}%
\bibitem [{\citenamefont {{Georgiev}}(2006)}]{qubitsdeeper}%
  \BibitemOpen
  \bibfield  {author} {\bibinfo {author} {\bibfnamefont {L.~S.}\ \bibnamefont
  {{Georgiev}}},\ }\href@noop {} {\bibfield  {journal} {\bibinfo  {journal}
  {Phys. Rev. B}\ }\textbf {\bibinfo {volume} {74}},\ \bibinfo {pages} {235112}
  (\bibinfo {year} {2006})}\BibitemShut {NoStop}%
\bibitem [{\citenamefont {{Hastings}}(2017)}]{Hastings17}%
  \BibitemOpen
  \bibfield  {author} {\bibinfo {author} {\bibfnamefont {M.~B.}\ \bibnamefont
  {{Hastings}}},\ }\href@noop {} {\bibfield  {journal} {\bibinfo  {journal}
  {arXiv:1703.00612}\ } (\bibinfo {year} {2017})}\BibitemShut {NoStop}%
\bibitem [{\citenamefont {{de Vries}}(2012)}]{deVries}%
  \BibitemOpen
  \bibfield  {author} {\bibinfo {author} {\bibfnamefont {A.}~\bibnamefont {{de
  Vries}}},\ }\href@noop {} {\emph {\bibinfo {title} {{Quantum Computation. An
  Introduction for Engineers and Computer Scientists}}}}\ (\bibinfo
  {publisher} {Books On Demand},\ \bibinfo {year} {2012})\BibitemShut {NoStop}%
\bibitem [{\citenamefont {{Bravyi}}\ and\ \citenamefont
  {{Kitaev}}(2002)}]{Bravyikitaevfermonic}%
  \BibitemOpen
  \bibfield  {author} {\bibinfo {author} {\bibfnamefont {S.}~\bibnamefont
  {{Bravyi}}}\ and\ \bibinfo {author} {\bibfnamefont {A.~Y.}\ \bibnamefont
  {{Kitaev}}},\ }\href@noop {} {\bibfield  {journal} {\bibinfo  {journal}
  {Annals of Physics}\ }\textbf {\bibinfo {volume} {298}},\ \bibinfo {pages}
  {210 } (\bibinfo {year} {2002})}\BibitemShut {NoStop}%
\bibitem [{\citenamefont {{Zilberberg}}\ \emph {et~al.}(2008)\citenamefont
  {{Zilberberg}}, \citenamefont {{Braunecker}},\ and\ \citenamefont
  {{Loss}}}]{Zilberberg}%
  \BibitemOpen
  \bibfield  {author} {\bibinfo {author} {\bibfnamefont {O.}~\bibnamefont
  {{Zilberberg}}}, \bibinfo {author} {\bibfnamefont {B.}~\bibnamefont
  {{Braunecker}}}, \ and\ \bibinfo {author} {\bibfnamefont {D.}~\bibnamefont
  {{Loss}}},\ }\href@noop {} {\bibfield  {journal} {\bibinfo  {journal} {Phys.
  Rev. A}\ }\textbf {\bibinfo {volume} {77}},\ \bibinfo {pages} {012327}
  (\bibinfo {year} {2008})}\BibitemShut {NoStop}%
\bibitem [{\citenamefont {{Hassler}}\ \emph {et~al.}(2010)\citenamefont
  {{Hassler}}, \citenamefont {{Akhmerov}}, \citenamefont {{Hou}},\ and\
  \citenamefont {{Beenakker}}}]{hassler}%
  \BibitemOpen
  \bibfield  {author} {\bibinfo {author} {\bibfnamefont {F.}~\bibnamefont
  {{Hassler}}}, \bibinfo {author} {\bibfnamefont {A.~R.}\ \bibnamefont
  {{Akhmerov}}}, \bibinfo {author} {\bibfnamefont {C.-Y.}\ \bibnamefont
  {{Hou}}}, \ and\ \bibinfo {author} {\bibfnamefont {C.~W.~J.}\ \bibnamefont
  {{Beenakker}}},\ }\href@noop {} {\bibfield  {journal} {\bibinfo  {journal}
  {New Journal of Physics}\ }\textbf {\bibinfo {volume} {12}},\ \bibinfo
  {pages} {125002} (\bibinfo {year} {2010})}\BibitemShut {NoStop}%
\bibitem [{\citenamefont {{Tiwari}}\ and\ \citenamefont
  {{Stroud}}(2007)}]{treejunfc}%
  \BibitemOpen
  \bibfield  {author} {\bibinfo {author} {\bibfnamefont {R.~P.}\ \bibnamefont
  {{Tiwari}}}\ and\ \bibinfo {author} {\bibfnamefont {D.}~\bibnamefont
  {{Stroud}}},\ }\href@noop {} {\bibfield  {journal} {\bibinfo  {journal}
  {Phys. Rev. B}\ }\textbf {\bibinfo {volume} {76}},\ \bibinfo {pages} {220505}
  (\bibinfo {year} {2007})}\BibitemShut {NoStop}%
\bibitem [{\citenamefont {{Hassler}}\ \emph {et~al.}(2011)\citenamefont
  {{Hassler}}, \citenamefont {{Akhmerov}},\ and\ \citenamefont
  {{Beenakker}}}]{transmons}%
  \BibitemOpen
  \bibfield  {author} {\bibinfo {author} {\bibfnamefont {F.}~\bibnamefont
  {{Hassler}}}, \bibinfo {author} {\bibfnamefont {A.~R.}\ \bibnamefont
  {{Akhmerov}}}, \ and\ \bibinfo {author} {\bibfnamefont {C.~W.~J.}\
  \bibnamefont {{Beenakker}}},\ }\href@noop {} {\bibfield  {journal} {\bibinfo
  {journal} {New Journal of Physics}\ }\textbf {\bibinfo {volume} {13}},\
  \bibinfo {pages} {095004} (\bibinfo {year} {2011})}\BibitemShut {NoStop}%
\bibitem [{\citenamefont {{Clarke}}\ and\ \citenamefont
  {{Wilhelm}}(2008)}]{naturepics}%
  \BibitemOpen
  \bibfield  {author} {\bibinfo {author} {\bibfnamefont {J.}~\bibnamefont
  {{Clarke}}}\ and\ \bibinfo {author} {\bibfnamefont {F.~K.}\ \bibnamefont
  {{Wilhelm}}},\ }\href@noop {} {\bibfield  {journal} {\bibinfo  {journal}
  {Nature Physics}\ }\textbf {\bibinfo {volume} {453}},\ \bibinfo {pages}
  {1031} (\bibinfo {year} {2008})}\BibitemShut {NoStop}%
\bibitem [{\citenamefont {{Budich}}\ \emph {et~al.}(2012)\citenamefont
  {{Budich}}, \citenamefont {{Walter}},\ and\ \citenamefont
  {{Trauzettel}}}]{trauzettel}%
  \BibitemOpen
  \bibfield  {author} {\bibinfo {author} {\bibfnamefont {J.~C.}\ \bibnamefont
  {{Budich}}}, \bibinfo {author} {\bibfnamefont {S.}~\bibnamefont {{Walter}}},
  \ and\ \bibinfo {author} {\bibfnamefont {B.}~\bibnamefont {{Trauzettel}}},\
  }\href@noop {} {\bibfield  {journal} {\bibinfo  {journal} {Phys. Rev. B}\
  }\textbf {\bibinfo {volume} {85}},\ \bibinfo {pages} {121405(R)} (\bibinfo
  {year} {2012})}\BibitemShut {NoStop}%
\bibitem [{\citenamefont {{Goldstein}}\ and\ \citenamefont
  {{Chamon}}(2011)}]{goldstein}%
  \BibitemOpen
  \bibfield  {author} {\bibinfo {author} {\bibfnamefont {G.}~\bibnamefont
  {{Goldstein}}}\ and\ \bibinfo {author} {\bibfnamefont {C.}~\bibnamefont
  {{Chamon}}},\ }\href@noop {} {\bibfield  {journal} {\bibinfo  {journal}
  {Phys. Rev. B}\ }\textbf {\bibinfo {volume} {84}},\ \bibinfo {pages} {205109}
  (\bibinfo {year} {2011})}\BibitemShut {NoStop}%
\bibitem [{\citenamefont {{Rainis}}\ and\ \citenamefont {{Loss}}(2012)}]{loss}%
  \BibitemOpen
  \bibfield  {author} {\bibinfo {author} {\bibfnamefont {D.}~\bibnamefont
  {{Rainis}}}\ and\ \bibinfo {author} {\bibfnamefont {D.}~\bibnamefont
  {{Loss}}},\ }\href@noop {} {\bibfield  {journal} {\bibinfo  {journal} {Phys.
  Rev. B}\ }\textbf {\bibinfo {volume} {85}},\ \bibinfo {pages} {174533}
  (\bibinfo {year} {2012})}\BibitemShut {NoStop}%
\bibitem [{\citenamefont {{Higginbotham}}\ \emph {et~al.}(2015)\citenamefont
  {{Higginbotham}}, \citenamefont {{Albrecht}}, \citenamefont {{Kir{\v
  s}anskas}}, \citenamefont {{Chang}}, \citenamefont {{Kuemmeth}},
  \citenamefont {{Krogstrup}}, \citenamefont {{Jespersen}}, \citenamefont
  {{Nyg{\aa}rd}}, \citenamefont {{Flensberg}},\ and\ \citenamefont
  {{Marcus}}}]{10ms}%
  \BibitemOpen
  \bibfield  {author} {\bibinfo {author} {\bibfnamefont {A.~P.}\ \bibnamefont
  {{Higginbotham}}}, \bibinfo {author} {\bibfnamefont {S.~M.}\ \bibnamefont
  {{Albrecht}}}, \bibinfo {author} {\bibfnamefont {G.}~\bibnamefont {{Kir{\v
  s}anskas}}}, \bibinfo {author} {\bibfnamefont {W.}~\bibnamefont {{Chang}}},
  \bibinfo {author} {\bibfnamefont {F.}~\bibnamefont {{Kuemmeth}}}, \bibinfo
  {author} {\bibfnamefont {P.}~\bibnamefont {{Krogstrup}}}, \bibinfo {author}
  {\bibfnamefont {T.~S.}\ \bibnamefont {{Jespersen}}}, \bibinfo {author}
  {\bibfnamefont {J.}~\bibnamefont {{Nyg{\aa}rd}}}, \bibinfo {author}
  {\bibfnamefont {K.}~\bibnamefont {{Flensberg}}}, \ and\ \bibinfo {author}
  {\bibfnamefont {C.~M.}\ \bibnamefont {{Marcus}}},\ }\href@noop {} {\bibfield
  {journal} {\bibinfo  {journal} {Nature Physics}\ }\textbf {\bibinfo {volume}
  {11}},\ \bibinfo {pages} {1017} (\bibinfo {year} {2015})}\BibitemShut
  {NoStop}%
\end{thebibliography}

%


\end{document}